\documentclass[aps,prl,onecolumn,superscriptaddress,notitlepage,secnumarabic]{revtex4-2}
\usepackage{amsmath}
\usepackage{amssymb}
\usepackage{graphicx}
\usepackage{subcaption}
\usepackage{color}
\usepackage{booktabs}

\usepackage[justification=raggedright,singlelinecheck=false]{caption}

\makeatletter
\def\section{%
  \@startsection{section}{1}{\z@}{-3.5ex \@plus -1ex \@minus -.2ex}%
  {2.3ex \@plus.2ex}{\normalfont\large\bfseries\raggedright}%
}
\def\subsection{%
  \@startsection{subsection}{2}{\z@}{-3.25ex\@plus -1ex \@minus -.2ex}%
  {1.5ex \@plus .2ex}{\normalfont\large\bfseries\raggedright}%
}
\def\subsubsection{%
  \@startsection{subsubsection}{3}{\z@}{-3.25ex\@plus -1ex \@minus -.2ex}%
  {1.5ex \@plus .2ex}{\normalfont\normalsize\bfseries\raggedright}%
}
\makeatother

\begin{document}

\title{The scales of disorder in perfect quasicrystals}

\author{Alan Rodrigo Mendoza Sosa}
\email{alanmendoza@iphy.ac.cn}
\affiliation{Key Laboratory of Soft Matter Physics, Institute of Physics, Chinese Academy of Sciences, Beijing 100190, China}

\author{Atahualpa S.~Kraemer}
\email{ata.kraemer@ciencias.unam.mx}
\affiliation{Departamento de F\'isica, Facultad de Ciencias, Universidad Nacional Aut\'onoma de M\'exico, Ciudad Universitaria 04510, Mexico City, Mexico}

\author{Michael Schmiedeberg}
\email{michael.schmiedeberg@fau.de}
\affiliation{Soft Matter Theory Group, Theoretical Physics: Lab for Emergent Phenomena, Friedrich-Alexander-Universit\"at Erlangen-N\"urnberg, 91058 Erlangen, Germany}

\author{Erdal C.~O\u{g}uz}
\email{ecoguz@iphy.ac.cn}
\affiliation{Key Laboratory of Soft Matter Physics, Institute of Physics, Chinese Academy of Sciences, Beijing 100190, China}

\begin{abstract}
The classical dichotomy between crystalline order and amorphous disorder is increasingly challenged by novel states that lack conventional crystalline symmetries while retaining crystal-like properties \cite{Fan2026,Corwin2026,Wang2025,Casiulis2025,Torquato2015,Klatt2022}. Quasicrystals occupy a distinctive position within this expanding framework by possessing long-range order without translational periodicity, thereby permitting arbitrary $N$-fold rotational symmetry~\cite{levine1984quasicrystals,shechtman1984metallic}. Paradoxically, far from their unique symmetry center, high-symmetry quasicrystals closely resemble disordered patterns \cite{sosa2023structural}, raising the question of how deterministic order can be detected. Here we show that increasing rotational symmetry progressively suppresses local statistical signatures of quasiperiodicity, while preserving its underlying exact long-range order. This order is thus concealed below an emergent crossover length that grows linearly with $N$. 
Therefore, as $N \rightarrow \infty$, the disorder-like regime expands without bound, defining a symmetry-controlled geometric critical point at which deterministic order and randomness become statistically indistinguishable over any finite observation window. For finite $N$, however, quasiperiodic order becomes detectable beyond this crossover, revealing a second emergent length scale that we identify as the size of a \textit{statistical unit cell} -- finite patches over which statistical properties recur despite the absence of conventional translational periodicity. In one dimension, the statistical-unit-cell size coincides with the crossover length, whereas in two dimensions it grows as $N^2$, remaining smaller than the size of typical approximants and establishing a hierarchy of emergent length scales. Together, the disorder-to-order crossover and statistical unit cells provide a quantitative framework connecting crystals, quasicrystals, and amorphous matter, showing how apparent disorder can emerge from purely deterministic geometry.
\end{abstract}

\maketitle
\let\oldaddcontentsline\addcontentsline
\renewcommand{\addcontentsline}[3]{}

\section{Introduction}\label{sec1}

As structures with long-range order but no translational periodicity, quasicrystals stand apart from both conventional periodic crystals and amorphous matter, embodying a distinct form of structural order.
They have been identified across a broad spectrum of systems: 
in metallic alloys~\cite{shechtman1984metallic},
in colloidal suspensions~\cite{fischer2011colloidal, Mirkin2024} stabilized by competing length scales~\cite{Dotera2014, Engel2015, Smallenburg2024} or preferred binding angles~\cite{Pinto2025, Noya2025}, in nanoparticle assemblies~\cite{talapin2009quasicrystalline, Marrows2018},
in superconducting quasicrystalline systems~\cite{Kamiya2018,Uri2023}, and even in works of arts and architecture~\cite{LU2007}. Quasicrystals exhibit a range of distinctive material properties, including exceptional  brittleness~\cite{Jang}, low-friction behavior~\cite{Park}, and fracture-repair mechanisms~\cite{Trebin1998}. Many of these properties have been linked to phasons~\cite{phason1}--additional degrees of freedom that are unique to quasicrystals and become increasingly important with increasing rotational symmetry $N$~\cite{baake2013, mikhael2010proliferation}. Since quasicrystals can, in principle, possess any rotational symmetry $N$~\cite{levine1984quasicrystals,baake2013}, the number of phason modes is likewise unbounded and can be increased systematically by constructing higher-symmetry quasicrystals.

Yet high-symmetry quasicrystals present a striking paradox: although perfectly ordered by construction, typical finite patches far from the global symmetry center closely resemble random tilings~\cite{sosa2022efficient}. This raises a fundamental question: can deterministic order become statistically indistinguishable from randomness, and if so, at what length scale does quasiperiodic order become detectable? Conventional rational approximants~\cite{goldman1993quasicrystals, matsubara2024aperiodic}, whose size grows exponentially with symmetry, provide limited insight, leaving these characteristic scales largely unresolved.

Here we show that high-symmetry quasicrystals exhibit a symmetry-controlled crossover between disorder-like and crystalline behavior governed by two emergent length scales: a crossover length and an additional statistical unit cell. For structures with $N$-fold rotational symmetry in two dimensions the crossover length grows linearly with $N$ while the statistical unit cell grows quadratically; for corresponding quasicrystals in one dimension the two coincide. Below the crossover length, density fluctuations mimic those of disordered systems; only beyond the statistical unit cell does crystal-like order become statistically detectable. In the limit $N\to\infty$, the disorder-like regime expands without bound and local statistics converge to those of a random, uncorrelated arrangement over any finite observation window, even as exact long-range order is preserved. High-symmetry quasicrystals thus approach a geometric critical point where randomness emerges as an asymptotic limit of deterministic geometry, establishing a symmetry-controlled framework that bridges periodic crystals, quasicrystals, and amorphous matter.

\begin{figure}[h]
    \centering
    \includegraphics[width=1\textwidth]{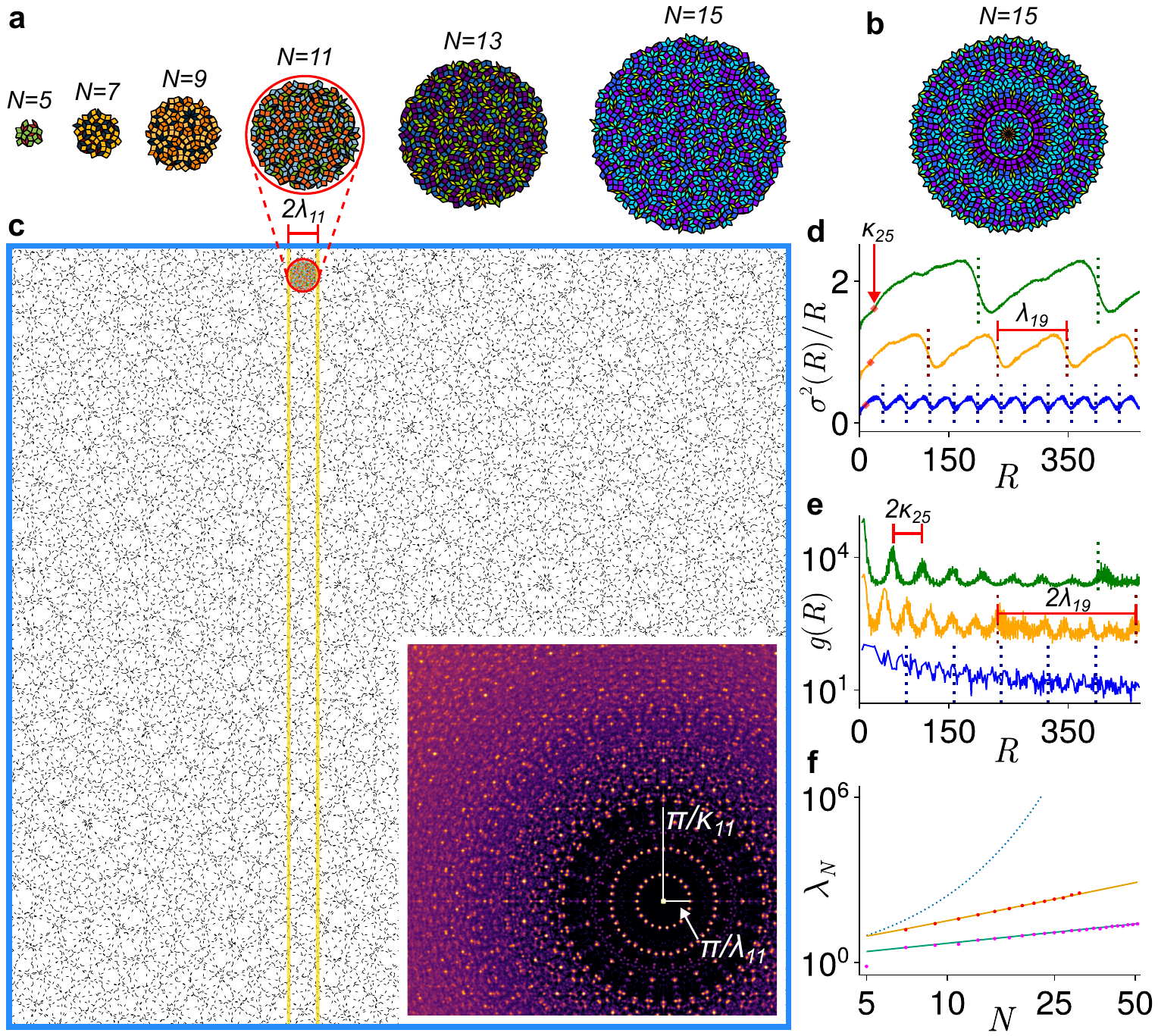}
    \caption{
    \textbf{Statistical unit cells and crossover length scale.}
    (a) Representative statistical unit cells of radius $R=\lambda_N$, extracted far from the exact symmetry center for two-dimensional quasiperiodic tilings with rotational symmetry $N$. (b) A same-sized neighborhood centered on the exact symmetry center for $N=15$, where rotational order remains visually apparent. (c) Portion of the $N=11$ quasicrystal showing only the smallest-area tiles. The blue box indicates the size of a classical approximant, while the red circle marks a statistical unit cell. The inset in the bottom-right corner displays the corresponding diffraction pattern, with characteristic wavevectors associated with $\lambda_N$ and the crossover scale $\kappa_N$ indicated. (d) Rescaled number variance $\sigma^2(R)/R$ and (e) pair correlation function $g(R)$ for two-dimensional quasiperiodic systems with rotational symmetry $N = 11$ (blue), $N = 19$ (orange), and $N = 25$ (green), with the vertical dotted lines indicating their $\lambda_N$, and diamonds in (d) marking the corresponding crossover scales $\kappa_N$. The curves are vertically offset for clarity. (f) Statistical-unit-cell scale $\lambda_N$ as a function of $N$ for two-dimensional quasicrystals (numerical: red filled circles; theoretical curve as given in the text without any fit parameters: orange solid line) and one-dimensional quasicrystals (numerical: purple filled circles; fitted curve proportional to $N$: green solid line), compared with the size of conventional rational approximants in two dimensions that approximately grows exponentially (blue dotted line).
    }
    \label{fig1}
\end{figure}
 
\section{Results}\label{sec2} 

\subsection{Crossover length and statistical recurrence}

To produce one- and two-dimensional $N$-fold quasicrystals, we employ the generalized dual method \cite{DeBruijn1981algebraic, socolar1986quasicrystals} through a high-performance implementation \cite{sosa2022efficient} capable of generating large-scale realizations with arbitrarily high rotational symmetry (Methods).

Typical quasicrytalline patches far from the exact symmetry center are show in Fig.~1a. Although each patch is perfectly deterministic, increasing symmetry progressively obscures local motifs, in sharp contrast to the neighborhood of the symmetry center (Fig.~1b), where rotational order remains visually apparent.

Figure~1c shows a conventional rational approximant of the $N=11$ quasicrystal visualized through its smallest tiles. The typical tile spacing along a line and the mesh size of the network are related to the two characteristic lengths, $\kappa_N$ and $\lambda_N$, discussed in the following. 

These length scales are also reflected in the density fluctuations (Figs.~1d,e). The rescaled number variance $\sigma^2(R)/R$, measured with a circular window of radius $R$, exhibits a crossover near $\kappa_N$ (diamonds in Fig. 1d): below this scale density fluctuations are weakly hyperuniform (Class III), while beyond it pronounced oscillations indicating statistical repetitions with a wavelength $\lambda_N$ emerge, marking the onset of statistical detectability of strongly hyperuniform (Class I), crystal-like order~\cite{torquato2018hyperuniform,oguz2017hyperuniformity, koga2024hyperuniformity}. 
Quantitatively, we extracted $\lambda_N$ as the dominant oscillation period of $\sigma^2(R)/R$ via a Fast Fourier Transform of the data. The envelope of the pair correlation function $g(R)$ develops corresponding oscillations on scales of order $2\kappa_N$, which we used to determine $\kappa_N$ (see SI), together with enhanced correlations at multiples of $2\lambda_N$ (Fig. 1e).

Both scales are also visible in the structure factor (inset of Fig.~1c). Within the central region of radius $\pi/\kappa_N$
only a few pronounced Bragg peaks appear, arranged on rings at integer multiples of $\pi/\lambda_N$, while weaker peaks remain indistinguishable from the background. This indicates an emerging order on length scales larger than $2\kappa_N$, dominated by a characteristic modulation with wavelength $2\lambda_N$. Outside this region the contrast between dominant and background peaks is substantially reduced, suggesting that below $2\kappa_N$, the underlying quasiperiodic order is much less readily discernible -- consistent with the disorder-like statistical regime identified from the density fluctuations. Further examples, including one-dimensional quasicrystals, are given in the Supplementary Information.

The patches in Fig.~1a are circular regions of radius $\lambda_N$ and are thus  statistically representative of the whole quasicrystal,  allowing us to identify $\lambda_N$ as the statistical-unit-cell size. As seen in Fig.~1c, such a statistical unit cell is substantially smaller than the full approximant, a size difference that becomes increasingly pronounced with $N$. Statistical recurrence therefore emerges far below the scales required for periodic rational approximation.

While the size of a typical rational approximant grows exponentially with $N$, the statistical unit cell scale grows only algebraically: $\lambda_N \approx 4\pi/p_N$ in two dimensions, where $p_N= 2 \big( 1-\cos (\pi/N) \big)$ is the probability to find the smallest-area tiles \cite{sosa2023structural}, giving $\lambda_N \propto N^2$ asymptotically (Fig. 1f), and $\lambda_N \propto N$ in one dimension. The crossover length scales as $\kappa_N \propto N$ in both dimensions, coinciding with $\lambda_N$ only in one dimension.

\subsection{Origin of emergent length scales}

To identify the microscopic origin of the emergent scales $\kappa_N$ and $\lambda_N$, and to motivate their dependence on $N$ as derived in the previous section, we decompose density fluctuations into contributions from different prototile families (Figs. 2a–d, $N=23$). Their spatial organization reflects the underlying dual-grid construction, in which all tiles lie on intersecting families of parallel lines. The mean separation between equivalent tiles along any such line equals $2\kappa_N$ by construction and is independent of the tile type (Fig.~2e), demonstrating that the onset of Class-I hyperuniform order is a global property of the dual geometry, with $\kappa_N$ scaling linearly with $N$~\cite{sosa2022efficient}.

By contrast, the long-range oscillatory structure depends strongly on tile rarity: oscillation wavelength and amplitude increase systematically as tile frequency decreases, with the rarest tiles -- the thinnest rhombi in two dimensions, the longest segments in one -- recovering the full-system wavelength $\lambda_N$, and setting therefore the statistical-unit-cell scale. The distinct microscopic origins of the two scales thus explain their different scaling: $\kappa_N$ reflects the universal line geometry of the dual construction, while $\lambda_N$ is set by the increasingly sparse distribution of the rarest prototiles.

\subsection{Routes to randomness}

The role of disorder is explored in Fig.~3 by comparing perfect quasicrystals with structures obtained after random local phasonic flips -- rearrangements of tiles at a vertex in two dimensions or interchanges of neighboring segments in one dimension (insets of Figs.~3a,b). In both cases, randomization progressively suppresses oscillations in the number variance (Figs.~3a,b), confirming their quasiperiodic origin.

The effect of randomization differs qualitatively between dimensions. In two dimensions, geometric compatibility conditions constrain random rearrangements, suppressing the long-range oscillatory structure while leaving the scaling behavior above $\kappa_N$ essentially unchanged, resulting in a \textit{hyperuniform random tiling}. In one dimension, repeated flipping produces a \textit{Poissonian-like random tiling} with $\sigma^2(R)\sim R$ (dotted line, Fig.~3b), providing an operational reference for the local statistics of high-symmetry quasicrystals.

A similar statistical evolution is observed as rotational symmetry increases (Fig. 3c,d). In both dimensions, increasing $N$ shifts the characteristic scales to larger distances and raises the local exponent $\beta(N)$ extracted from the density fluctuations below the crossover $\kappa_N$. Fitting $\beta(\epsilon)$, where $\epsilon=1/N$,  
to $a_0\exp(a_1\epsilon^{a_2})$ yields $a_0\approx d$, demonstrating convergence toward the Poissonian-like limit $\beta(\epsilon\to0)\rightarrow d$ (insets of Fig. 3c,d); alternative fit functions in the SI confirm this asymptotic behavior. The Poissionian-like behavior is rooted in the growing number of prototiles required as rotational symmetry increases, driving a concomitant increase in local structural complexity~\cite{sosa2023structural}.

These results reveal two distinct pathways toward apparent randomness. Phasonic flips produce hyperuniform random tilings in two dimensions and Poissonian-like random tilings in one. Increasing $N$ follows a qualitatively different route, driving local statistics toward Poissonian-like behavior in both dimensions alike
without any tile rearrangements. The resulting structures remain perfectly deterministic -- not random tilings -- yet their local statistics become indistinguishable from a Poisson process statistics at all accessible scales, defining a regime inaccessible through conventional randomization in two dimensions.

\subsection{Infinite-symmetry scaling collapse}

The approach to the infinite-symmetry limit is governed by a symmetry-controlled scaling collapse. Introducing $\epsilon=1/N$, the limit $\epsilon \rightarrow 0$ defines a geometric critical point at which the statistical-unit-cell scale diverges. The number variance becomes invariant under $R \rightarrow R/\lambda_N$ and $\sigma^2 \rightarrow \sigma^2 / \Lambda_{\infty}\!(N)$, where $\Lambda_\infty(N)$ is the $N$-dependent asymptotic fluctuation amplitude. As shown in Fig. 4a, this rescaling produces an increasingly accurate collapse with growing $N$, already well developed for the modest symmetries studied here. The resulting scaling function connects the weakly-hyperuniform disorder-like regime at small rescaled distances to the asymptotic crystal-like regime at larger scales, with oscillations of wavelength $\lambda_N$ encoding statistical recurrence (Methods).

Both $\lambda_N$ and $\kappa_N$ diverge algebraically, $\lambda_N \sim \epsilon^{-\nu}$, with $\nu=1$ in one dimension and $\nu \approx 2$ in two. The pair-correlation function mirrors this collapse through an algebraic envelope decay (Fig.~4b), with pronounced oscillations on scale $2\kappa_N$ superimposed on enhanced correlations at integer multiples of $2\lambda_N$, identifying $\lambda_N$ as the fundamental long-range correlation scale of the point process, consistent with the statistical-unit-cell interpretation inferred from density fluctuations.

Together, these results establish $N \rightarrow \infty$ as a  
geometric limit in which deterministic quasiperiodic order becomes statistically indistinguishable from a random uncorrelated system over any finite observation window, with the statistical-unit-cell scale providing a quantitative measure of the length at which intrinsic quasiperiodic order becomes resolvable. 

\section{Discussion}

Since their discovery, quasicrystals have occupied an intermediate conceptual position between periodic crystals and statistically random structures. Our results place this intuition on a quantitative footing by revealing two coexisting statistical regimes separated by a symmetry-controlled crossover length $\kappa_N$. Below this scale, density fluctuations display weakly suppressed, random-tiling-like correlations; beyond it, quasiperiodic order becomes statistically detectable through strongly hyperuniform oscillatory fluctuations.

The oscillations originate from the sparse network formed by the thinnest tiles and occur on a length scale $\lambda_N$, defining a statistical unit cell: a finite region sufficient to recover the bulk statistical properties of the quasicrystal. Unlike a conventional unit cell, whose translational repetition generates the entire periodic structure, the statistical unit cell encodes quasiperiodic organization through statistical recurrence rather than geometric repetition, providing a structural descriptor with no counterpart in either periodic or amorphous systems. 

Unlike conventional rational approximants, whose size grows exponentially with rotational symmetry, $\lambda_N$ grows only algebraically, so that observables such as mean density, pair correlations, and number fluctuations converge far below the scales required for periodic approximation~\cite{goldman1993quasicrystals, matsubara2024aperiodic, socolar1986quasicrystals}. Statistical recurrence is therefore both conceptually distinct from, and far more accessible than, conventional approximant constructions.

We note that Steinhardt and Jeong introduced a quasi-unit-cell for specific low rotational symmetries quasicrystals~\cite{Steinhardt1996}, defined geometrically through a single overlapping cluster. Our statistical unit cell is defined instead through the recurrence of bulk statistical properties; whether the quasi-unit cell generalizes to arbitrary $N$ and how the two concepts relate remains an open question.

Both $\kappa_N$ and $\lambda_N$ diverge as $N \to \infty$, establishing a purely geometric critical point -- controlled by rotational symmetry rather than thermodynamic parameters -- at which 
deterministic quasiperiodic order becomes statistically indistinguishable from a Poisson process over any finite observation window, even as exact long-range order is preserved for all finite $N$.

High-symmetry quasicrystals may also serve as controlled models for low-temperature quantum glass anomalies. Localized phasonic flips connect nearly degenerate structural configurations, furnishing deterministic realizations of the double-well potentials underlying two-level systems in tunneling models of glasses~\cite{Anderson72, Phillips72}. The explicit tiling geometry allows, in principle, barrier heights, asymmetry energies, and elastic couplings to be computed exactly, while rotational symmetry provides a tunable parameter for the degree of disorder-like local statistics, potentially enabling systematic studies of how two-level-system spectra evolve as quasiperiodic structures approach the infinite-symmetry limit.

Beyond localized excitations, quasicrystals may offer insight into collective relaxation near the glass transition~\cite{glass_biroli, glass_ediger2000, glasses_Lerner, glass_spectrum_franz}, including its relationship to marginal stability and low-energy excitations~\cite{marginal_liu1, marginal_corwin1, marginal_wyart}. 
Unlike amorphous systems, quasicrystals possess well-defined phonon and phason basis~\cite{phason1} that remains explicitly calculable even as local statistics become Poissonian-like, making it possible to decompose collective rearrangements into phononic and phasonic contributions and quantify their spectral content systematically — positioning high-symmetry quasicrystals as analytically tractable surrogates for amorphous matter.

High-symmetry quasicrystals are experimentally accessible in optically induced colloidal systems~\cite{burns1990optical, mikhael2008archimedean, mikhael2010proliferation}, Bose–Einstein condensates~\cite{PhysRevLett.79.3363, Jagannathan_2013}, photosensitive metamaterials~\cite{Freedman2006}, and quantum moiré materials~\cite{Ahn2018}. The algebraic growth of $\lambda_N$  implies that disorder-like statistics can be realized at accessible system sizes while preserving exact quasiperiodic order, enabling direct experimental study of the crossover between regimes.

More broadly, our results show that deterministic order and statistical randomness need not be mutually exclusive: high-symmetry quasicrystals emerge as an analytically controllable framework for understanding how apparent disorder arises from deterministic geometry, opening new connections between order, randomness, and collective phenomena beyond the traditional boundaries of crystalline and amorphous matter.

\section{Methods}\label{}

\subsection{Construction of quasiperiodic point sets}

Two-dimensional quasiperiodic tilings with $N$-fold rotational symmetry were generated using a highly efficient implementation of the generalized dual method developed by some of the present authors \cite{sosa2022efficient}. 
This approach enables the construction of quasiperiodic lattices with arbitrary rotational symmetry $N$ at arbitrary positions. Large tiling patches were generated around randomly chosen positions sufficiently far from the global symmetry center to avoid artifacts associated with special symmetry points (cf.\ Fig. 1b).

\begin{figure}[h]
    \centering
    \includegraphics[width=1\textwidth]{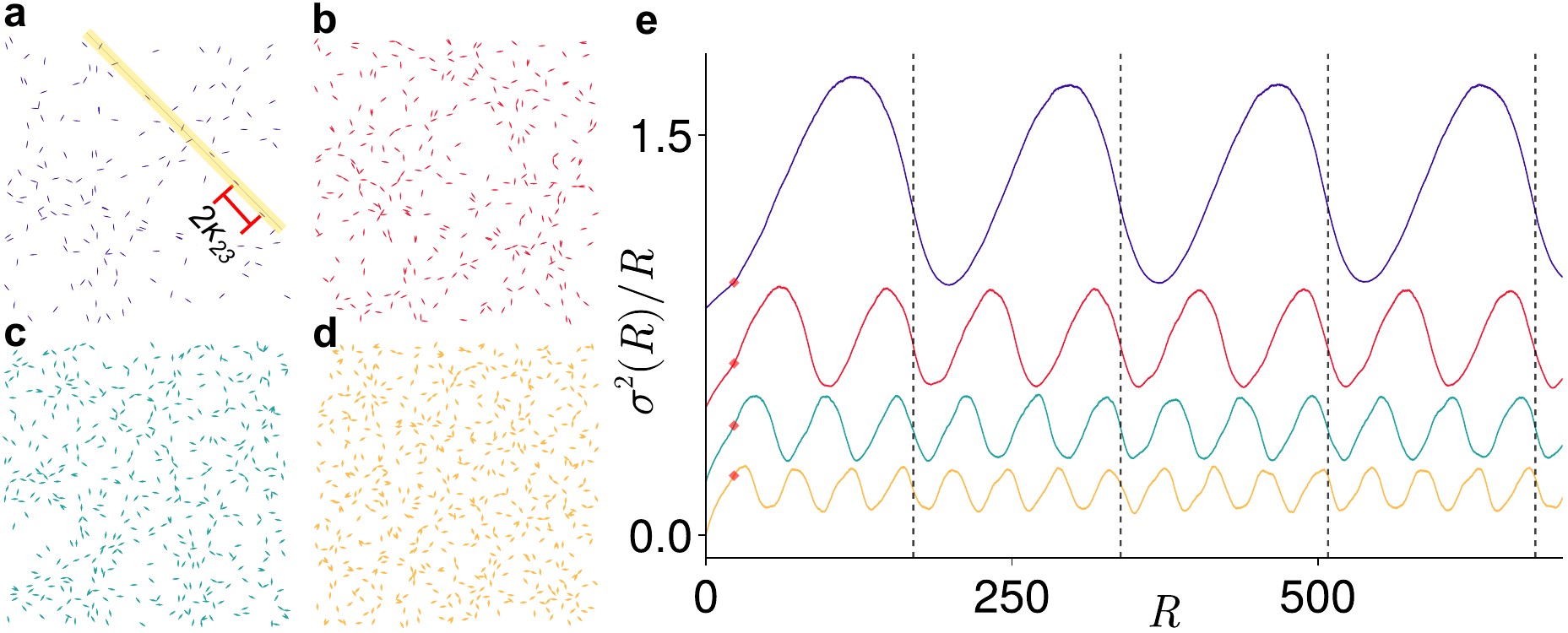}
    \caption{
    \textbf{Hierarchical tile contributions to statistical recurrence.}
    (a–d) Tile subsets of a two-dimensional quasiperiodic tiling with rotational symmetry $N=23$, separated according to increasing tile area. The sparsest subset in (a), formed by the smallest-area tiles, generates an extended network that dominates long-range correlations. (e) Number variance computed from centroid decorations of the corresponding subsets. Vertical dashed lines indicate integer multiples of the statistical-unit-cell scale $\lambda_{23}$, while diamonds mark the common crossover scale $\kappa_{23}$. All subsets exhibit a change in concavity near $\kappa_{23}$, indicating a shared onset of the long-range quasiperiodic order. Oscillation wavelength and amplitude increase systematically with tile rarity. The smallest-area tiles already reproduce the full statistical-unit-cell wavelength of the complete quasicrystal. Curves are vertically offset for clarity.
    }
    \label{Fig2}
\end{figure}

Point sets were obtained by placing particles at the vertices of the tilings. We additionally analyzed centroid decorations, in which particles are placed at the tile centroids, and found no qualitative differences in any of the statistical observables reported here. All results shown in the main text correspond to vertex-decorated tilings, except for the analysis of tile-specific number variance (cf.\ Fig.~\ref{Fig2}), for which centroid decorations were used to assign particles uniquely to individual tile classes.

One-dimensional quasiperiodic point sets were generated using the same construction principle. We first define $N$ star vectors and project them onto a reference axis, coinciding with one of the star-vector directions. System sizes were chosen such that the linear extent exceeded the length $\lambda_N$ studied for each symmetry.

By construction, the unit of length is set by the tile side length and is used for all distances analyzed throughout this work and the Supplementary Information. In two dimensions, distances are additionally rescaled by a factor of $2 \sqrt{\pi \rho}$, originally introduced in \cite{LinCorrigendum2017}, where $\rho$ denotes the number density of tile vertices. This normalization ensures a constant mean density across different quasicrystalline symmetries.

\subsection{Number variance and hyperuniformity}

Density fluctuations were quantified using the variance $\sigma^2(R)$, associated with the number of points contained within an observation window of size $R$. In one dimension, windows were intervals of length $2R$, while in two dimensions circular windows of radius $R$ were used. Window centers were chosen randomly within large quasiperiodic patches sufficiently far from the global symmetry center.

The number variance provides a standard criterion for hyperuniformity. In spatial dimension $d$, systems for which $\sigma^2(R)$ grows more slowly than the window volume $R^d$ are termed hyperuniform. Periodic crystals represent the strongest form of hyperuniformity (Class I), for which $\sigma^2(R)\sim R^{d-1}$, whereas uncorrelated (Poisson) point sets satisfy $\sigma^2(R)\sim R^d$. Intermediate scaling behaviour with exponents between $d-1$ and $d$ correspond to a weaker form of hyperuniformity (Class III). 

For each value of $R$, $\sigma^2(R)$ was computed by averaging over $10^4–10^5$ randomly placed windows. The rescaled quantity $\sigma^2(R)/R^{d-1}$ was used to identify deviations from asymptotic Class-I behavior and to determine the characteristic length scales discussed in the main text. Statistical uncertainties were estimated using block averaging over independent window samples. In addition, the characteristic oscillation wavelength was determined independently from Fourier analysis of the number variance (see Supplementary Information).

\subsection{Pair correlation function}

The pair correlation function $g(R)$ was computed from quasiperiodic point sets constructed within large circular regions whose radii substantially exceed $\lambda_N$ for the corresponding value of $N$. The regions were centered at randomly chosen points sufficiently far from the global symmetry center. In each realization, distances between a reference point near the center of the region and all other points were accumulated into a radial histogram.

Because the pair correlation function of a quasicrystal formally consists of a dense set of sharp peaks, we focus on its large-scale statistical structure by averaging over many independent realizations and analyzing the envelope of the resulting peak distribution. The envelope was extracted by constructing the
concave hull (see Supplementary Information). Histograms were computed using a fixed radial bin width and averaged over $10^4$ independent center positions to ensure convergence. The resulting distributions were normalized by the mean density to obtain $g(R)$. The large-scale scaling behavior of the envelope and its hierarchical organization are discussed in the main text and illustrated in Fig.~4b.

\subsection{Random tilings and local disorder}

Random tilings were generated by introducing local disorder through tile-flip operations that preserve tile shapes and global density. These flips locally rearrange the tiling connectivity without changing the prototile set, providing a controlled way to interpolate between perfect quasiperiodic and disordered configurations.

\begin{figure}[h]
    \centering
    \includegraphics[width=1\textwidth]{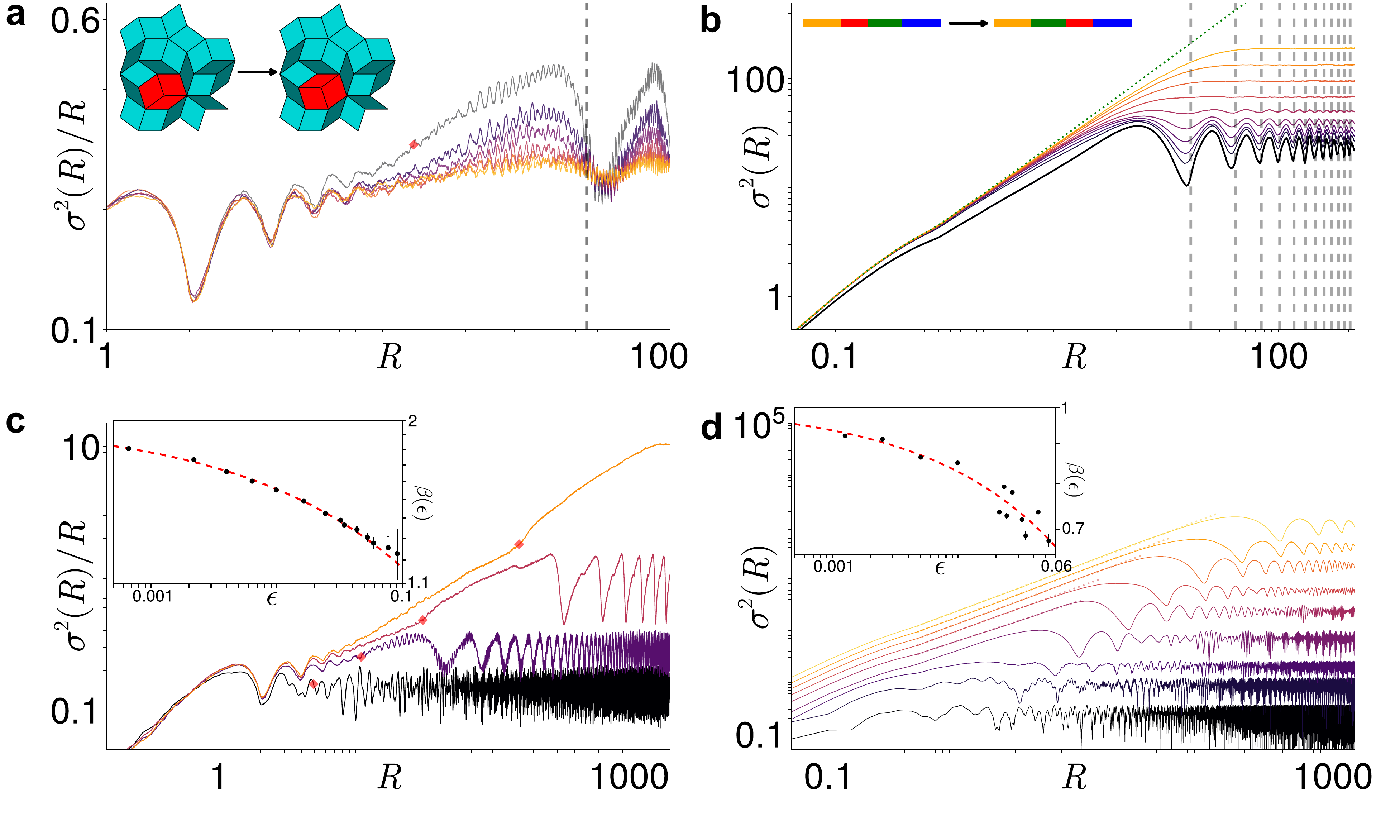}
    \caption{
    \textbf{Approaching Poissonian statistics through randomization and in the infinite-symmetry limit.}
    Log-log plots of number variance as a function of window radius $R$, comparing randomization-induced disorder (a,b) with increasing rotational symmetry (c,d).
    (a) Two-dimensional quasicrystal with $N=13$ subjected to increasing iterations of random local flips, ranging from unperturbed lattice (black) to maximally flipped state (yellow). 
    (b) One-dimensional quasicrystal with $N=51$ under repeated randomization. The dashed green curve denotes the fully randomized limit, which approaches Poissonian scaling while preserving the prototile distribution and provides an operational definition of a random tiling. Randomization leads in both dimensions to progressively suppressed long-range oscillations above the crossover scale.
    (c,d) Number variance for increasing rotational symmetry $N$ in two and one dimensions, respectively. In both cases, the local scaling exponent below the crossover scale increases systematically toward the Poissonian value $d$ for $N\to\infty$, reproducing the statistical trend observed under randomization in one dimension. Insets show fits of $\beta(\epsilon)$, with $\epsilon=1/N$, extrapolating to $\beta\to2$ in two dimensions and $\beta\to1$ in one dimension. Curves are vertically offset for clarity.
    }
    \label{Fig3}
\end{figure}

In one dimension, a tile flip consists of selecting a vertex and exchanging the two segments meeting at that vertex (inset of Fig.~\ref{Fig3}b). During a single sweep, $V$ flip attempts are performed, where $V$ is the total number of vertices. For each attempt, a vertex is selected independently and uniformly at random from the vertices $2,\ldots,V-1$, and the corresponding tile exchange is carried out. Consequently, a given sweep does not guarantee that every vertex is visited exactly once: some vertices may be selected multiple times, whereas others may not be selected at all. Successive sweeps progressively randomize the tiling while conserving the tile composition. For each symmetry $N$, up to $10^4$ sweeps were performed, and the number variance was evaluated at selected intervals to monitor the evolution of density fluctuations.

In two dimensions, flips are restricted by geometric compatibility and can only be performed at vertices where three edges meet (inset of Fig.~\ref{Fig3}a), corresponding to local rearrangements of rhombic tiles that preserve the tiling rules. As a result, only a subset of vertices can be flipped, and the randomization remains partial even after many sweeps.

To construct a fully randomized reference state in one dimension (dotted line in Fig.~\ref{Fig3}b), we additionally generated sequences of tiles by random sampling from the prototile frequency distribution measured in a quasiperiodic segment. These tiles were concatenated to produce configurations with identical composition but without quasiperiodic correlations. This procedure produces  point sets with Poisson-like number fluctuations, i.e., $\sigma^2(R)\sim R$. 

For all cases, the number variance $\sigma^2(R)$ was computed to characterize the evolution of density fluctuations. The local scaling exponent $\beta(\epsilon=1/N)$, defined via $\sigma^2(R)\sim R^{\beta}$ below the crossover scale $\kappa_N$, was extracted from power-law fits (Supplementary Information).

\subsection{Scaling analysis}

The scaling behavior is extracted from observables that exhibit pronounced oscillations as a function of distance. In particular, the number variance $\sigma^2(R)$ does not follow a simple power law with a constant prefactor. Instead, for window sizes larger than the crossover length $\kappa_N$, it can be expressed as $\sigma^2(R;N)\sim A(R;N)\,R^{d-1}$, where $A(R;N)$ is a bounded function that oscillates with wavelength $\lambda_N$.

To characterize the large-scale behavior, we define $\Lambda_\infty(N)$ as the cycle-averaged amplitude of $\sigma^2(R)/R^{d-1}$ for $R\gtrsim \lambda_N$, obtained by averaging over several oscillation periods. This quantity measures the amplitude of boundary-dominated density fluctuations at large scales and is evaluated numerically for different values of $N$:  
$\Lambda_\infty(N)\sim N^{\gamma}$, with $\gamma \approx 1.658$ in two dimensions and $\gamma \approx 1.425$ in one dimension.

The length scale $\lambda_N$ is determined independently from the oscillation period of $\sigma^2(R)/R^{d-1}$, its Fourier spectrum, and the pair correlation function. Using the symmetry-control parameter $\epsilon = 1/N$, we find that $\lambda_N \sim \epsilon^{-\nu}$, with the asymtotic exponent $\nu = 2$ in two dimensions and $\nu = 1$ in one dimension. 

Combining these observations, the number variance obeys the scaling form
\begin{equation}
    \sigma^2(R;N)=R^{d-1}\,\Lambda_\infty(N)\, f\!\left(\frac{R}{\lambda_N}\right) 
    = R^{d-1}\,\epsilon^{-\gamma} f\!\left(\frac{R}{\epsilon^{-\nu}}\right)
\end{equation}
where the scaling function $f(x)$ is oscillatory and bounded, reflecting quasiperiodic order. For $x\gg1$, $f(x)$ approaches a constant mean value upon averaging over oscillations, while for $x\ll1$ it captures the disordered-like regime below the stochastic unit-cell scale.

\begin{figure}[h]
    \centering
    \includegraphics[width=1\textwidth]{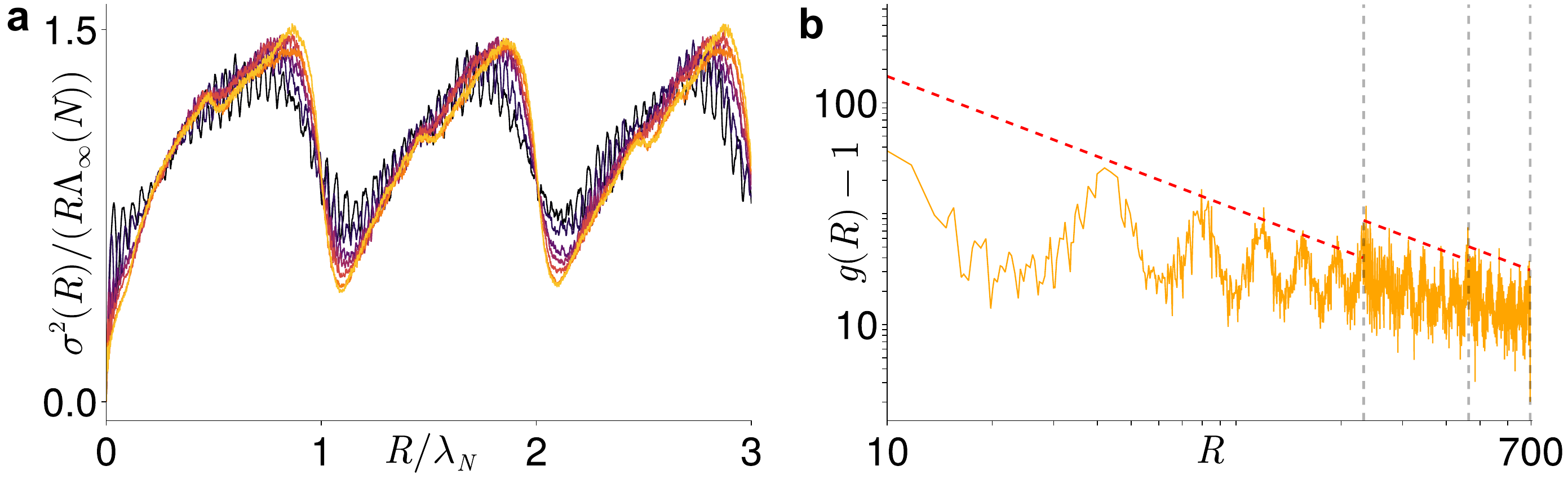}
    \caption{
    \textbf{
    Data collapse and scaling of the hierarchical statistical recurrence.}
    (a) Rescaled number variance $\sigma^2(R)/(R \Lambda_\infty(N))$ for two-dimensional quasicrystals with prime rotational symmetries $11\leq N\leq31$, plotted as a function of the scaled radius $R/\lambda_N$. Curves are colored from black to yellow, corresponding to increasing symmetry. The collapse onto a common master curve demonstrates symmetry-controlled scaling invariance. The quality of the collapse improves systematically with increasing $N$, while remaining accurate already for the finite symmetries considered here.
    (b) Envelope of $g(R)-1$ shown on log-log scales for $N=19$. 
    Peaks separated by $2\kappa_N$ exhibit algebraic decay, as indicated by the red dashed guide lines. The same decay law appears to repeat across successive intervals of width $2\lambda_N$, revealing hierarchical statistical recurrence on the scale of the statistical unit cell.
    }
    \label{Fig4}
\end{figure}

\section*{Acknowledgements}\label{Acknowledgements}
We thank Ke Chen and Hartmut Löwen for illuminating discussions. E.C.O. acknowledges financial support from the National Natural Science Foundation of China under Grant Nos. 12374221 and 12361141816. M.S. acknowledges support from the Deutsche Forschungsgemeinschaft (DFG, Grant No. 541211648). 
A.S.K. acknowledges support from the PAPIIT project IN113923 and from the Chinese Academy of Sciences through the President’s International Fellowship Initiative for Visiting Scientists (Grant No. 2026PVA0249).  
A.R.M.S. acknowledges support from the Beijing Natural Science Foundation through the International Scientists Project (Grant No. IS25042).
A.R.M.S. and A.S.K. further acknowledge access to the high-performance computing facilities of the Laboratorio de Cómputo de Alto Rendimiento, coordinated by the Departamento de Matemáticas, Facultad de Ciencias, National Autonomous University of Mexico.

\section*{Additional Information}

\subsection*{Competing interests}
The authors declare no competing interests.

\section*{Data availability and reproducibility}
All simulations were performed using independently written codes and verified by cross-checking results obtained from different implementations of the generalized dual method. Parameters used in the simulations are provided in the Supplementary Information. Data supporting the findings of this study are available from the corresponding author upon reasonable request.
The code used in the present work has been written in JULIA and can be downloaded under https://github.com/AlanRodrigoMendozaSosa/Quasiperiodic-Tiles

\clearpage
\newpage

\begin{center}
  \textbf{\large Supplementary Information for: \\[0.5em] 
  The scales of disorder in perfect quasicrystals}
\end{center}

\setcounter{equation}{0}
\setcounter{figure}{0}
\setcounter{table}{0}
\setcounter{page}{1}
\setcounter{section}{0}

\makeatletter
\renewcommand{\theequation}{S\arabic{equation}}
\renewcommand{\thefigure}{S\arabic{figure}}
\renewcommand{\thetable}{S\arabic{table}}
\renewcommand{\thesection}{S\arabic{section}}
\renewcommand{\theHfigure}{S\arabic{figure}}
\renewcommand{\theHequation}{S\arabic{equation}}
\renewcommand{\theHtable}{S\arabic{table}}
\makeatother

\let\addcontentsline\oldaddcontentsline

\tableofcontents

\section{Two-dimensional quasicrystalline systems}

\subsection{Construction principles}

Two-dimensional quasiperiodic tilings with rotational symmetry $N$ are generated using our implementation~\cite{sosa2022efficient} of the Generalized Dual Method (GDM), with the set of star vectors $\mathbf{S}_{2D} = \{\vec{S}_i\}$, where
$$\vec{S}_{i} = \left( \cos \left( \frac{2 \pi (i - 1)}{N}  \right), \sin \left( \frac{2 \pi (i - 1)}{N}  \right) \right), \ i \in \{1, 2, \dots, N\}.$$
The GDM offset parameter is set to $\alpha_{i} = 0$ throughout, and all length scales are reported in the units defined in the main text. Unless stated otherwise, tilings are decorated by placing a single point at each vertex; these points are referred to as \emph{sites}.

\subsection{Computing the number variance $\sigma^2(R)$}
\label{sec:Variance_Algorithm_2D}

The number variance $\sigma^2(R)$ is computed for two-dimensional quasiperiodic systems as follows. An arbitrary point $\vec{P} \in \mathbb{R}^2$ is drawn uniformly at random from a square of half-length $10^6$ centered at the origin, and a local circular neighborhood of radius $R_\text{max}$ is constructed around $\vec{P}$ from a quasiperiodic tiling with rotational symmetry $N$.

A circular window $\mathbf{W}$ centered at $\vec{P}$ is then expanded in steps of $\Delta R$, from  $R = \Delta R$ to $R_{max}$. This procedure is repeated for $10^{4}$ independent realizations of $\vec{P}$, yielding ensemble averages $\langle n(R) \rangle$ and $\langle n(R)^2 \rangle$, from which the number variance is obtained as $\sigma^{2}(R) = \langle n(R)^2 \rangle  - \langle n(R) \rangle^{2}$. 

\subsection{Extraction of the local scaling exponent $\beta(\epsilon)$}

To quantify the local scaling behavior of density fluctuations, we determine the exponent $\beta(\epsilon)$, $\epsilon=1/N$ from the growth of the number variance,
$\sigma^2(R)\sim R^{\beta}$, within the disorder-like regime below the crossover scale $\kappa_N$.

Because the crossover at $\kappa_N$ is gradual, including data too close to $\kappa_N$ can bias the fitted exponent through the onset of curvature in the log–log representation of $\sigma^2(R)$. We therefore restrict the analysis to $R<0.8\,\kappa_N$. At very small distances, the number variance is dominated by nearest-neighbor structure and deviates strongly from asymptotic scaling; the fitting procedure is consequently begun at $\sigma^2(R)\approx 2$. All power-law fits are  performed within the interval $\sigma^2(R)\approx 2$ to $R=0.8\,\kappa_N$.

Within this interval, we perform linear fits in the $\log \sigma^2$-$\log R$ representation using a sliding fitting window of fixed width, with the window spanning from the third minimum of the short-range oscillatory structure (see Fig. 3 of main text) to the upper fitting boundary at $0.8\,\kappa_N$. Sliding this window through the fitting interval yields a distribution of fitted exponents; $\beta$ is reported as their mean, with the standard deviation as the uncertainty. The analysis was repeated for three different window widths: larger windows produce somewhat larger uncertainties, but the resulting values of $\beta$ remain quantitatively consistent and do not affect any of the conclusions reported in this work.

The same procedure is applied in both one and two dimensions, with fitting intervals adapted to the respective crossover scales.  Figure~S1a shows $\beta(\epsilon)$ versus $\epsilon=1/N$ for prime (black) and non-prime (orange) values of $N$ in two dimensions, both converging toward $\beta = 2$ as $\epsilon\to 0$. Replotting the deviation $2-\beta$ against the prime-counting function $\pi(N)$, i.e., the number of prime integers up to $N$, in Fig.~S1b reveals a clear power-law convergence. Introducing $n=\pi(N)$ as the prime index, the data for prime symmetries are well described by $2-\beta(N)\sim n^{-0.379}$, providing direct evidence for $\beta\to 2$ 
in the infinite-symmetry limit -- a convergence guaranteed by the infinitude of primes, which ensures that the random limit is approached along a dense sequence of symmetries.

\begin{figure}[h!]
    \centering
    \includegraphics[width=0.8\linewidth]
    {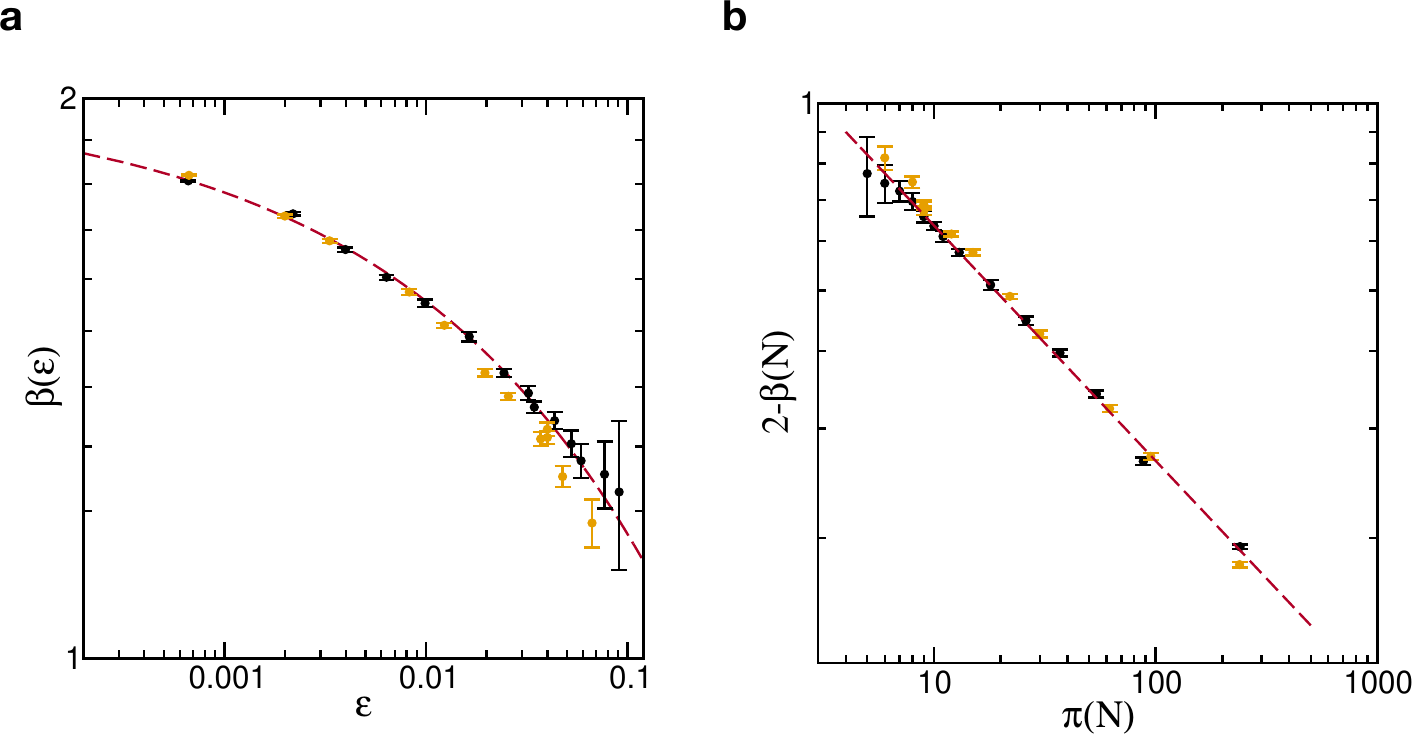}
    \caption{
    \textbf{Local scaling exponent $\beta$ in two-dimensional quasicrystals.}
    (a) Local exponent $\beta(\epsilon)$, obtained from power-law fits $\sigma^2(R)\sim R^{\beta}$, plotted as a function of the symmetry-control parameter $\epsilon=1/N$. Results for prime and non-prime rotational symmetries are shown in black and orange, respectively. Error bars correspond to the standard deviation of the distribution of fitted exponents obtained from the sliding-window procedure described in the text.
    (b) $2-\beta$, plotted as a function of the prime-counting function $\pi(N)$, where $\pi(N)$ denotes the number of prime numbers not exceeding $N$. The data exhibit a power-law decay, $2-\beta \rightarrow 0$ as $\pi(N) \rightarrow 0$, providing further evidence that $\beta \rightarrow 2$ in the infinite-symmetry limit. The dashed line represents a power-law fit to the prime-symmetry data.
    }
    \label{fig:FFT_Sigma2_N19}
\end{figure}

\subsection{Computing the radial distribution function $g(R)$} \label{sec:2D_gR_algorithm}

A circular neighborhood of radius $R_{max}$ is constructed around an arbitrary point $\vec{P} \in \mathbb{R}^{2}$ following the procedure described in Section~\ref{sec:Variance_Algorithm_2D}. Among the sites within this neighborhood, the site $\vec{C}$ closest to $\vec{P}$
is selected as the reference point. Distances $d_{i} = \lvert \vec{C} - \vec{p}_{i} \rvert$ are computed for all sites $\vec{p}_{i}$ in the neighborhood and accumulated into a histogram $H$ with bin width $\Delta R$ over the range $[0, R_\text{max}]$, using a right-closed interval convention. This procedure is repeated independently $10^4$ times, yielding the ensemble-averaged histogram $\langle H \rangle$, from which the pair correlation function is obtained as  
$$g(R) = \frac{\langle H \rangle_{R}}{(2 \pi R \Delta R) \rho},$$
where $\rho$ is the number density of the quasiperiodic system and $\langle H \rangle_{R}$ is the bin height at $(R - \Delta R, R]$.

\subsection{Additional examples of $\sigma^2(R)$ and $g(R)$}

\begin{figure}
    \centering
    \begin{subfigure}{0.45\textwidth}
        \centering
        \includegraphics[width=\linewidth]{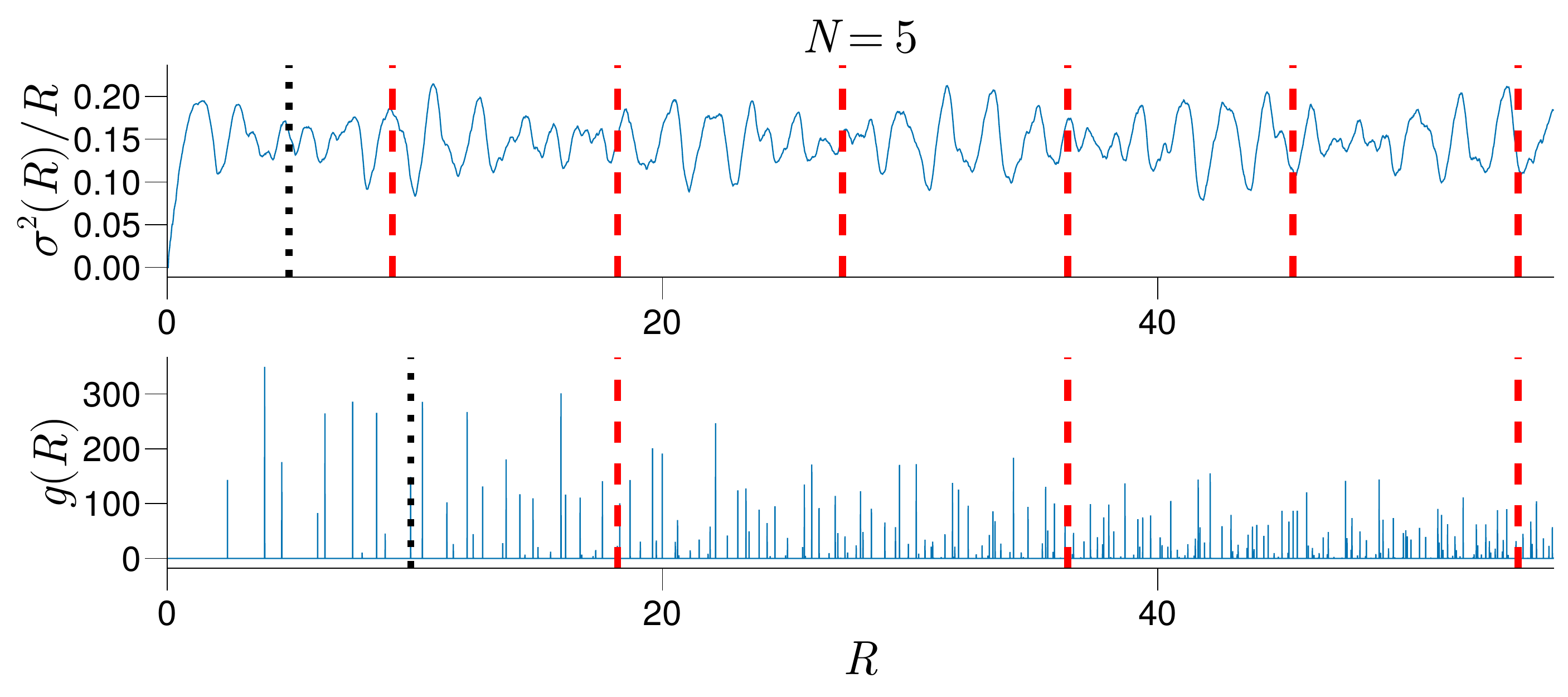}
    \end{subfigure}%
    \hfill
    \begin{subfigure}{0.45\textwidth}
        \centering
        \includegraphics[width=\linewidth]{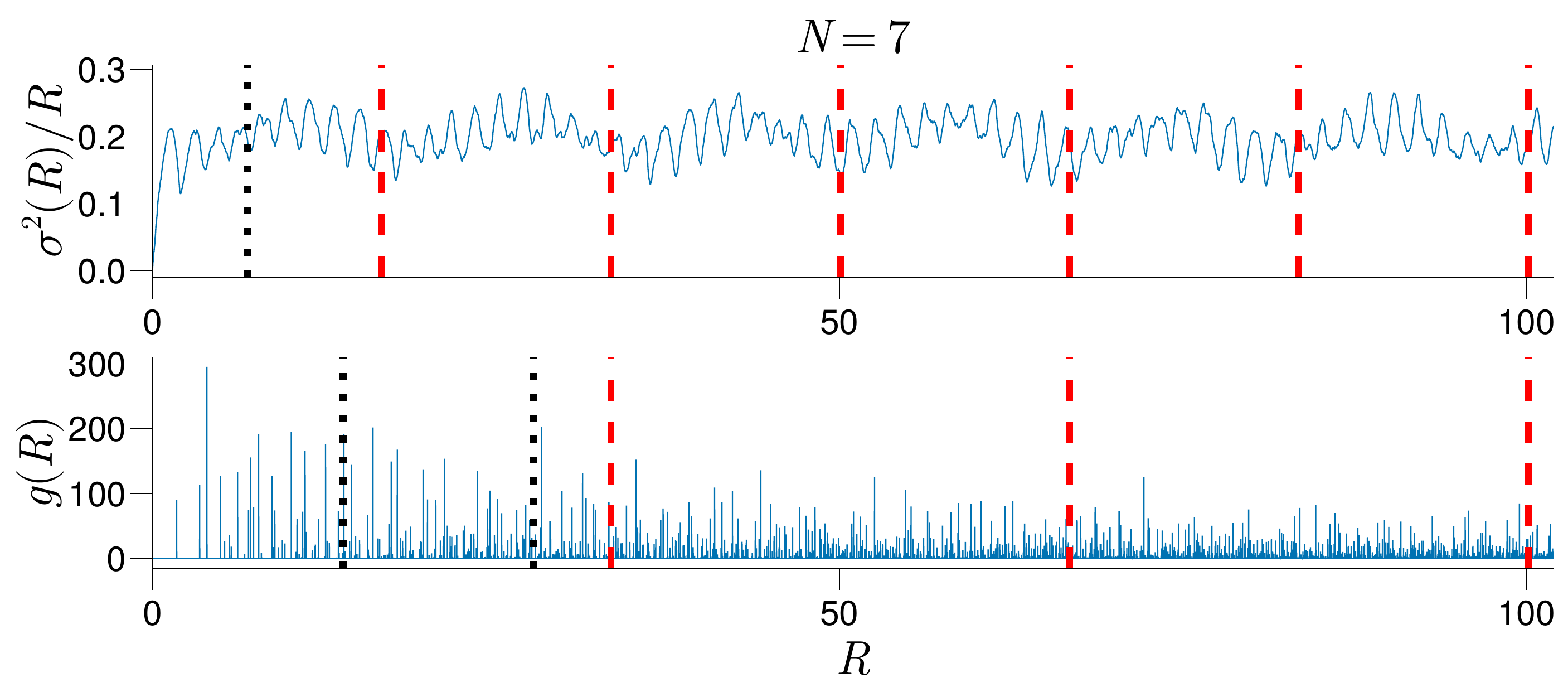}
    \end{subfigure}
    \begin{subfigure}{0.45\textwidth}
        \centering
        \includegraphics[width=\linewidth]{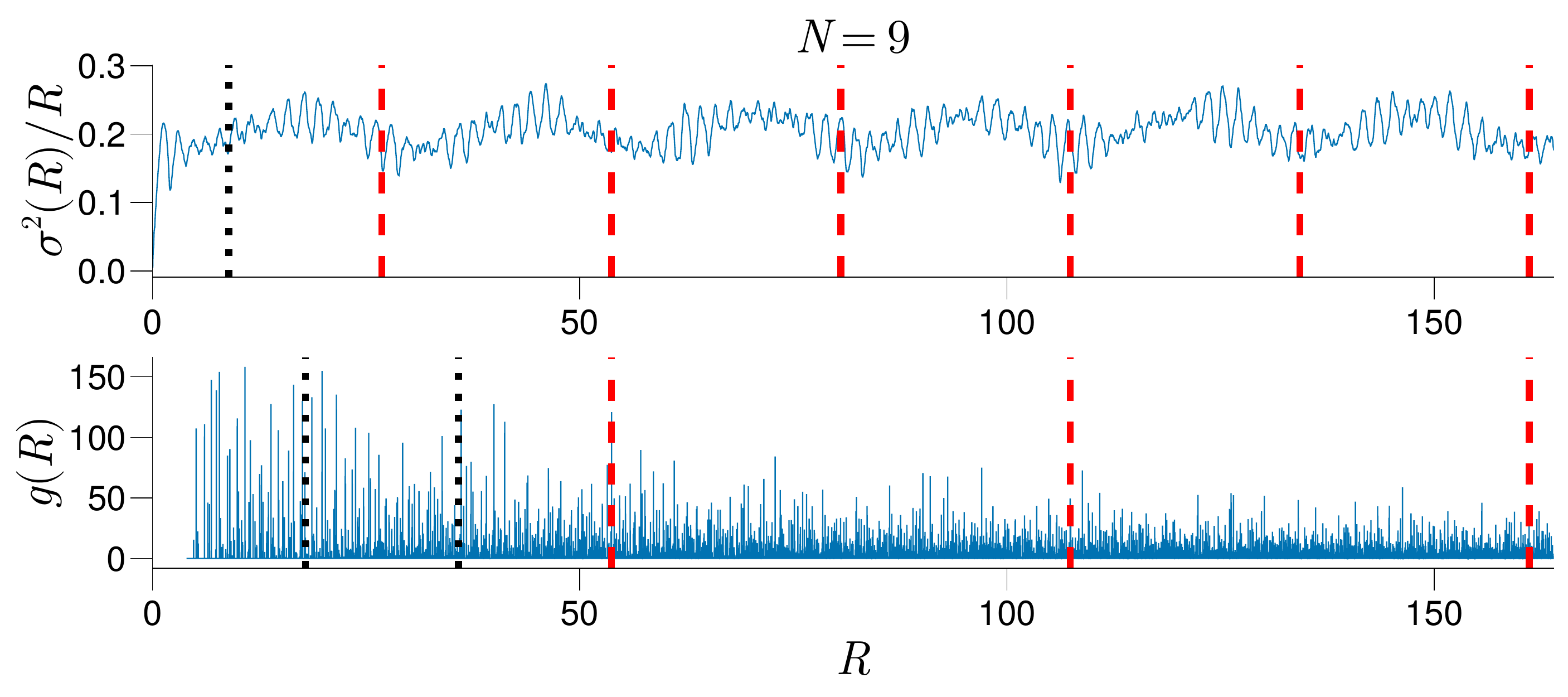}
    \end{subfigure}%
    \hfill
    \begin{subfigure}{0.45\textwidth}
        \centering
        \includegraphics[width=\linewidth]{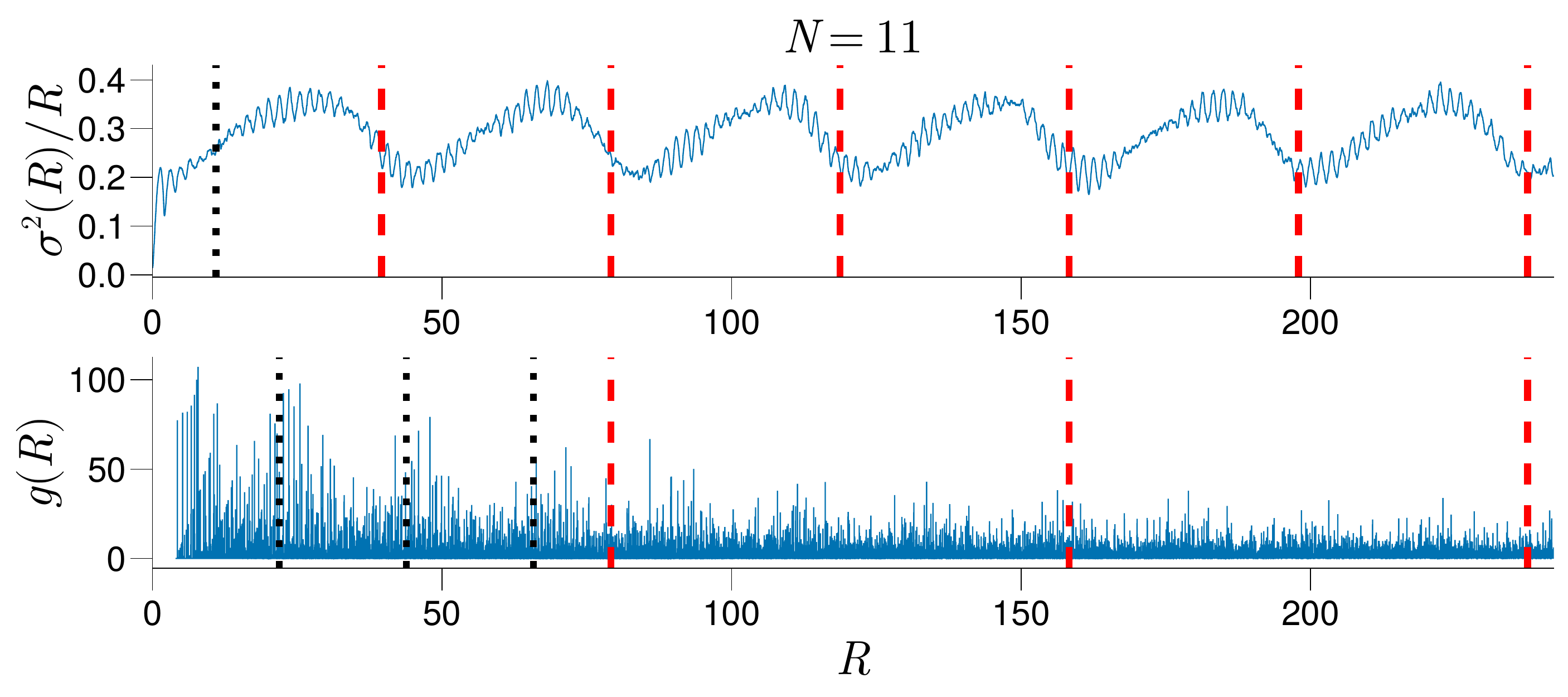}
    \end{subfigure}
    \begin{subfigure}{0.45\textwidth}
        \centering
        \includegraphics[width=\linewidth]{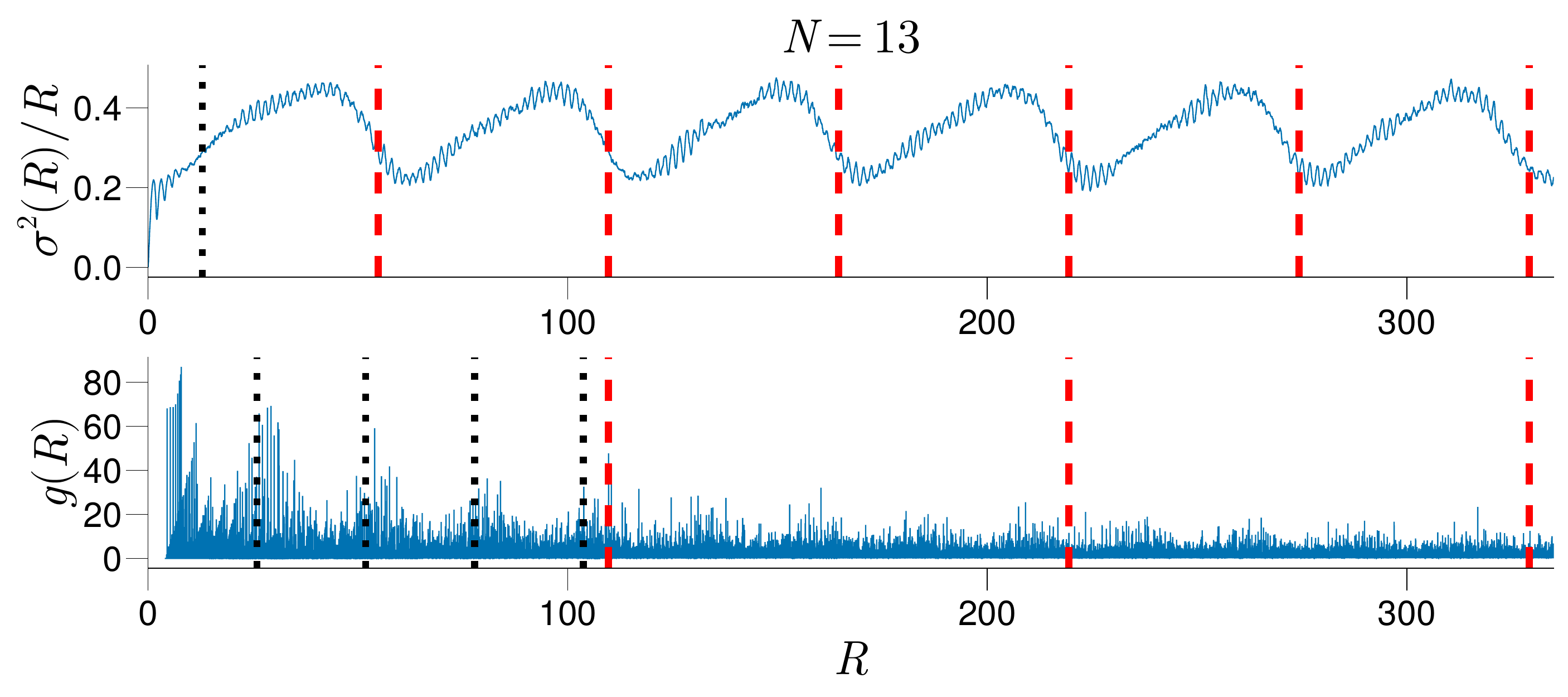}
    \end{subfigure}%
    \hfill
    \begin{subfigure}{0.45\textwidth}
        \centering
        \includegraphics[width=\linewidth]{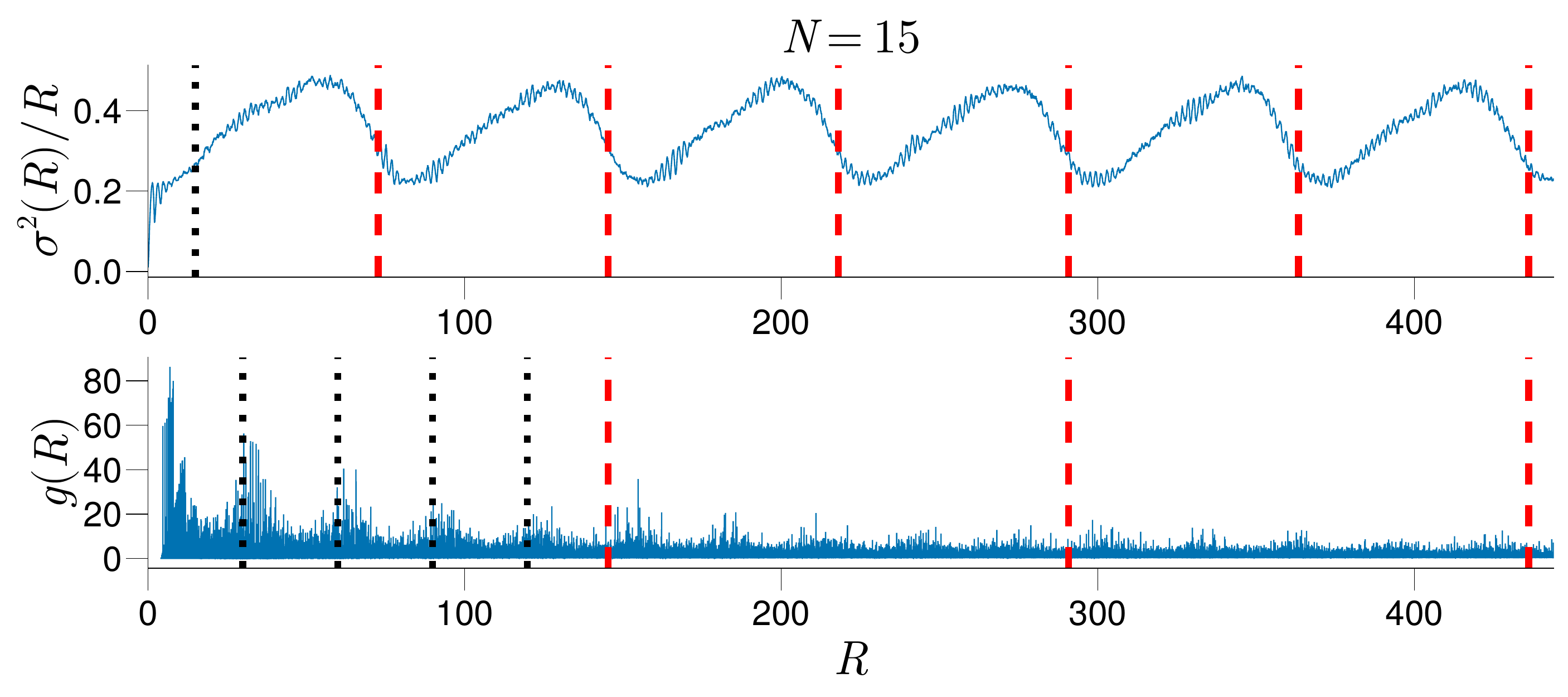}
    \end{subfigure}
    \begin{subfigure}{0.45\textwidth}
        \centering
        \includegraphics[width=\linewidth]{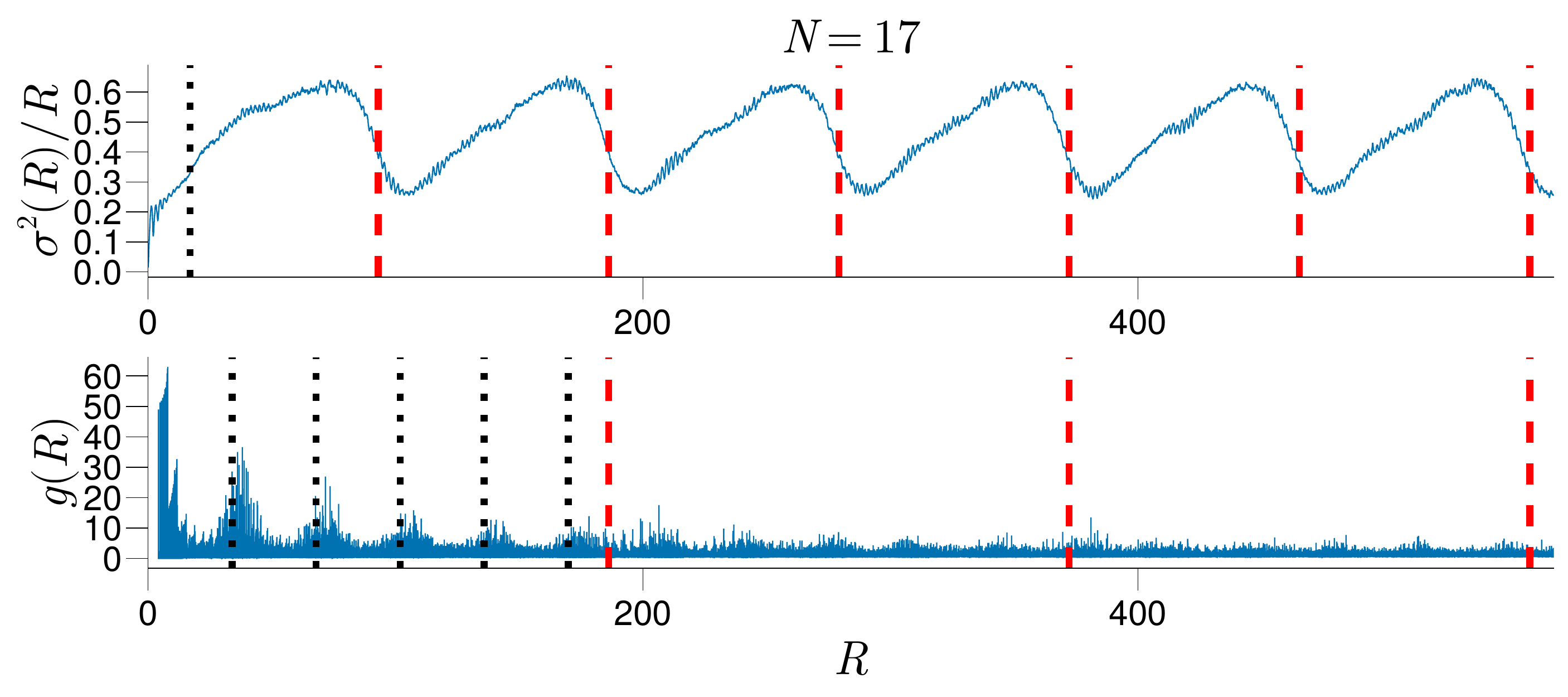}
    \end{subfigure}%
    \hfill
    \begin{subfigure}{0.45\textwidth}
        \centering
        \includegraphics[width=\linewidth]{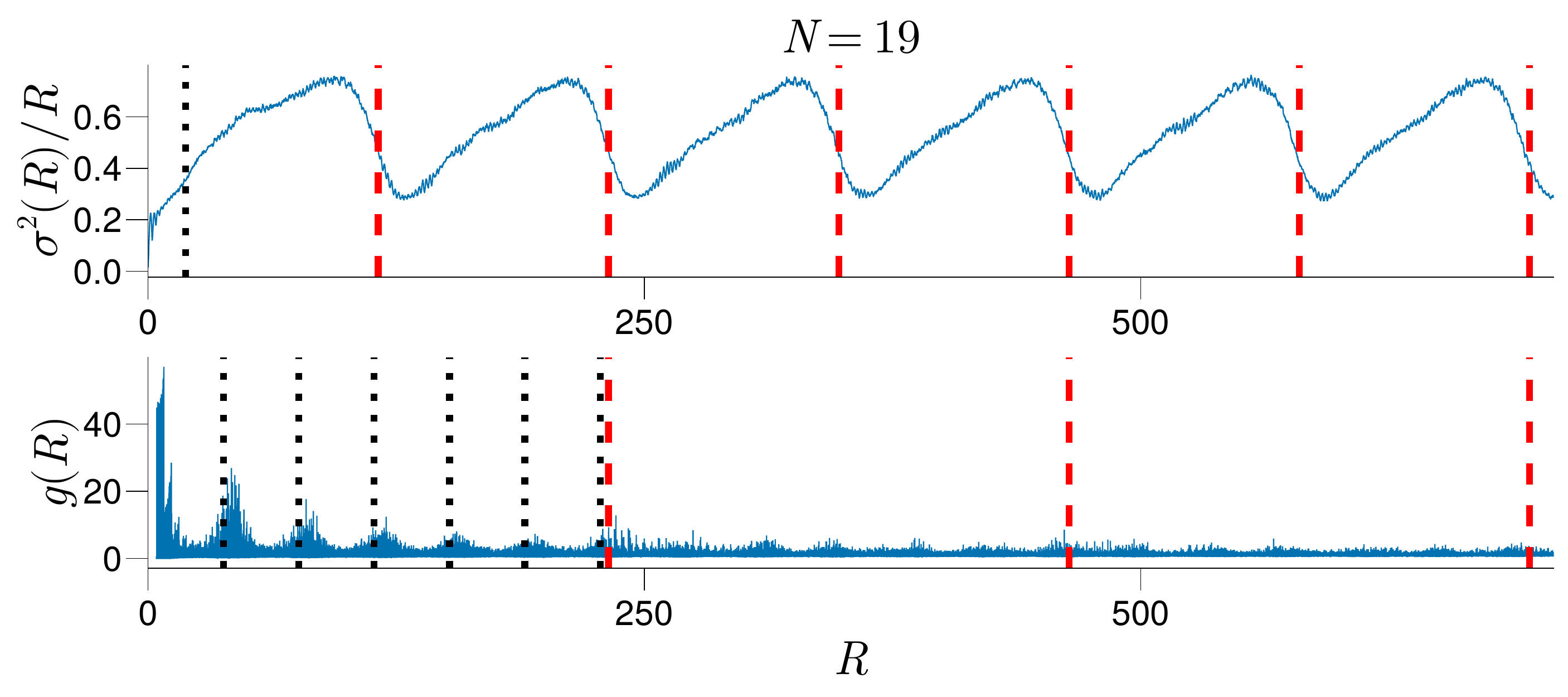}
    \end{subfigure}
    \caption{\textbf{Density fluctations and pair correlations across rotational symmetries in two dimensions.} 
    Rescaled number variance $\sigma^{2}(R) / R$ and pair correlation function $g(R)$ for two-dimensional quasicrystals with $N$-fold rotational symmetry, $5 \leq N \leq 19$. Red dashed lines mark integer multiples of $\lambda_{N}$ in the $\sigma^{2}(R) / R$ panels and of $2 \lambda_{N}$ in the $g(R)$ panels; black dotted lines mark $\kappa_{N}$ and its multiples of $2 \kappa_{N}$, respectively. Radial step size of $\Delta R = 0.005$ and $\Delta R = 0.001$ was used for $\sigma^{2}(R) / R$  and $g(R)$, respectively, both quoted prior to rescaling by $2 \sqrt{\pi \rho}$.}
    \label{fig:2D_Sigma2_gR}
\end{figure}

Figure~\ref{fig:2D_Sigma2_gR} extends the density fluctuation and pair correlation analysis of the main text to a broader range of rotational symmetries $N$. In all panels, red and black dashed vertical lines mark integer multiples of $\lambda_N$ and $\kappa_N$, respectively; the procedures for extracting these scales are detailed in Sections~\ref{sec:length-scales_2D-lambda} and~\ref{sec:length-scales_2D-kappa}. For $N \geq 9$, the initial region $R \leq 5$
is omitted from the $g(R)$ panels, as it contains a near-contact peak that exceeds the remaining data by several orders of magnitude and would otherwise obscure the long-range structure of interest.

\subsection{Determination of the statistic-unit-cell size  $\lambda_{N}$} \label{sec:length-scales_2D-lambda}

\begin{figure}
    \centering
    \includegraphics[width=1.0\linewidth]{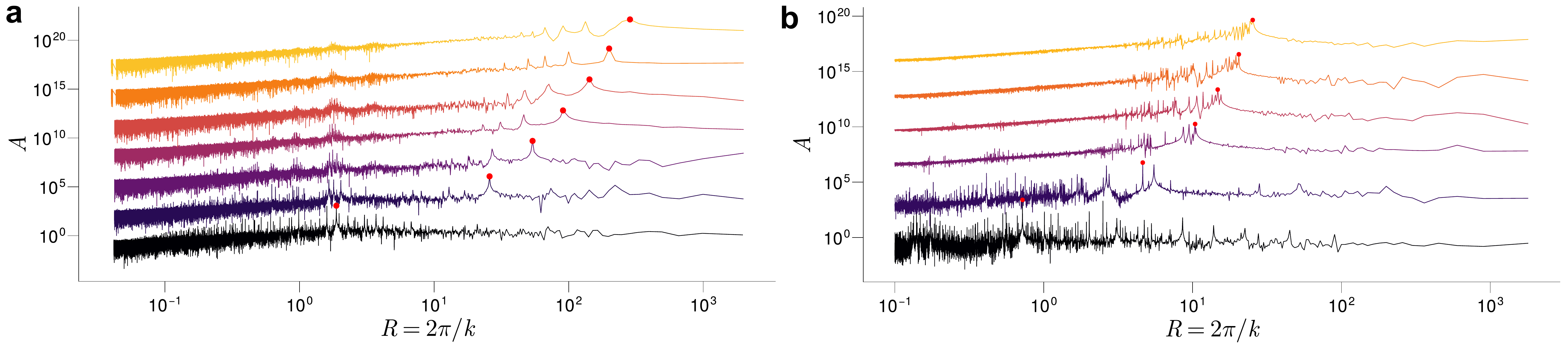}
    \caption{\textbf{Fourier extraction of the statistical-unit-cell scale $\lambda_N$.} Fast Fourier Transform of $\sigma^2(R) / R$ for two-dimensional quasicrystals (a) and of $\sigma^2(R)$ for one-dimensional quasicrystals (b), for $N=5, 9, 13, 17, 21, 25, 29$ and $N=5, 11, 21, 31, 41, 51$, respectively. Curves progress from black to yellow with increasing $N$ and are vertically offset for clarity. 
    Red dots mark the peak amplitudes identifying the statistical-unit-cell scale $\lambda_N$.
    }
    \label{fig:FFT_Sigma2_N5-2-29}
\end{figure}

The rescaled number variance $\sigma^2(R)/R$ exhibits clear oscillatory behavior with a dominant period $\lambda_{N}$. 
To extract this period precisely, we apply a Fast Fourier Transform (FFT) to $\sigma^2(R)/R$, converting the radial dependence to a length spectrum via $R = 2\pi/k$, where $k$ is the wavevector. The statistical-unit-cell size is then identified as $\lambda_N = 2\pi/k_\text{peak}$, where $k_\text{peak}$ corresponds to the peak amplitude in the FFT spectrum. Figure~\ref{fig:FFT_Sigma2_N5-2-29} displays the resulting FFT spectra for representative values of $N$ in both one and two dimensions, with peak amplitudes marked by red dots.  Curves are vertically offset by $10^{3i}$ for $i=0, \ldots, 5$, with $i=0$ corresponding to the smallest value $N=5$. The values of $\lambda_N$ obtained through this procedure correspond to those shown in Fig.~1f of the main text.

\subsection{Determination of the length scale $\kappa_{N}$} 
\label{sec:length-scales_2D-kappa}

The pair correlation function $g(R)$ exhibits a series of peaks at regular intervals of $2\kappa_N$, with amplitudes that decrease progressively from $R = 0$ to $R = 2 \lambda_{N}$ and become increasingly difficult to resolve at larger distances. A direct FFT of $g(R)$ is unsuitable for extracting this periodicity, since the dense fluctuations in the data dominate the spectrum and obscure the signal associated with the peak spacing. To isolate the relevant periodicity, we compute the concave hull of $g(R)$, extract its upper envelope, and apply a Lomb-Scargle periodogram~\cite{VanderPlas2018} to the resulting points, identifying $2\kappa_N = 2\pi/k_\text{peak}$ as the length associated with the local maximum. The Lomb-Scargle periodogram is necessary here because the points forming the concave hull are non-uniformly spaced, precluding a standard FFT.

\subsection{Hyperuniformity and $g(R)$ decomposed by prototile}

\begin{figure}
    \centering
    \begin{subfigure}{0.45\textwidth}
        \centering
        \includegraphics[width=\linewidth]{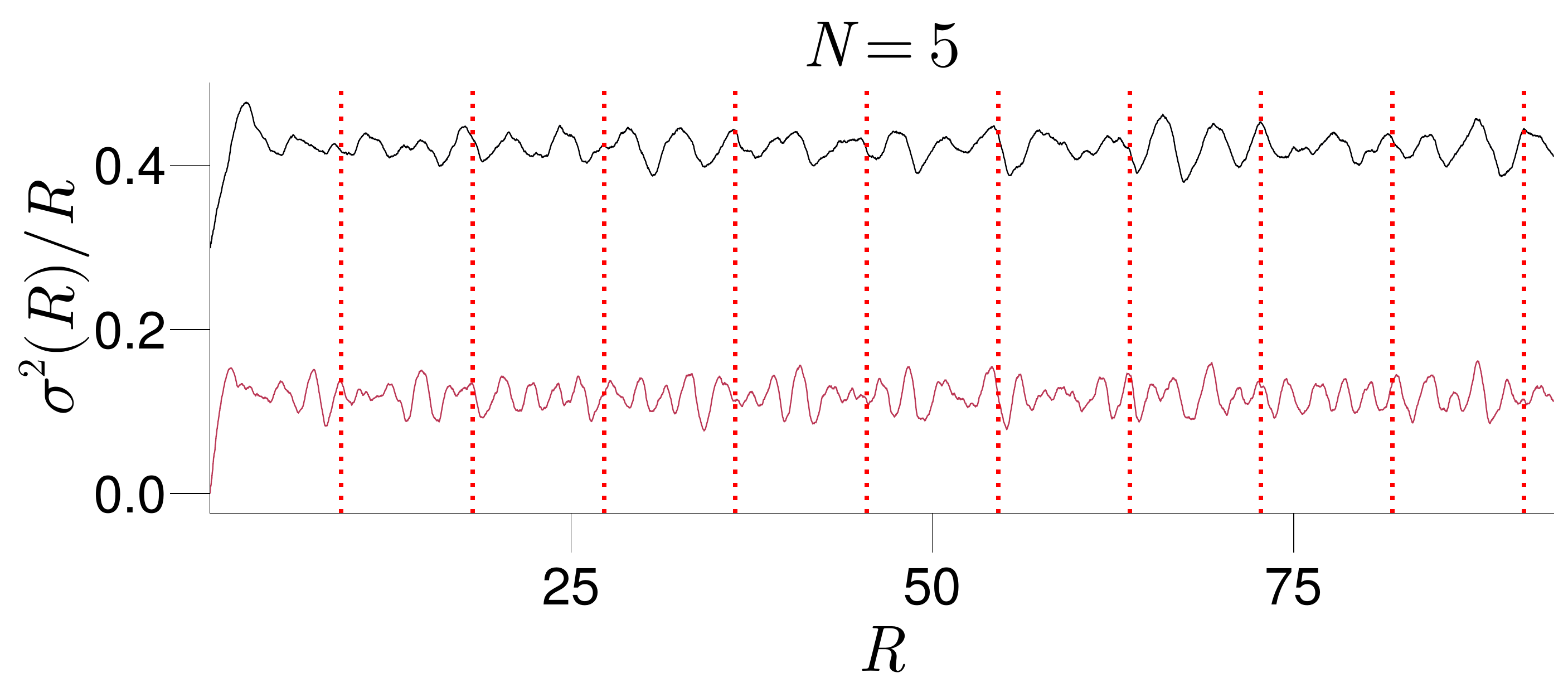}
    \end{subfigure}%
    \hfill
    \begin{subfigure}{0.45\textwidth}
        \centering
        \includegraphics[width=\linewidth]{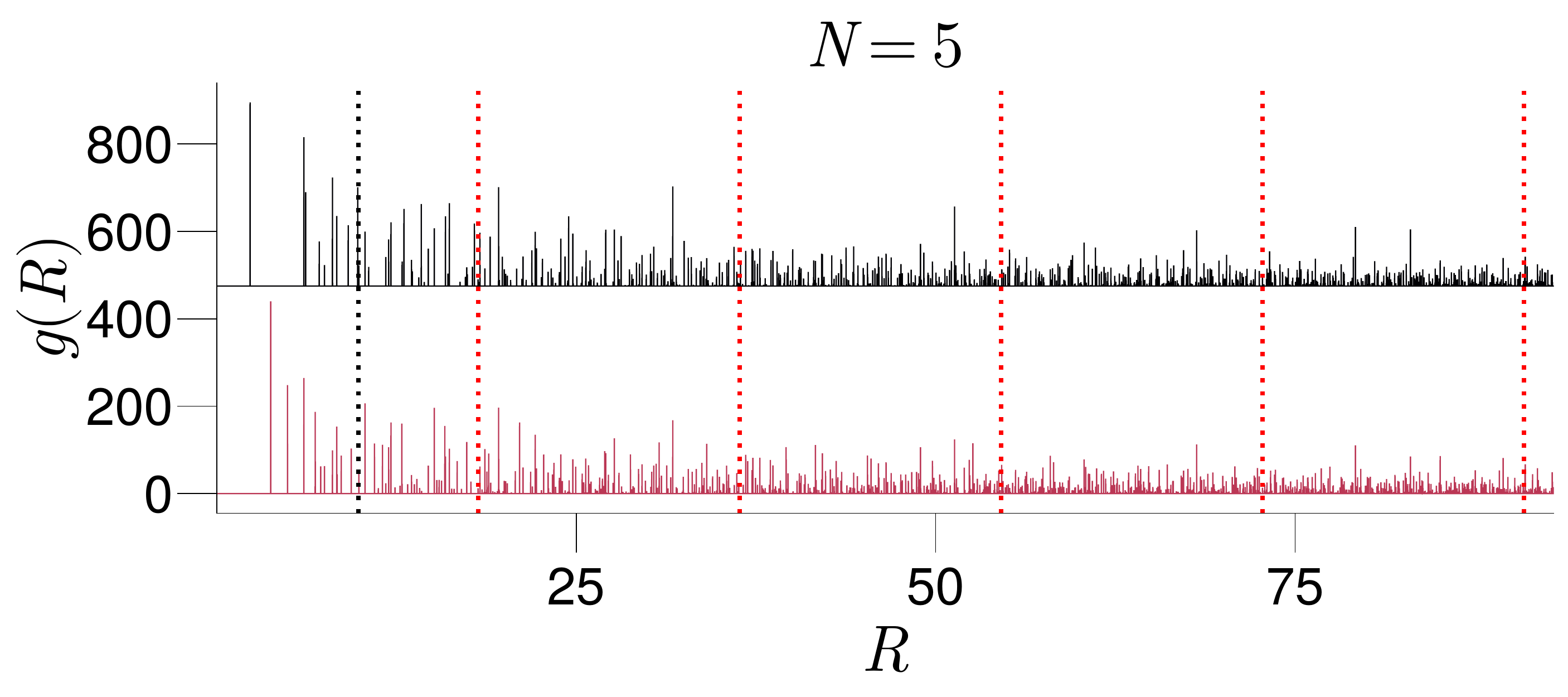}
    \end{subfigure}
    \begin{subfigure}{0.45\textwidth}
        \centering
        \includegraphics[width=\linewidth]{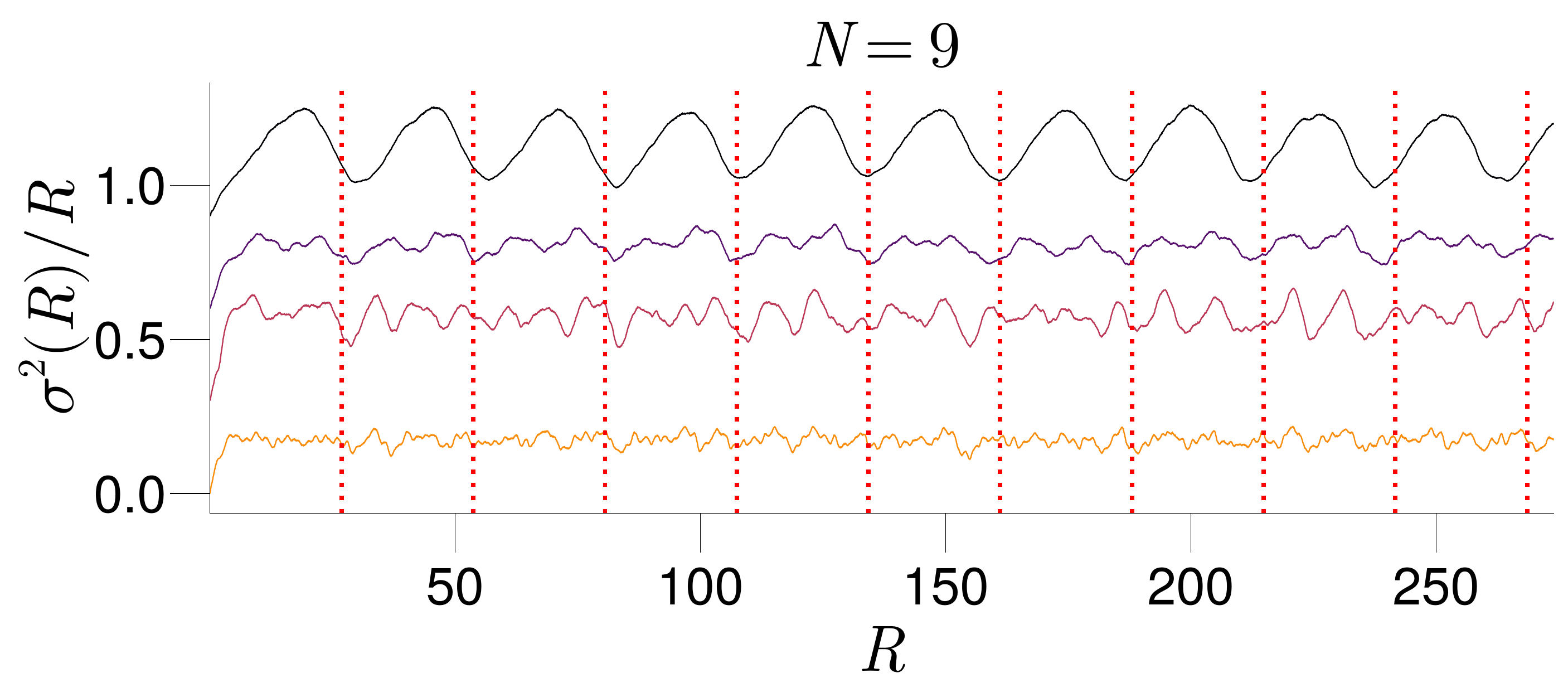}
    \end{subfigure}%
    \hfill
    \begin{subfigure}{0.45\textwidth}
        \centering
        \includegraphics[width=\linewidth]{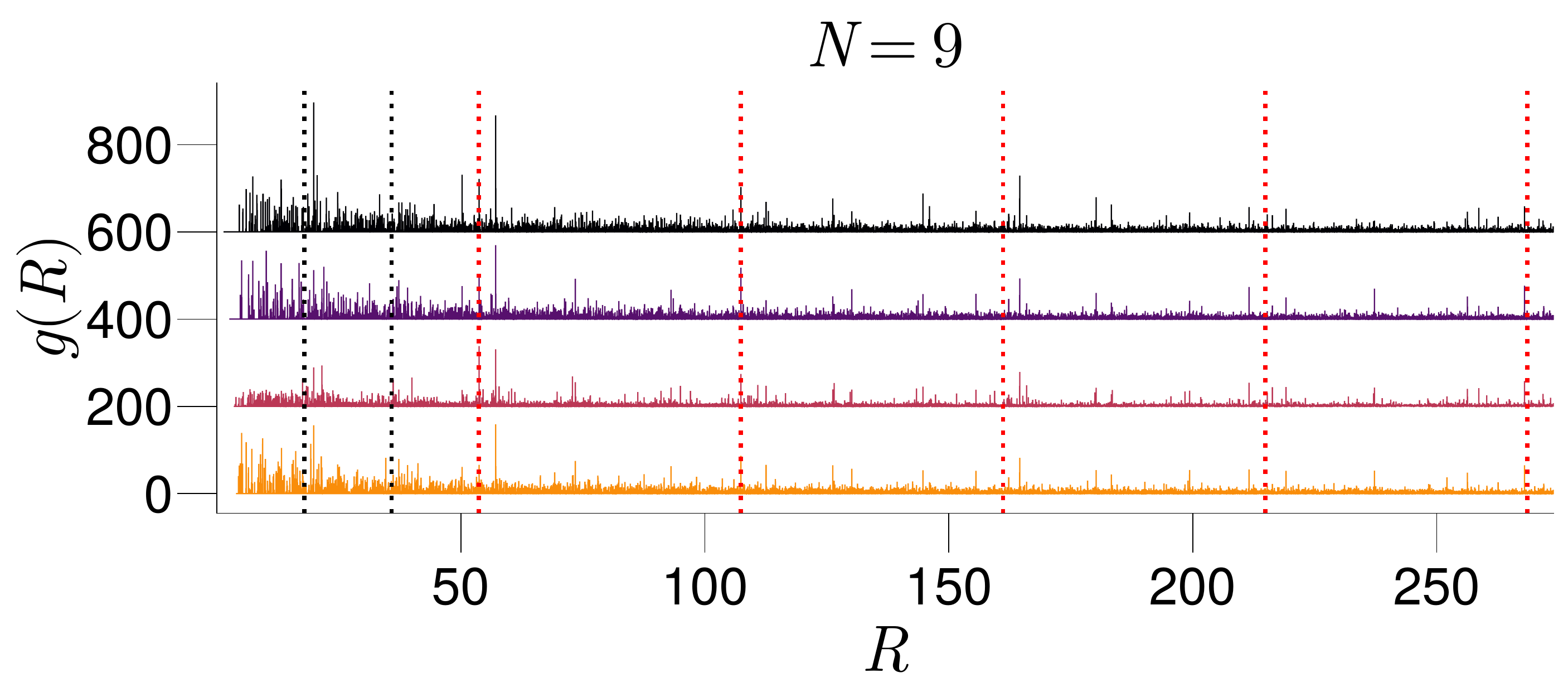}
    \end{subfigure}
    \begin{subfigure}{0.45\textwidth}
        \centering
        \includegraphics[width=\linewidth]{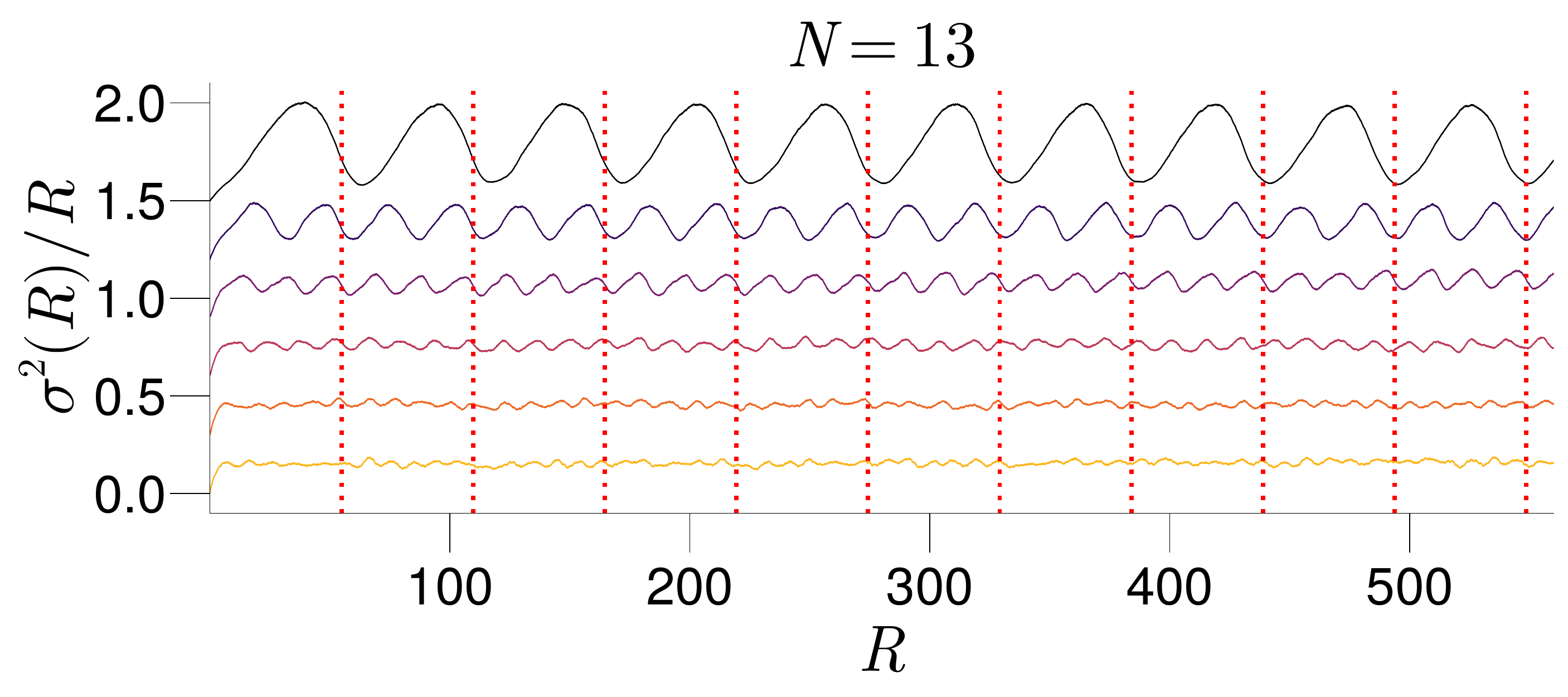}
    \end{subfigure}%
    \hfill
    \begin{subfigure}{0.45\textwidth}
        \centering
        \includegraphics[width=\linewidth]{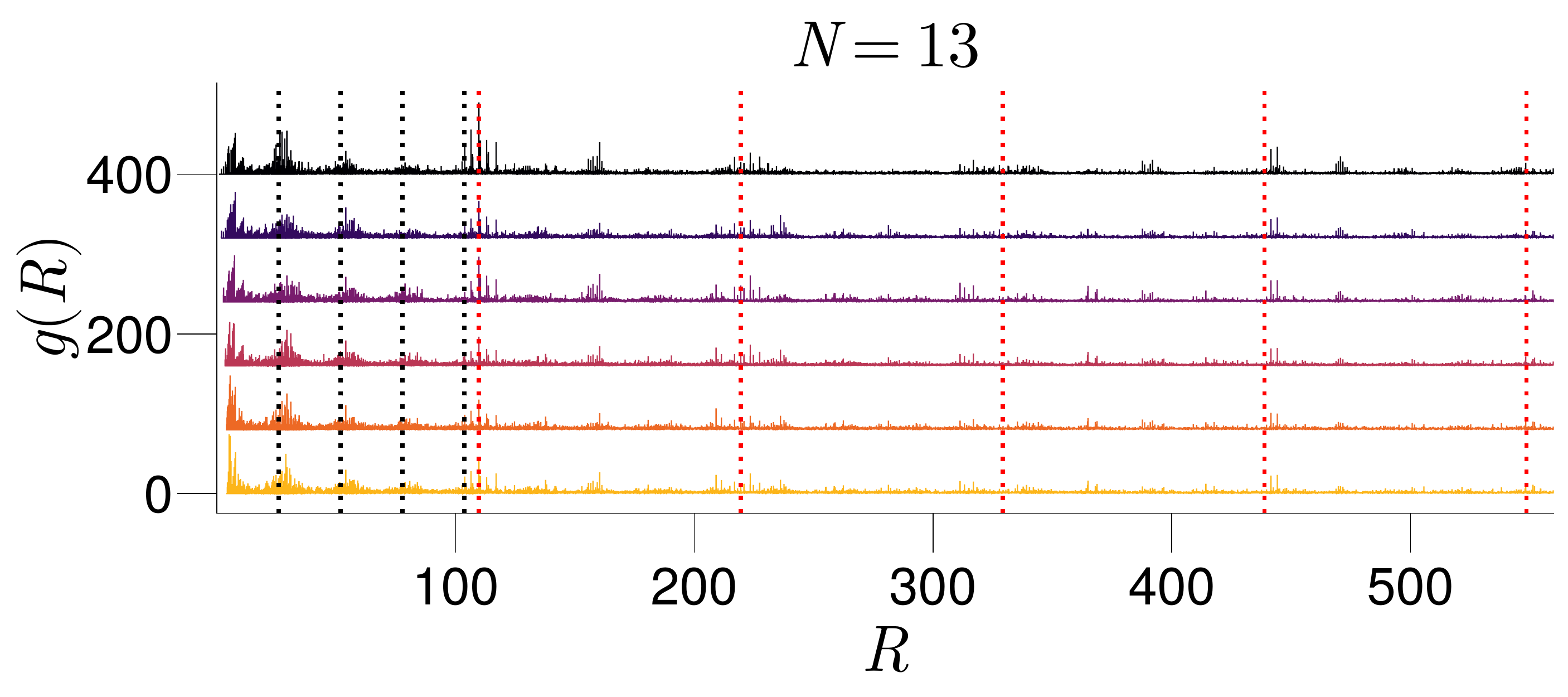}
    \end{subfigure}
    \begin{subfigure}{0.45\textwidth}
        \centering
        \includegraphics[width=\linewidth]{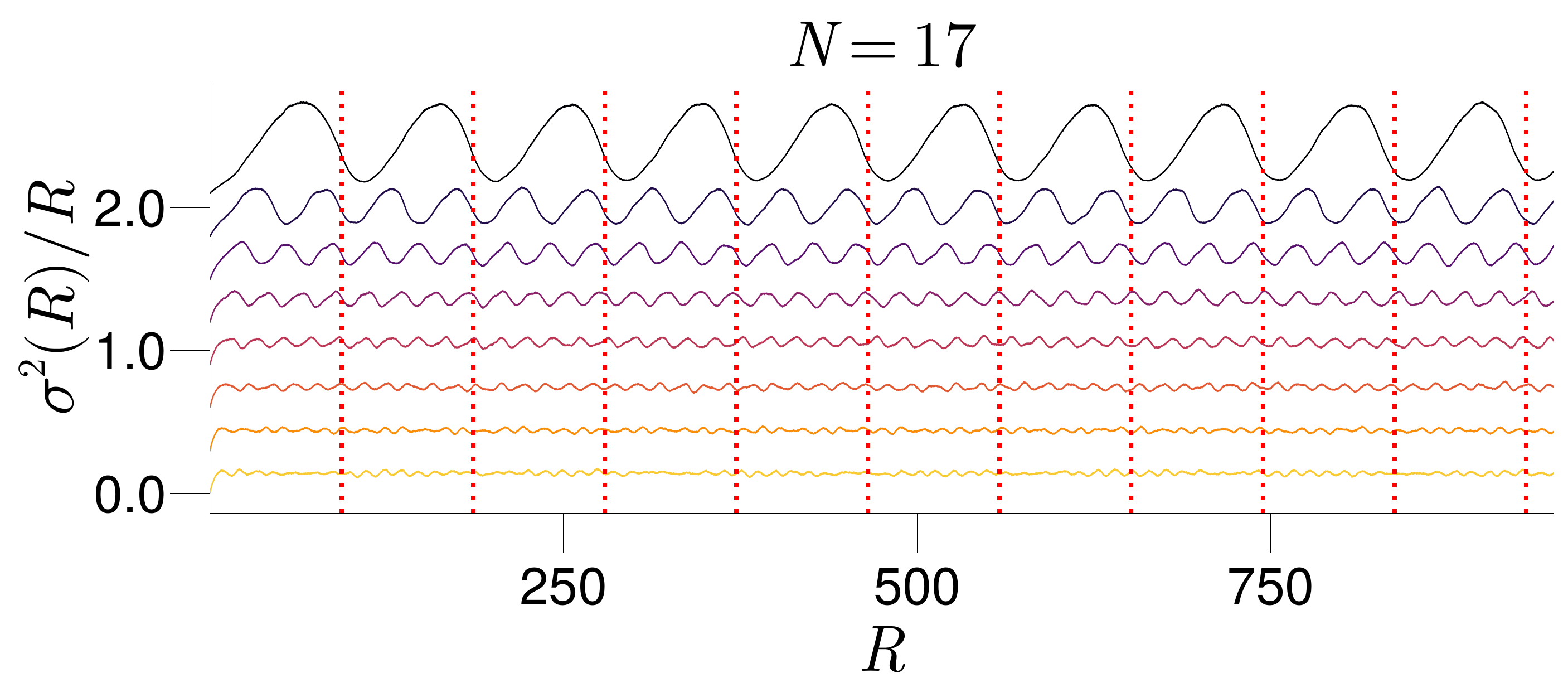}
    \end{subfigure}%
    \hfill
    \begin{subfigure}{0.45\textwidth}
        \centering
        \includegraphics[width=\linewidth]{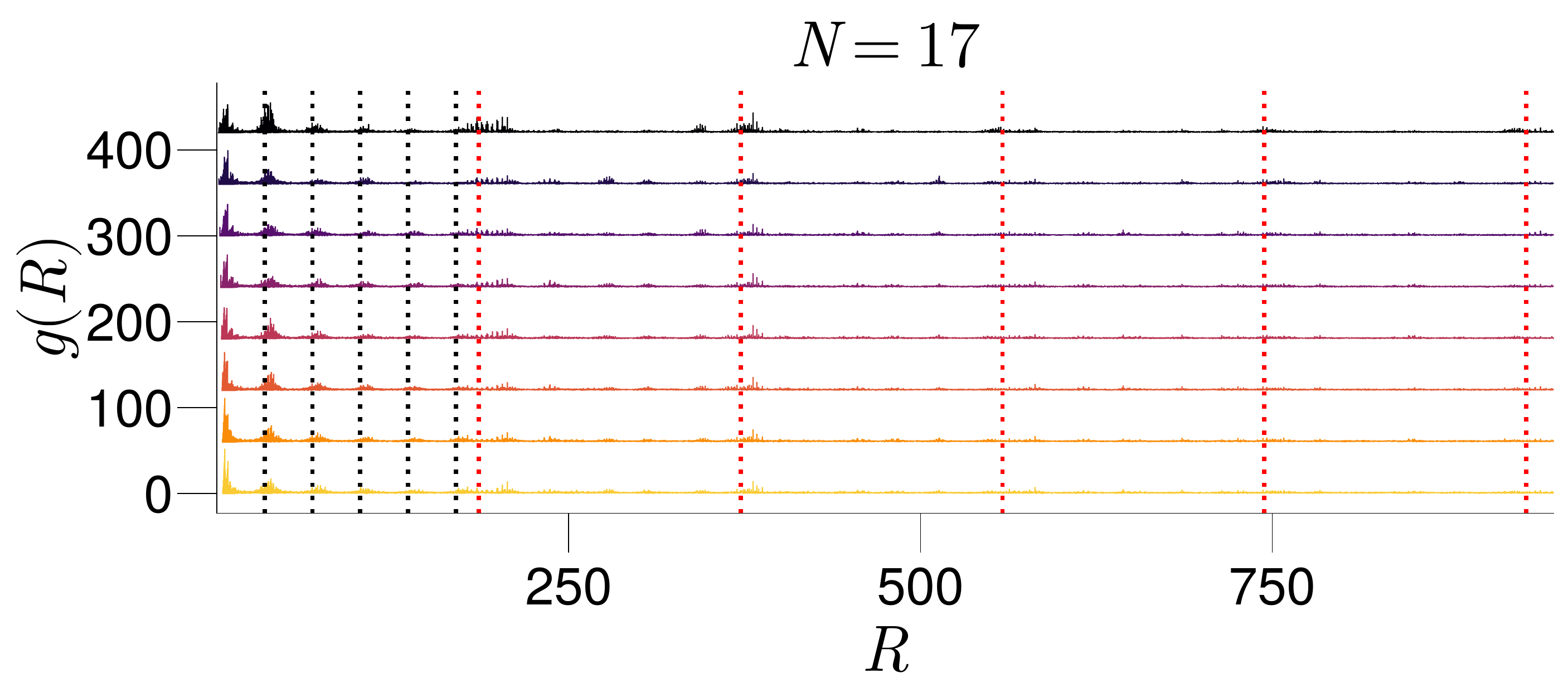}
    \end{subfigure}
    \caption{
    \textbf{Density fluctuations and pair correlations in two dimensions, resolved by prototile.}
    Rescaled number variance $\sigma^2(R)/R$ and pair correlation function $g(R)$ for two-dimensional quasicrystals with $N = 5,9,13,17$, decomposed by prototile. Red and black dashed lines 
    mark integer multiples of $\lambda_N$ and $\kappa_N$, respectively, with the same conventions as Fig.~\ref{fig:2D_Sigma2_gR}. 
    The color gradient from black to yellow corresponds to tile indices $T_{i} = 1$, $2$, $\dots$, $\lfloor N/2 \rfloor$, ordered by increasing tile area. Radial step sizes of $\Delta R = 0.005$ and $\Delta R = 0.001$ are used for $\sigma^2(R)/R$ and $g(R)$, respectively, both quoted prior to rescaling by $2\sqrt{\pi\rho}$.
    }
    \label{fig:2D_Sigma2_gR_PAG}
\end{figure}

In addition to the vertex-decorated analysis of the main text, we repeat the computation of $\sigma^2(R)$ and $g(R)$ using a centroid decoration, where a single point is placed at the center of each tile according to its prototile type. The main text shows $\sigma^2(R) / R$ for the $N=23$ quasicrystal decomposed into its four smallest-area prototiles; Fig.~\ref{fig:2D_Sigma2_gR_PAG} extends this analysis to a broader range of symmetries and prototile families.

For clarity, the $\sigma^2(R) / R$ curves are vertically offset by $0.3 \ (\lfloor N/2 \rfloor - T_{i})$, where $T_{i} \in \mathbb{Z}^{+}$ is the tile index with $T_i=1$ denoting the smallest-area tile and $T_{i} = \lfloor N/2 \rfloor$ the largest. 
The $g(R)$ curves are offset by $I (\lfloor N/2 \rfloor - T_{i})$, where $I \in \mathbb{Z}^{+}$ is an $N$-dependent constant. As in Fig.~\ref{fig:2D_Sigma2_gR}, the initial region $R\leq 5$ is  omitted for $N \geq 9$.

\subsection{Determination of $\Lambda_{\infty}$}

\begin{figure}
    \centering
    \includegraphics[width=1.0\linewidth]{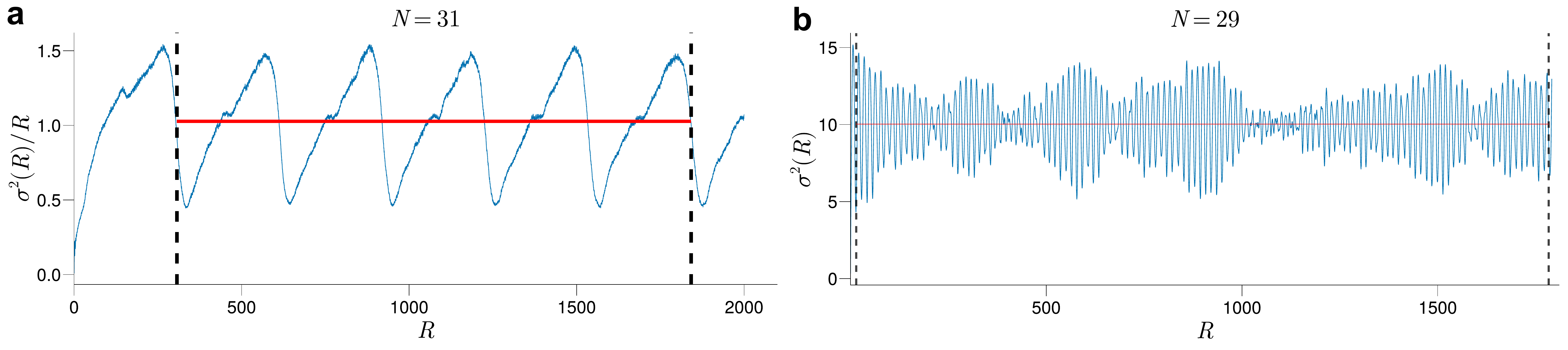}
    \caption{\textbf{Determination of the cycle-averaged amplitude $\Lambda_\infty$.}
    (a) $\sigma^2(R)/R$ for a two-dimensional quasicrystal with $N=31$, yielding $\Lambda_\infty = 1.0267$ over the interval $[\lambda_N,\, 6\lambda_N]$ (b) $\sigma^2(R)$ for a one-dimensional quasicrystal with $N = 29$, yielding $\Lambda_\infty = 10.029$ over the interval $[\lambda_N,\, 124\lambda_N]$. In both panels, the horizontal red line marks $\Lambda_\infty$ and vertical black dashed lines indicate the boundaries of the fitting interval.
    }
    \label{fig:LambdaInf_N31}
\end{figure}

The oscillatory behavior observed in the $\sigma^2(R) / R$ raises the question of how its cycle-averaged amplitude $\Lambda_{\infty}$ depends on rotational symmtery $N$. We compute $\Lambda_{\infty}$ as the mean of  $\sigma^2(R) / R$ over the interval $[R_i, R_f] = [\lambda_N,\, M\lambda_N]$, where $M = \lfloor R_\text{max}/\lambda_N \rfloor$ is the number of complete oscillation cycles within the available data. This procedure is illustrated for $N = 31$ in Fig.~\ref{fig:LambdaInf_N31}a.

\begin{figure}
    \centering
    \includegraphics[width=1.0\linewidth]{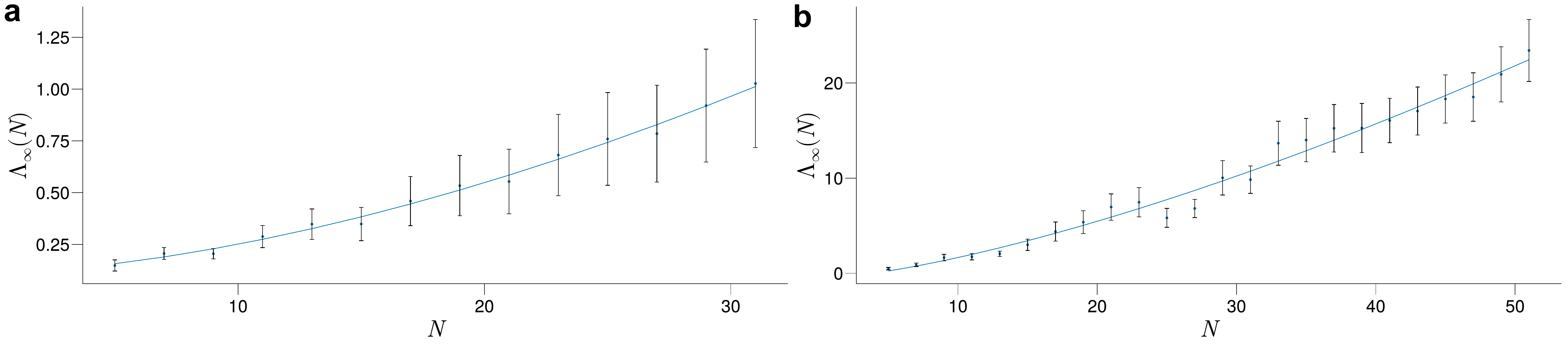}
    \caption{\textbf{Scaling of the cycle-averaged amplitude $\Lambda_\infty$ with rotational symmetry.}
    Cycle-averaged amplitude $\Lambda_{\infty}(N)$ as a function of rotational symmetry $N$ for two-dimensional (a) and one-dimensional (b) quasicrystals. Solid lines show the least-squares fits $\Lambda_{\infty}(N) = 0.113 + 0.003 N^{1.658}$ and $\Lambda_{\infty}(N) = -0.608 + 0.0848 N^{1.425}$, respectively. 
    Error bars indicate the standard deviation of each data point.
    }    
    \label{fig:LambdaInf_vs_N_2D}
\end{figure}

Applying this procedure to all odd rotational symmetries from $N=5$ to $N = 31$ yields the results shown in Fig.~\ref{fig:LambdaInf_vs_N_2D}a. The data are well described by the least-squares fit $\Lambda_\infty(N) = 0.113 + 0.003\,N^{1.658}$, with error bars indicating the standard deviation associated with each data point.

\begin{figure}[p]
    \centering
    \includegraphics[width=0.65\linewidth]{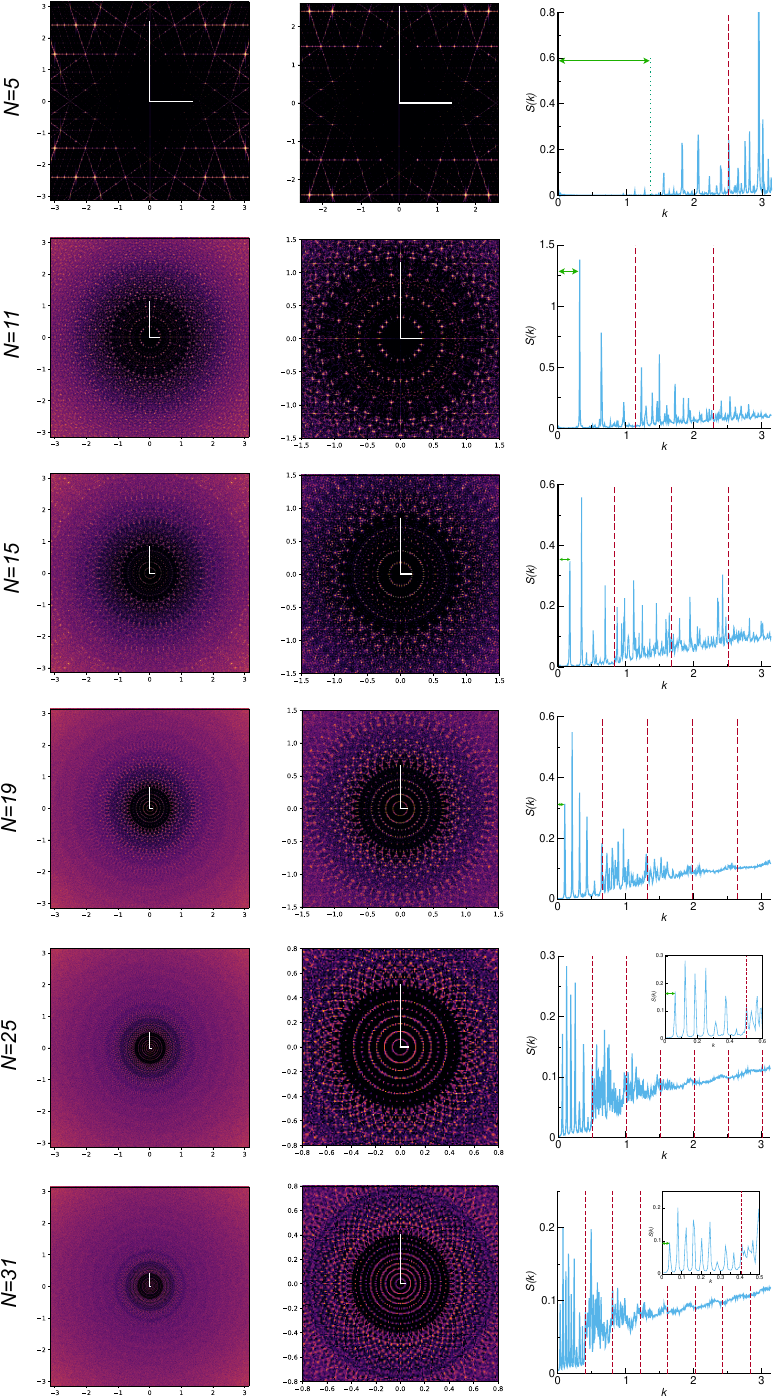}
    \caption{\textbf{Diffraction analysis for two-dimensional quasicrystals}. Rows correspond to rotational symmetries $N=5,11,15,19,25,31$ (from top to bottom). Left column: diffraction intensity $S(\mathbf{k})$ over the reciprocal-space range $k_x,k_y\in[-\pi,\pi]$. Middle column: magnified view of the central region, highlighting the small-wavevector Bragg-peak structure. The vertical and horizontal white lines indicate the characteristic reciprocal-space scales $\pi/\kappa_N$ and $\pi/\lambda_N$, respectively. Right column: radially averaged structure factor $S(k)$. Vertical dashed lines mark integer multiples of $\pi/\kappa_N$, and horizontal arrows $\pi/\lambda_N$.  
    }
    \label{fig:diffpatt1}
\end{figure}

\subsection{Diffraction pattern analysis of two-dimensional quasicrystals}

The diffraction patterns shown throughout this work (see Fig.~\ref{fig:diffpatt1}) were obtained from the static structure factor 
\begin{equation}
S(\mathbf{k}) = \frac{1}{m} \left \lvert \sum_{j=1}^m e^{i\mathbf{k} \cdot
\mathbf{r}_j} \right \rvert^2
\end{equation}
computed directly from the particle coordinates $\mathbf{r}_j=(x_j,y_j)$, where $m$ is the number of particles and $\mathbf{k}=(k_x,k_y)$ denotes a wave vector in reciprocal space. This direct evaluation yields the diffraction intensity without requiring an intermediate real-space density field or pair correlation function, thereby avoiding additional resolution limitations associated with such procedures.

\section{One-dimensional quasicrystalline systems}

\subsection{Construction principles}

One-dimensional quasiperiodic systems have no direct analogue of rotational symmetry. To enable meaningful comparison with two-dimensional $N$-fold systems, we construct one-dimensional quasiperiodic sequences that inherit a connection to $N$-fold symmetry via the Generalized Dual Method (GDM). Specifically, the 2D star vectors $\mathbf{S}_{2D}$ are projected onto the line $l_p$ 
defined by one of these vectors, yielding a set of 1D star vectors $\mathbf{S}_{1D}$ from which the one-dimensional tiling is generated. Throughout this work, we restrict to odd values of $N$.

This procedure yields $\lfloor N/2 \rfloor + 1$ prototiles, each associated with a star vector $\vec{S}_i \in \mathbf{S}_{2D}$, where $\vec{S}_1$ defines $l_p$ and subsequent vectors are indexed counter-clockwise. For $i > 1$, each prototile also accounts for the reflection of $\vec{S}_i$ across $l_p$. The mean separation between adjacent tiles of type $i$ is
$$ N \left\lvert \vec{S}_{ix} \right\rvert = N \left\lvert \cos \left( \frac{2 \pi (i - 1)}{N} \right) \right\rvert,$$
and knowledge of these separations allows the 2D neighborhood construction algorithm to be generalized directly to the 1D case.

The offset parameters $\alpha_i$ associated with $\vec{S}_i \in \mathbf{S}_{1D}$ are held fixed throughout all calculations, and are chosen initially at random as 
$$\alpha_{i} = A_{i} \cos \left( \frac{2 \pi (i - 1)}{N} \right),$$
where $A_i \in [0,1]$ is drawn randomly from a uniform distribution. As in the two-dimensional case, tilings are decorated at tile vertices unless stated otherwise, and these points are referred to as \emph{sites}.

\subsection{Computing the number variance $\sigma^{2}(R)$ }
\label{sec:Variance_Algorithm_1D}

The number variance $\sigma^2(R)$ within an interval of half-width $R$ is computed as follows. An arbitrary point $P \in \mathbb{R}$ is drawn uniformly at random from $[-SL, SL]$ with $SL = 10^{10}$, and a local neighborhood of half-length $R_\text{max}$ is constructed around $P$ from the one-dimensional quasiperiodic tiling with rotational symmetry $N$.

A window $\mathbf{W}$ centered at $P$ is then expanded in steps of $\Delta R = 0.05$, from $R = \Delta R$ to $R = R_\text{max}$. At each step, the number of sites $n(R)$ falling within $\mathbf{W}$ is recorded. This procedure is repeated for $10^5$ independent realizations of $P$, yielding ensemble averages $\langle n(R) \rangle$ and $\langle n(R)^2 \rangle$, from which the number variance is obtained as $\sigma^2(R) = \langle n(R)^2 \rangle - \langle n(R) \rangle^2$.

\subsection{Computing the radial distribution function $g(R)$}

A local neighborhood of half-width $R_\text{max}$ is constructed around an arbitrary point $P \in \mathbb{R}$ following the procedure described in Section~\ref{sec:Variance_Algorithm_1D}. The site $C$ closest to $P$ is selected as the reference point, and distances $d_{i} = \lvert C - p_{i} \rvert$ are computed for all sites $p_i$ in the neighborhood. These distances are accumulated into a histogram $H$ with bin width $\Delta R$ over the range $[0, R_\text{max}]$, using a right-closed interval convention. This procedure is repeated independently $10^5$ times, yielding the ensemble-averaged histogram $\langle H \rangle$, from which the pair correlation function is obtained as
$$g(R) = \frac{\langle H \rangle_{R}}{ 2 \rho \Delta R},$$
where $\rho$ is the number density of the quasiperiodic system and $\langle H \rangle_R$ is the bin height at $(R - \Delta R, R]$.

\subsection{Additional examples of $\sigma^{2}(R)$ and $g(R)$}

\begin{figure}
    \centering
    \begin{subfigure}{0.4\textwidth}
        \centering
        \includegraphics[width=\linewidth]{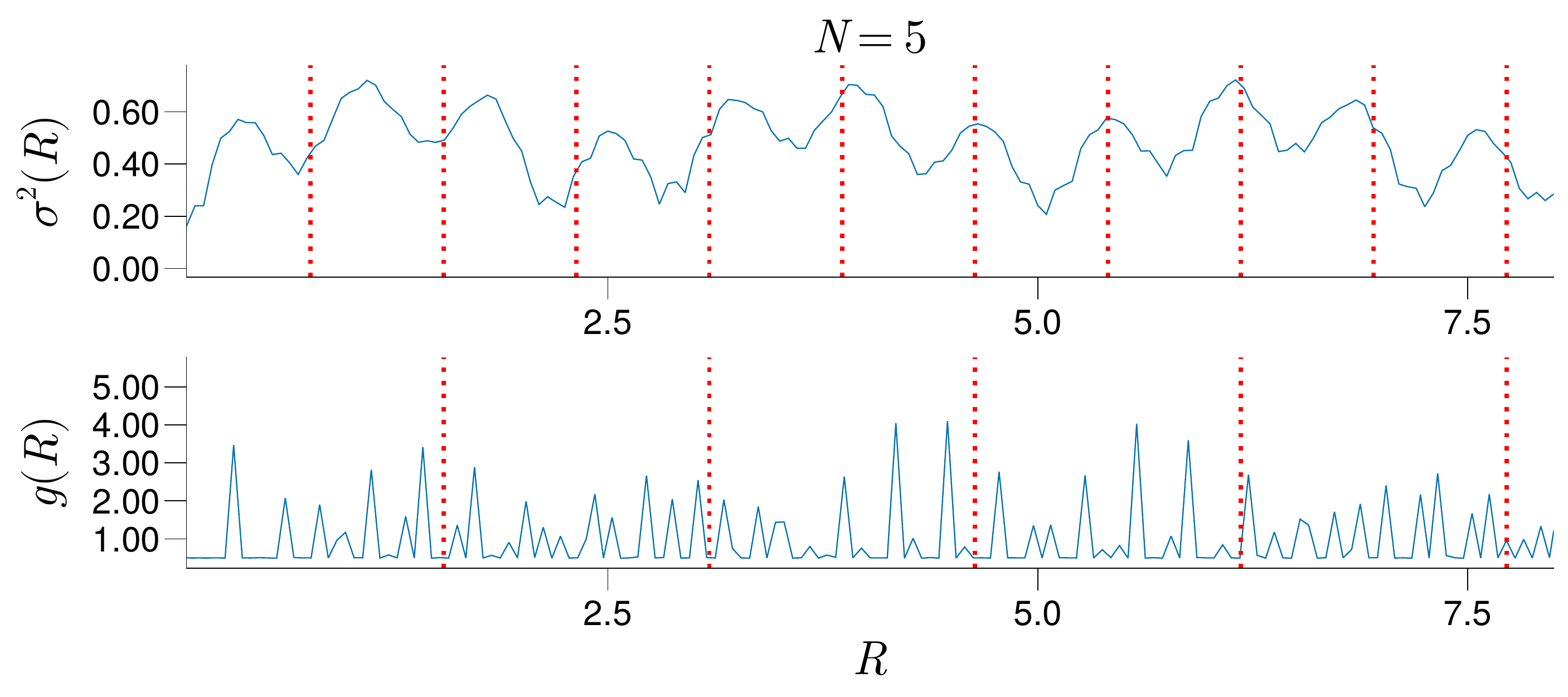}
    \end{subfigure}%
    \hfill
    \begin{subfigure}{0.4\textwidth}
        \centering
        \includegraphics[width=\linewidth]{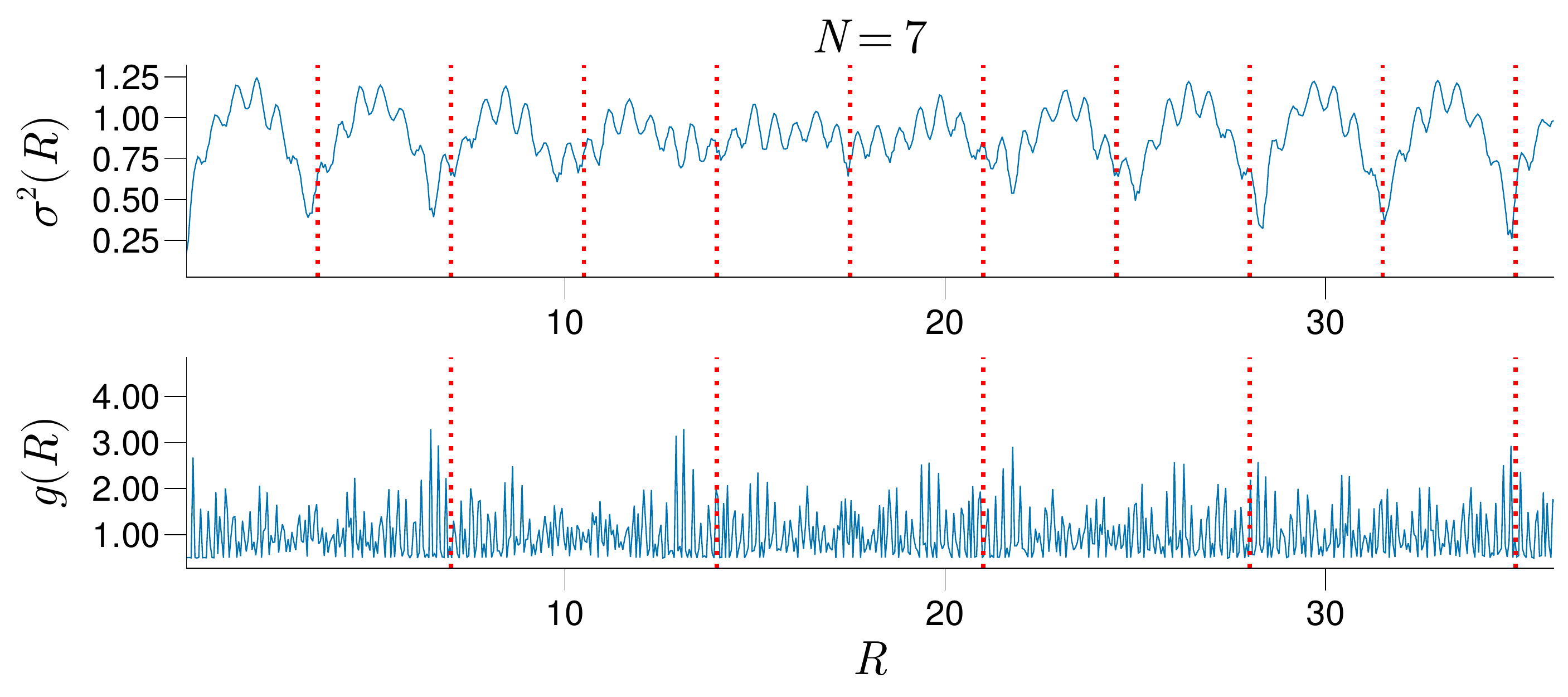}
    \end{subfigure}
    \begin{subfigure}{0.4\textwidth}
        \centering
        \includegraphics[width=\linewidth]{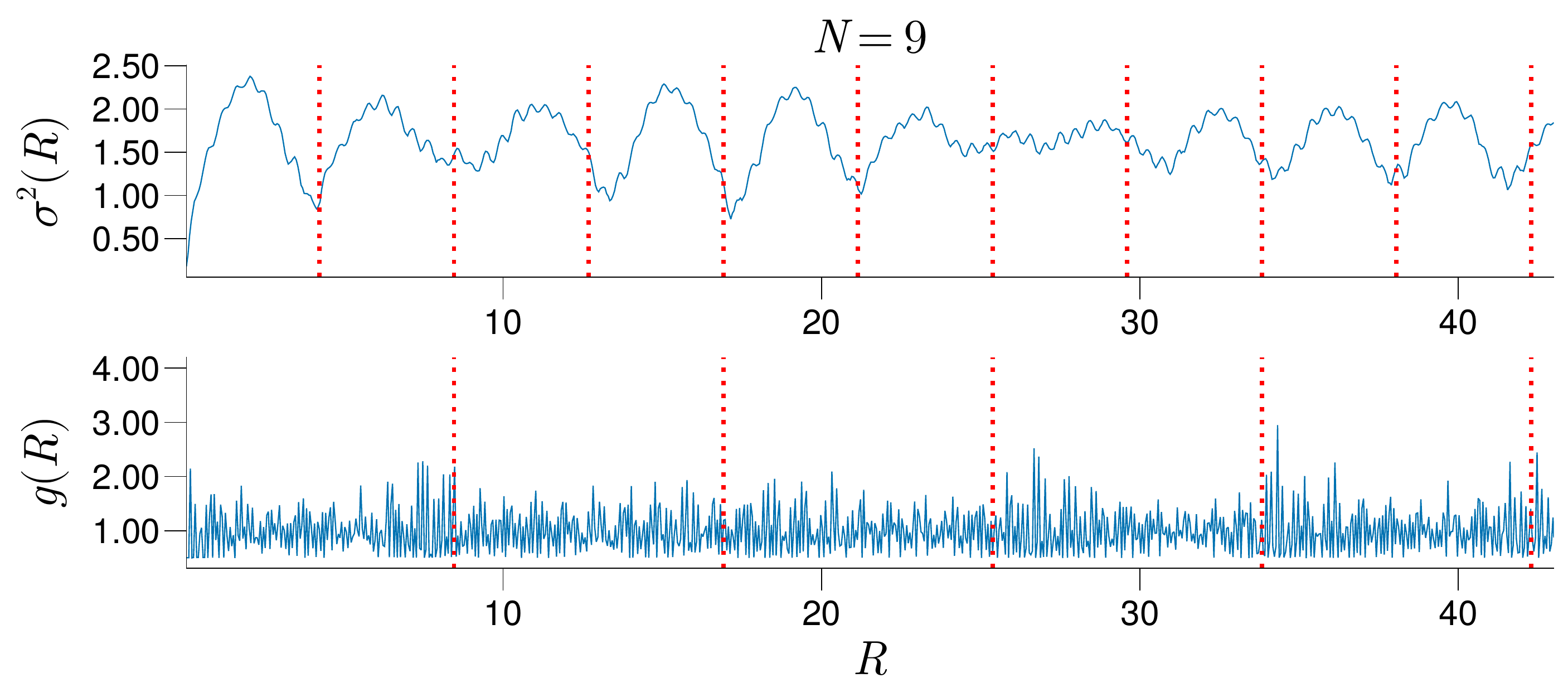}
    \end{subfigure}%
    \hfill
    \begin{subfigure}{0.4\textwidth}
        \centering
        \includegraphics[width=\linewidth]{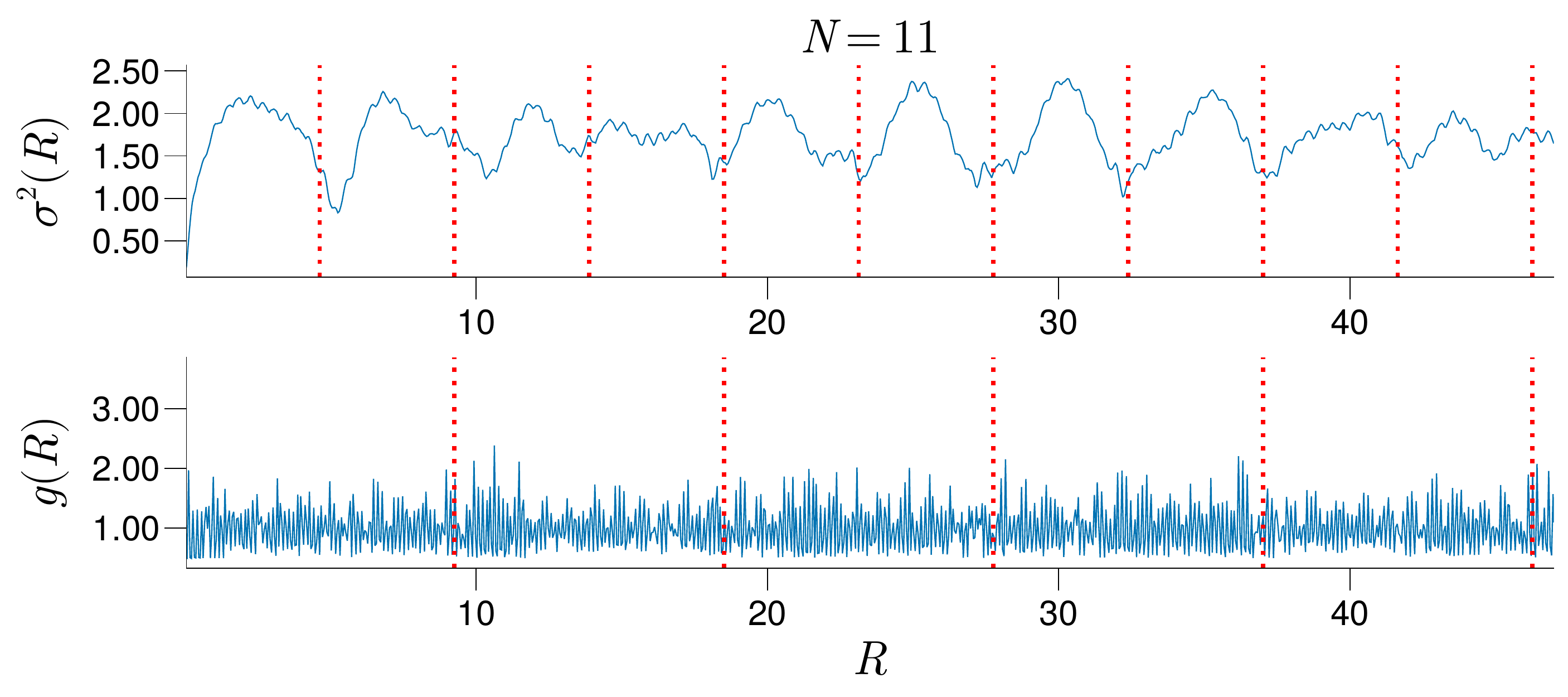}
    \end{subfigure}
    \begin{subfigure}{0.4\textwidth}
        \centering
        \includegraphics[width=\linewidth]{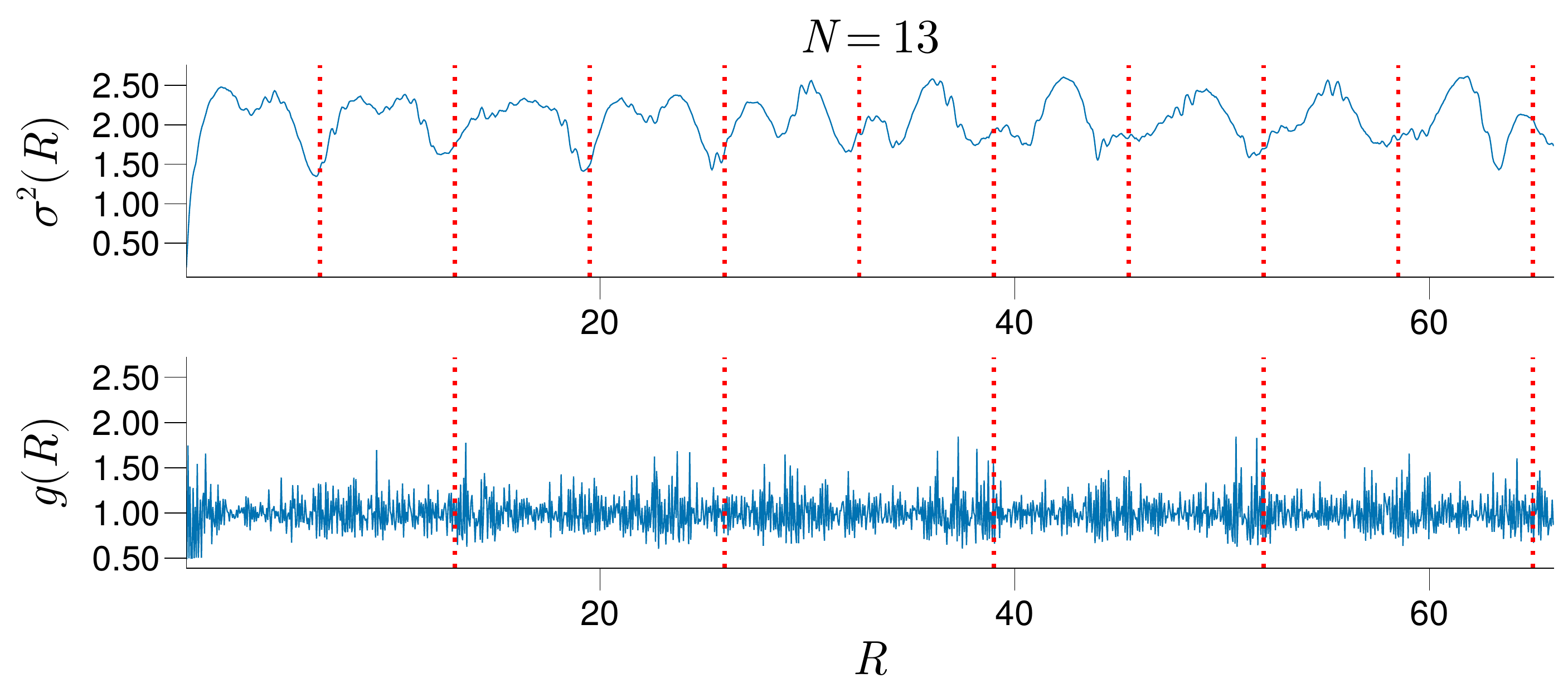}
    \end{subfigure}%
    \hfill
    \begin{subfigure}{0.4\textwidth}
        \centering
        \includegraphics[width=\linewidth]{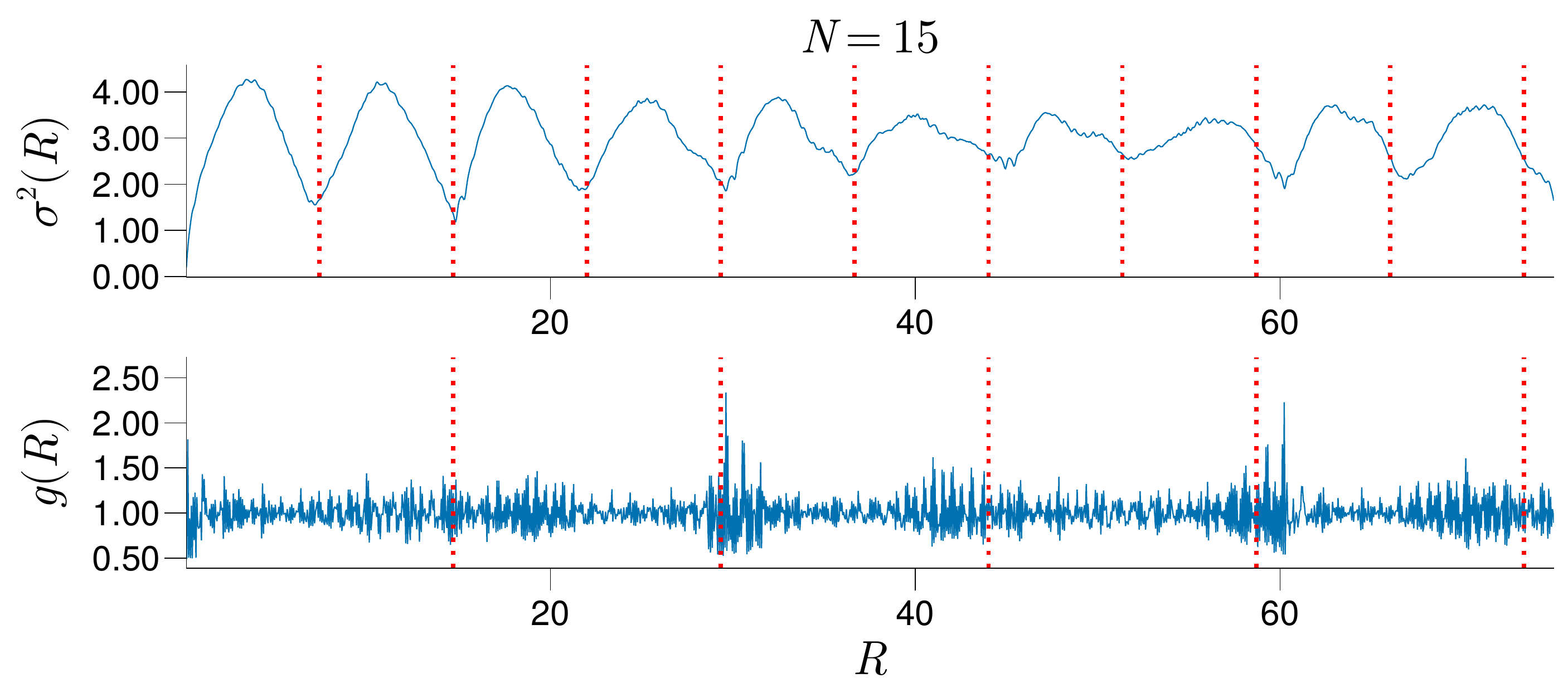}
    \end{subfigure}
    \begin{subfigure}{0.4\textwidth}
        \centering
        \includegraphics[width=\linewidth]{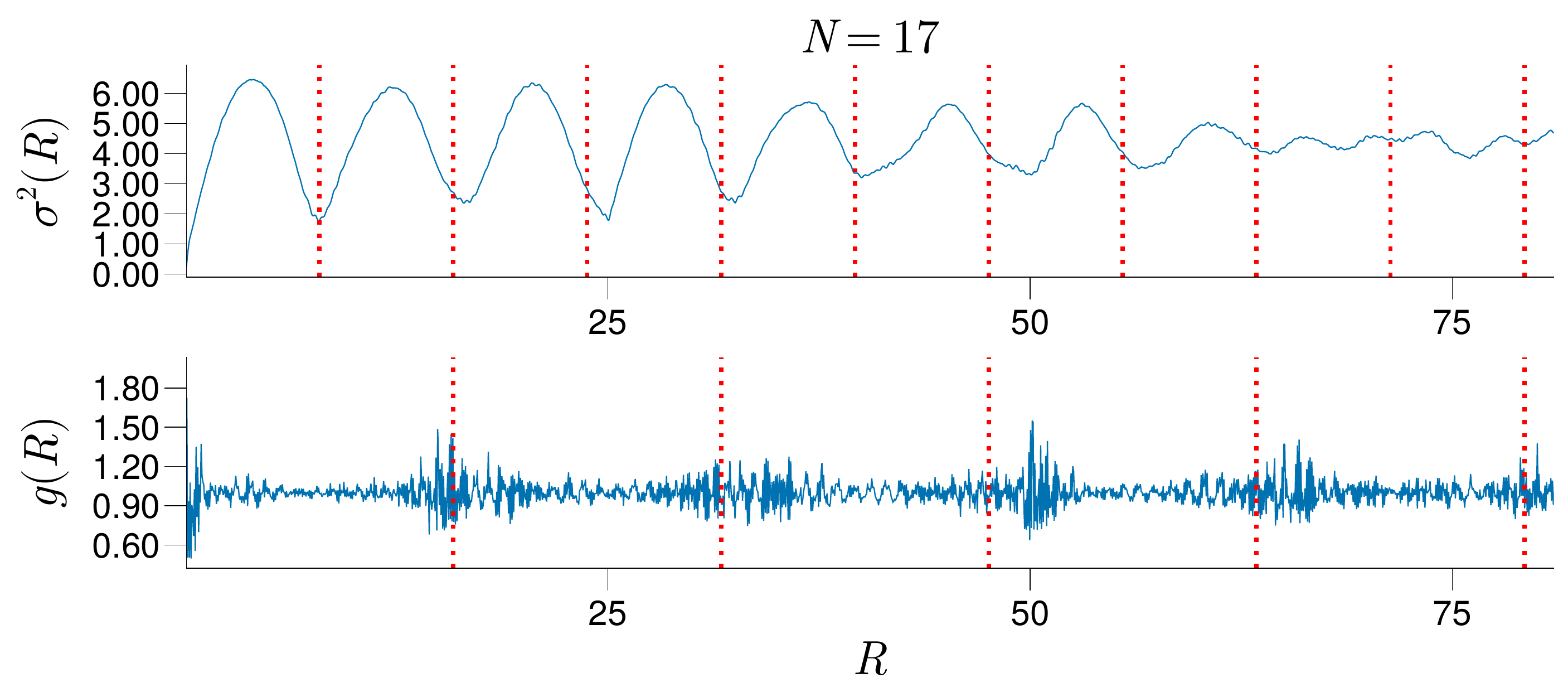}
    \end{subfigure}%
    \hfill
    \begin{subfigure}{0.4\textwidth}
        \centering
        \includegraphics[width=\linewidth]{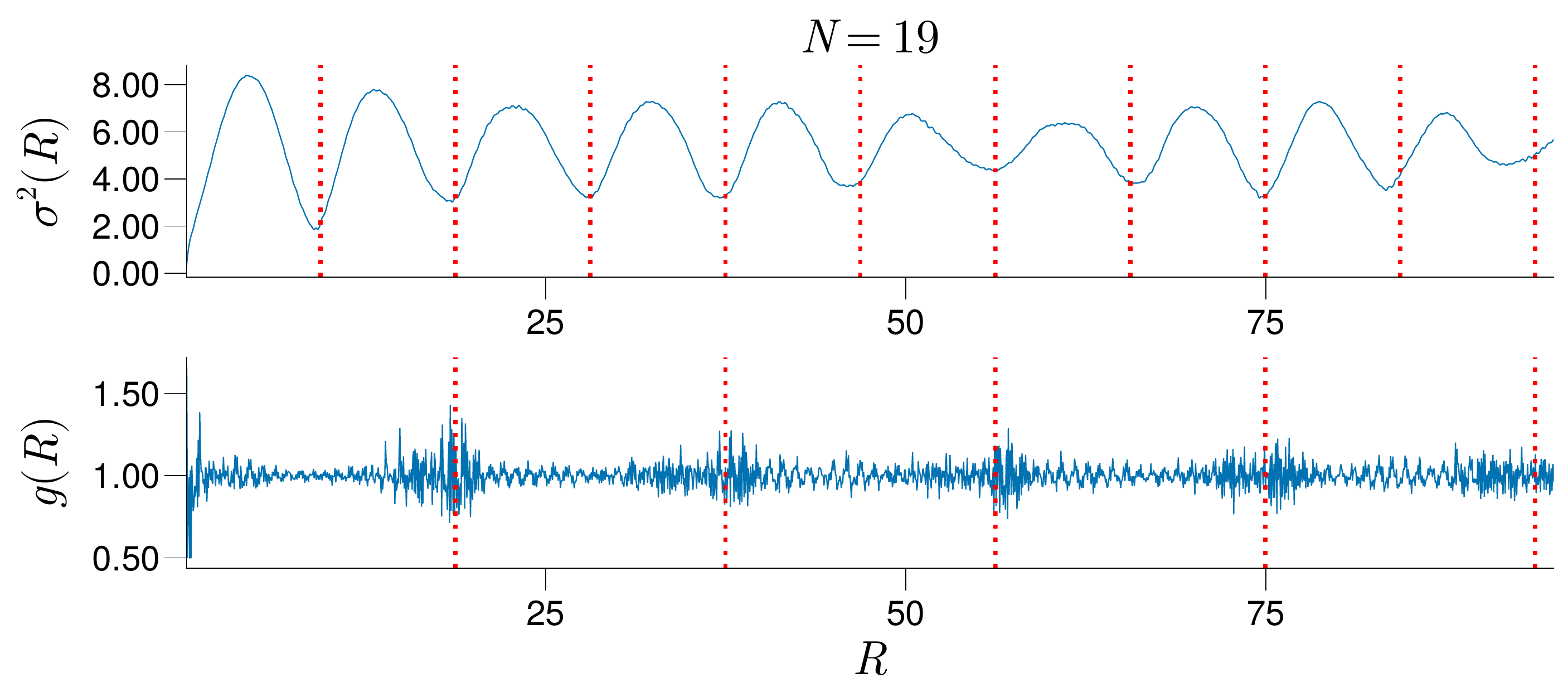}
    \end{subfigure}
    \begin{subfigure}{0.4\textwidth}
        \centering
        \includegraphics[width=\linewidth]{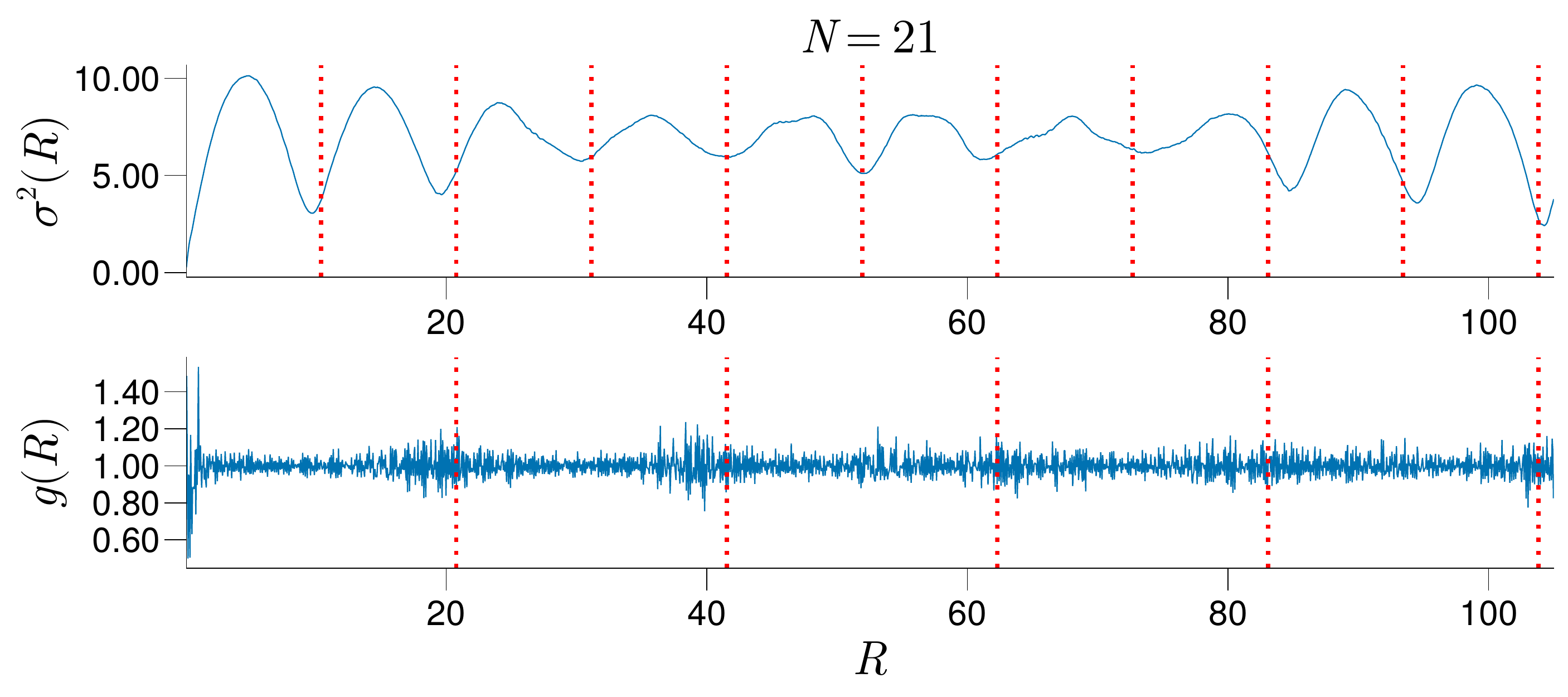}
    \end{subfigure}%
    \hfill
    \begin{subfigure}{0.4\textwidth}
        \centering
        \includegraphics[width=\linewidth]{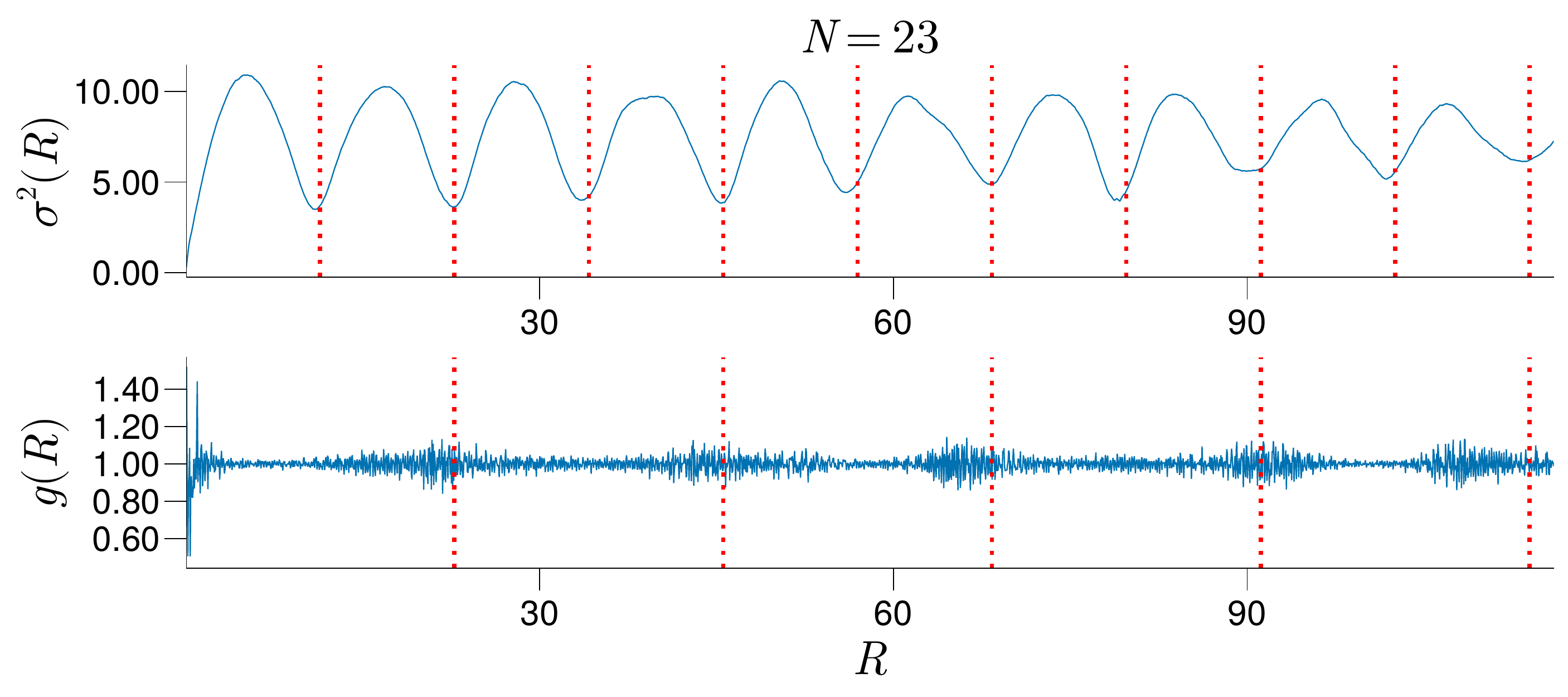}
    \end{subfigure}
    \begin{subfigure}{0.4\textwidth}
        \centering
        \includegraphics[width=\linewidth]{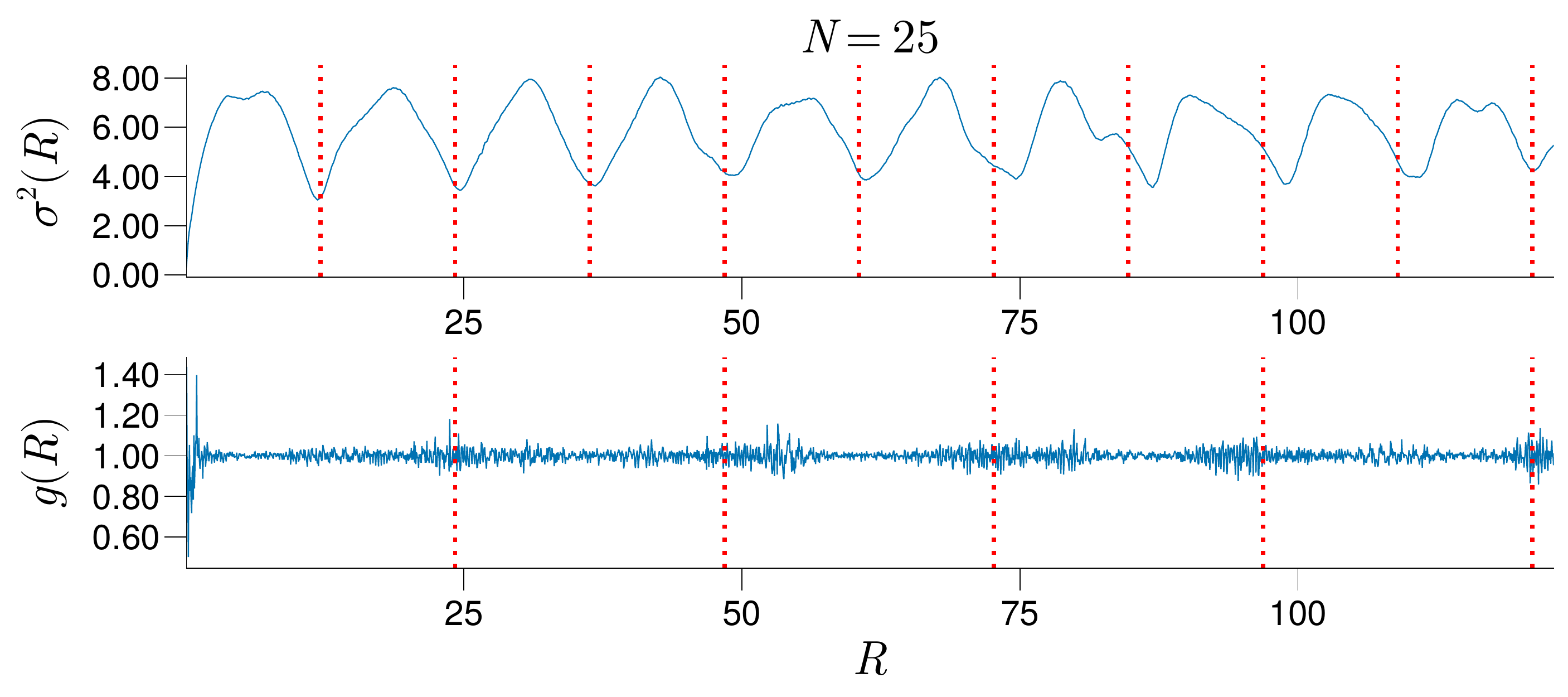}
    \end{subfigure}%
    \hfill
    \begin{subfigure}{0.4\textwidth}
        \centering
        \includegraphics[width=\linewidth]{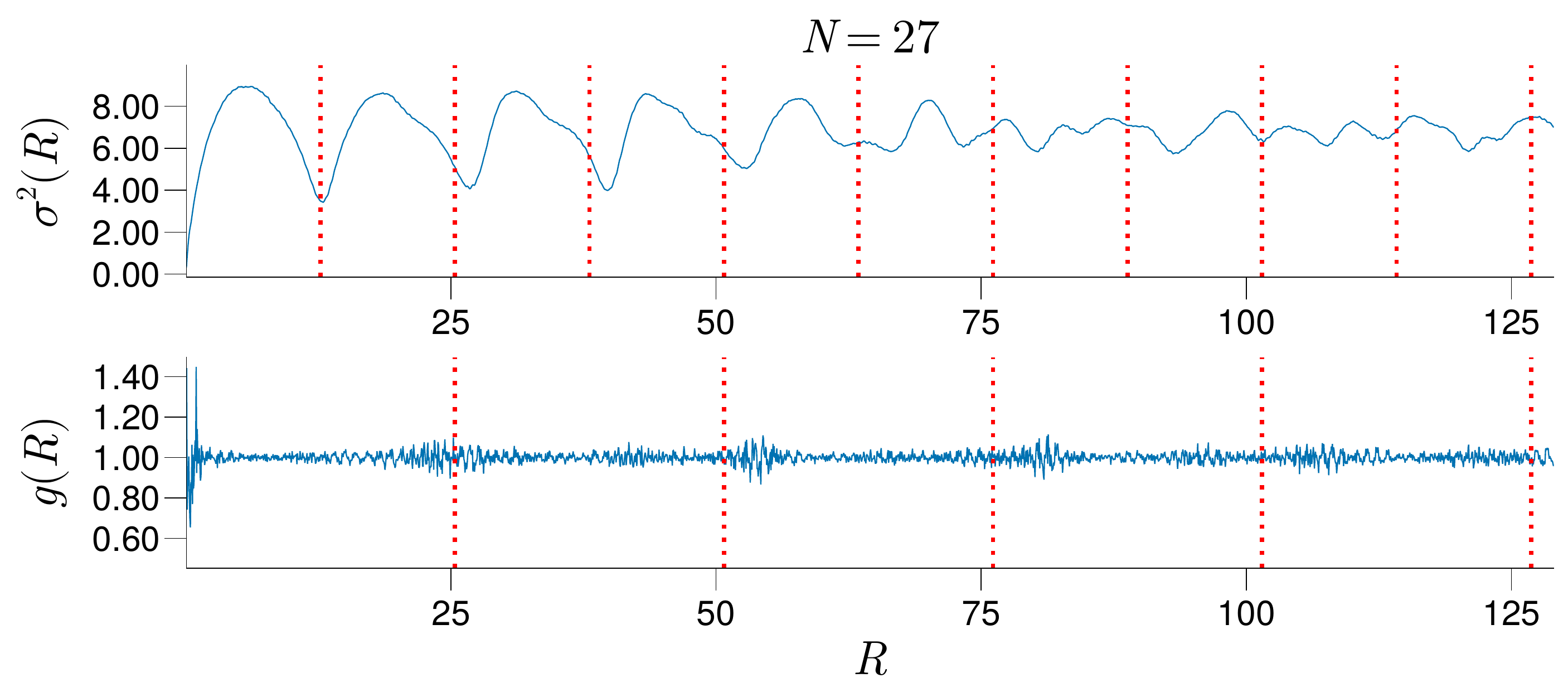}
    \end{subfigure}
    \begin{subfigure}{0.4\textwidth}
        \centering
        \includegraphics[width=\linewidth]{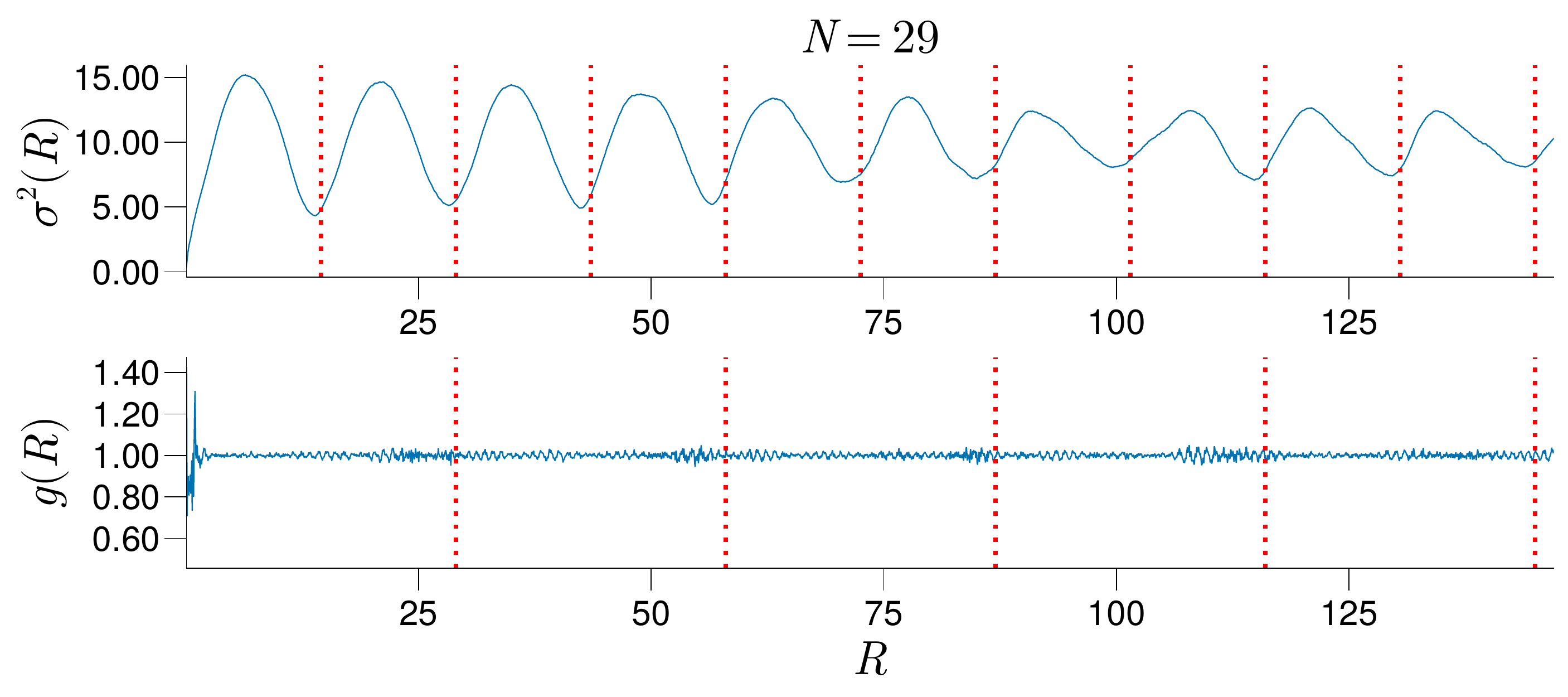}
    \end{subfigure}%
    \hfill
    \begin{subfigure}{0.4\textwidth}
        \centering
        \includegraphics[width=\linewidth]{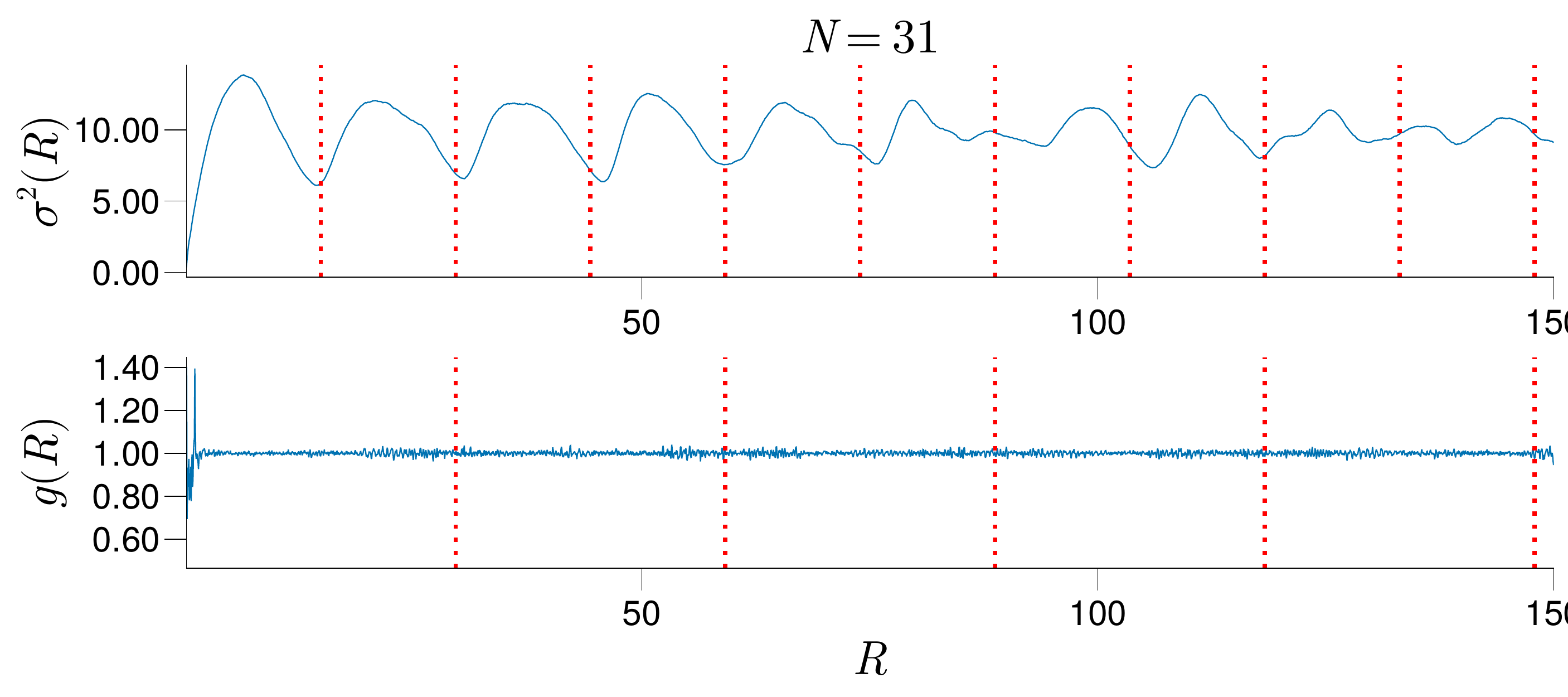}
    \end{subfigure}
    \caption{
    \textbf{Density fluctuations and pair correlations in one dimension.}
    Number variance $\sigma^2(R)$ and pair correlation function $g(R)$ for one-dimensional quasiperiodic systems with $N$-fold rotational symmetry. Dashed lines mark integer multiples of $\lambda_N = \kappa_N$ in the $\sigma^2(R)$ panels and of $2\lambda_N$ in the   $g(R)$ panels. A step size of $\Delta R = 0.05$ was used throughout.
    }
    \label{fig:1D_Sigma2_gR}
\end{figure}

Figure~\ref{fig:1D_Sigma2_gR} shows density fluctuation analysis for a broad range of one-dimensional quasiperiodic systems. In all panels, vertical dashed lines mark $\lambda_N$ and its integer multiples; note that in one dimension $\lambda_N = \kappa_N$. The procedure for extracting $\lambda_N$ is detailed in Section~\ref{sec:length-scales_1D-lambda}.

\subsection{Determination of the length scale $\lambda_{N} = \kappa_N$} \label{sec:length-scales_1D-lambda}

The number variance $\sigma^2(R)$ exhibits clear oscillatory behavior with dominant period $\lambda_N$. As in two dimensions, this period is extracted via a Fast Fourier Transform (FFT) of the data, identifying $\lambda_N = 2\pi/k_\text{peak}$ as the length associated with the peak amplitude. Figure~\ref{fig:FFT_Sigma2_N5-2-29}b shows the resulting FFT spectra for $N = 5$, $11$, $21$, $31$, $41$ and $51$ on a log-log scale, with curves vertically offset by $10^{3i}$ for $i = 0, \ldots, 5$ with $i = 0$ corresponding to $N = 5$.

Fitting the extracted values of $\lambda_N$ to a linear form $\lambda_N = aN + b$ by least squares yields $\lambda_N = 0.5009N - 0.4934$. Excluding the first data point $N = 5$ as an outlier gives $\lambda_N = 0.4933N - 0.2153$, both consistent with the theoretical expectation $\lambda_N \propto N$. The values of $\lambda_N$ and the fitted curve agree with those shown in Fig.~1f of the main text.

\subsection{Hyperuniformity and $g(R)$ decomposed by prototile} \label{sec:Sigma2_gR_PAG}

\begin{figure}
    \centering
    \begin{subfigure}{0.4\textwidth}
        \centering
        \includegraphics[width=\linewidth]{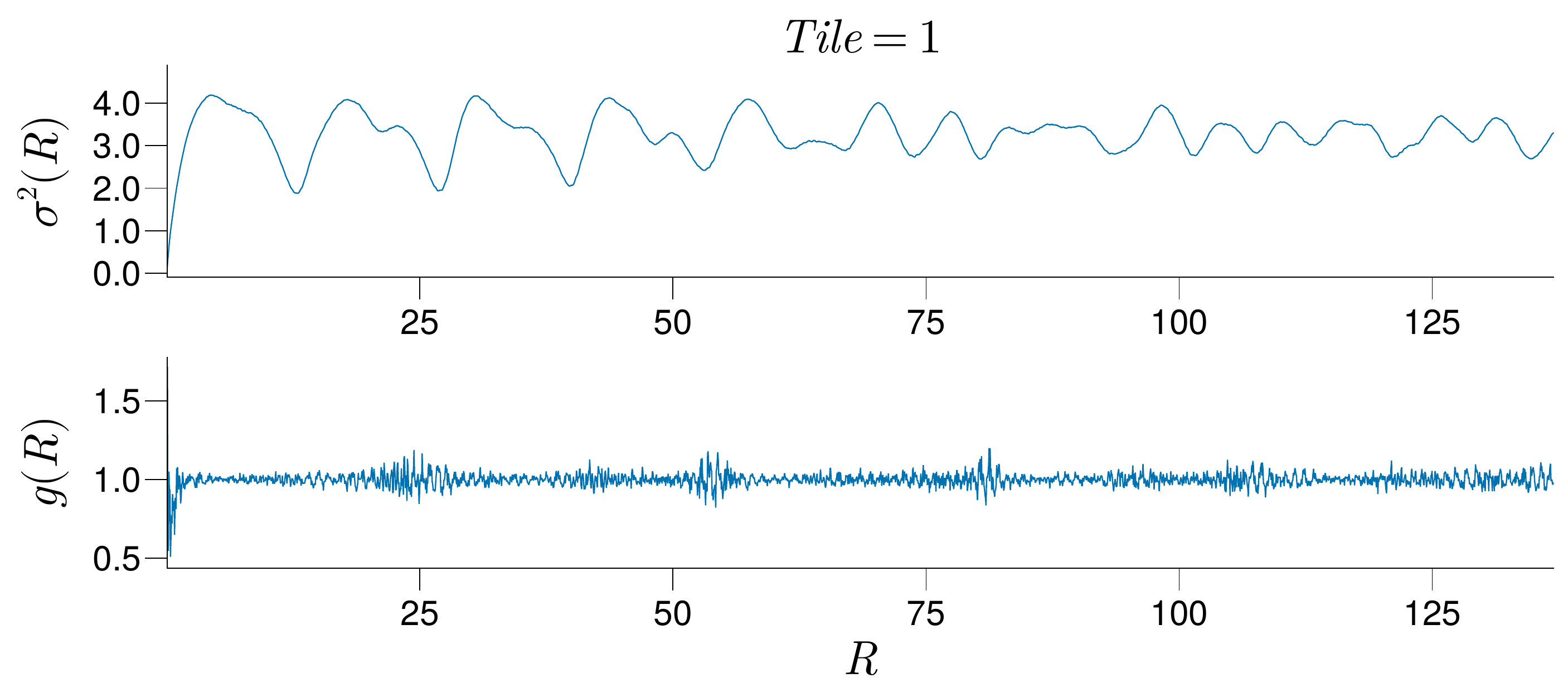}
    \end{subfigure}%
    \hfill
    \begin{subfigure}{0.4\textwidth}
        \centering
        \includegraphics[width=\linewidth]{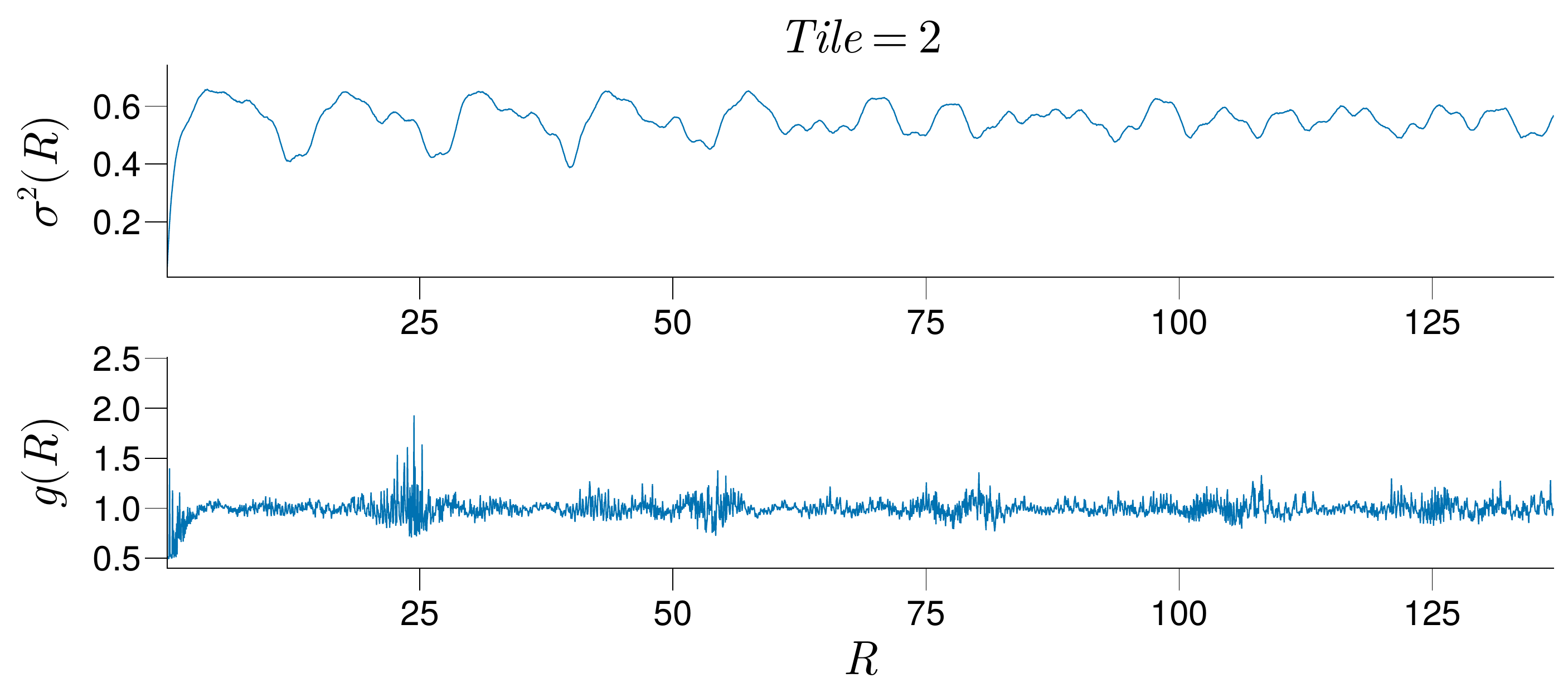}
    \end{subfigure}
    \begin{subfigure}{0.4\textwidth}
        \centering
        \includegraphics[width=\linewidth]{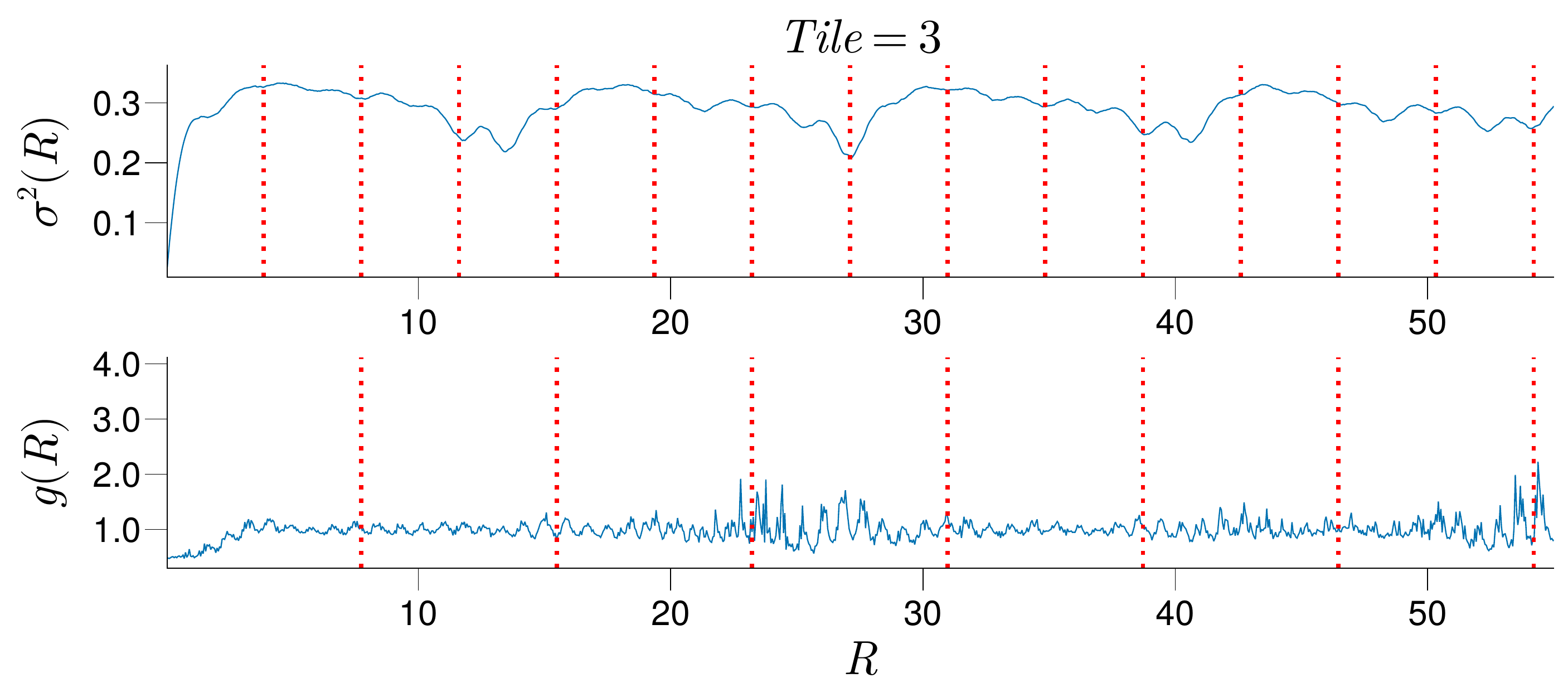}
    \end{subfigure}%
    \hfill
    \begin{subfigure}{0.4\textwidth}
        \centering
        \includegraphics[width=\linewidth]{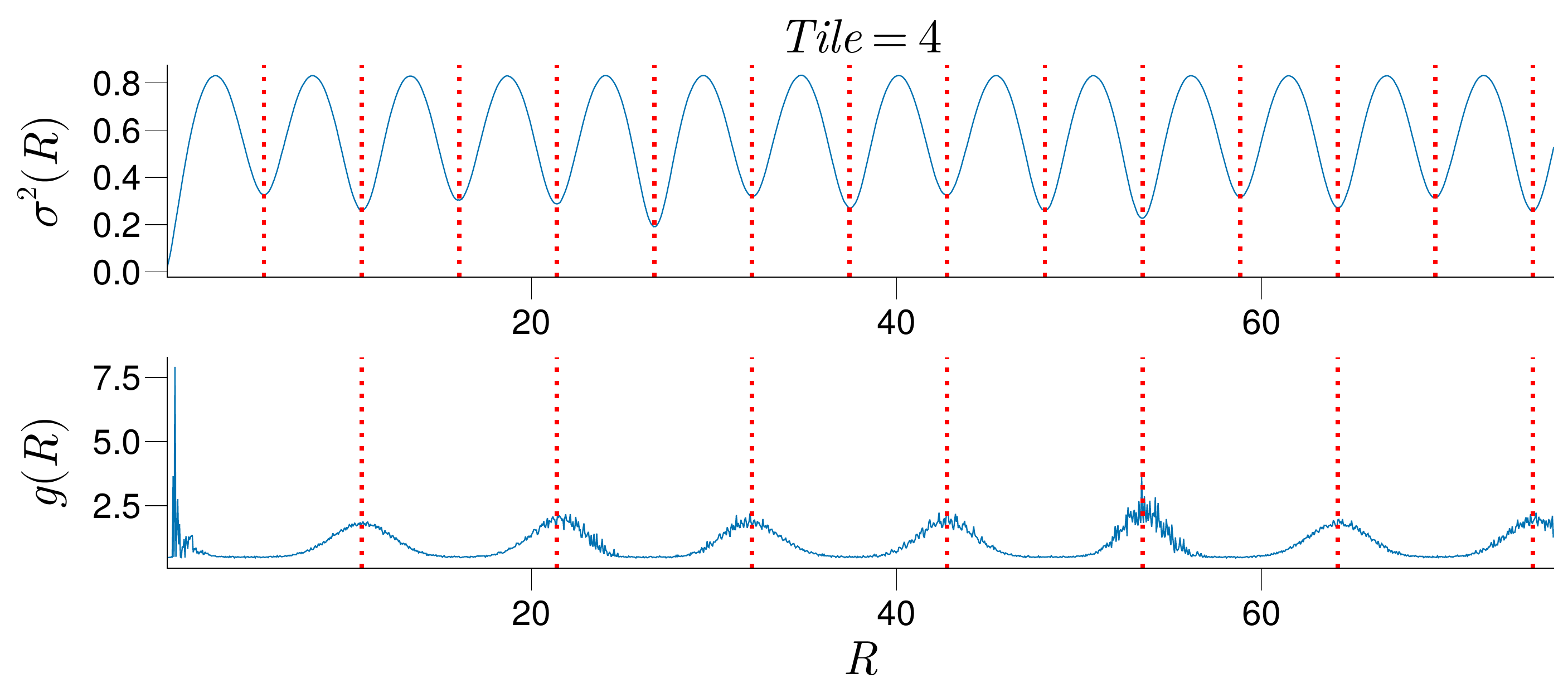}
    \end{subfigure}
    \begin{subfigure}{0.4\textwidth}
        \centering
        \includegraphics[width=\linewidth]{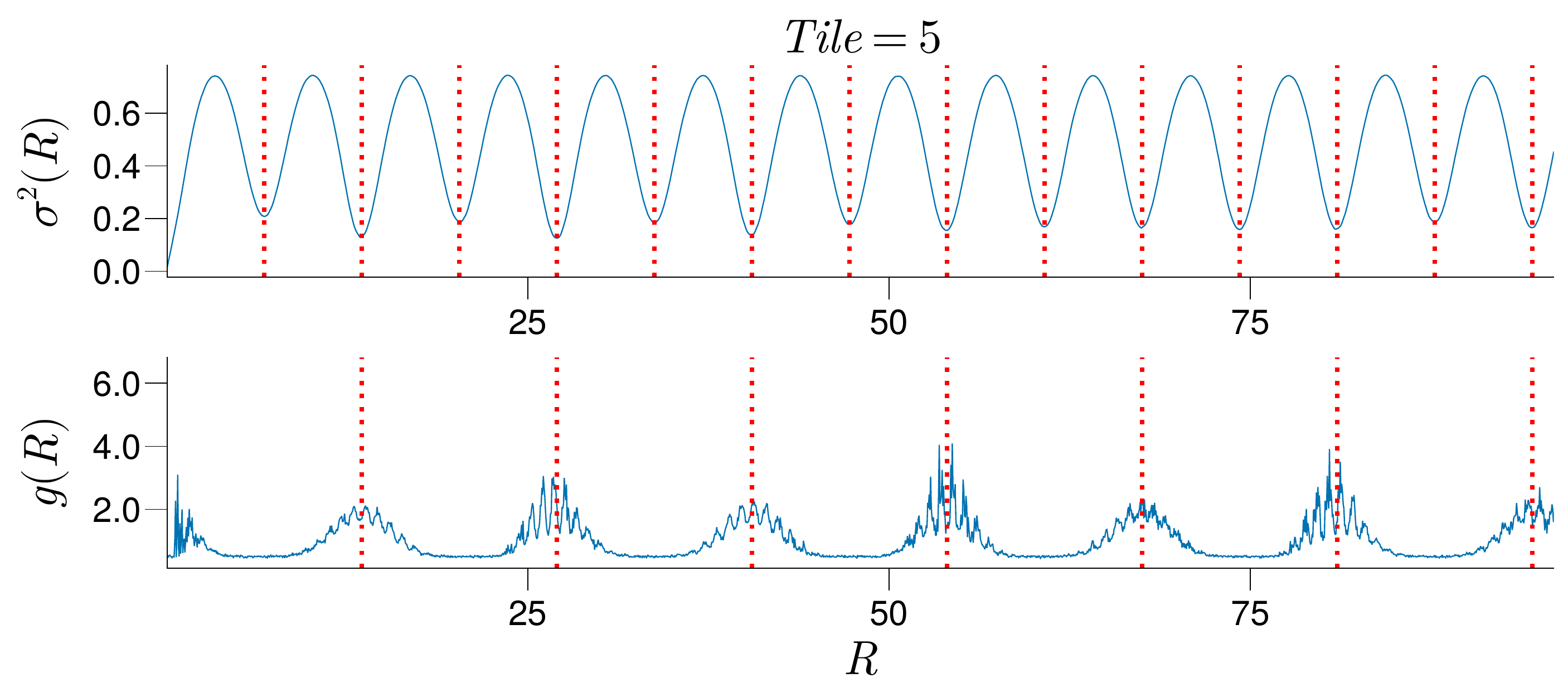}
    \end{subfigure}%
    \hfill
    \begin{subfigure}{0.4\textwidth}
        \centering
        \includegraphics[width=\linewidth]{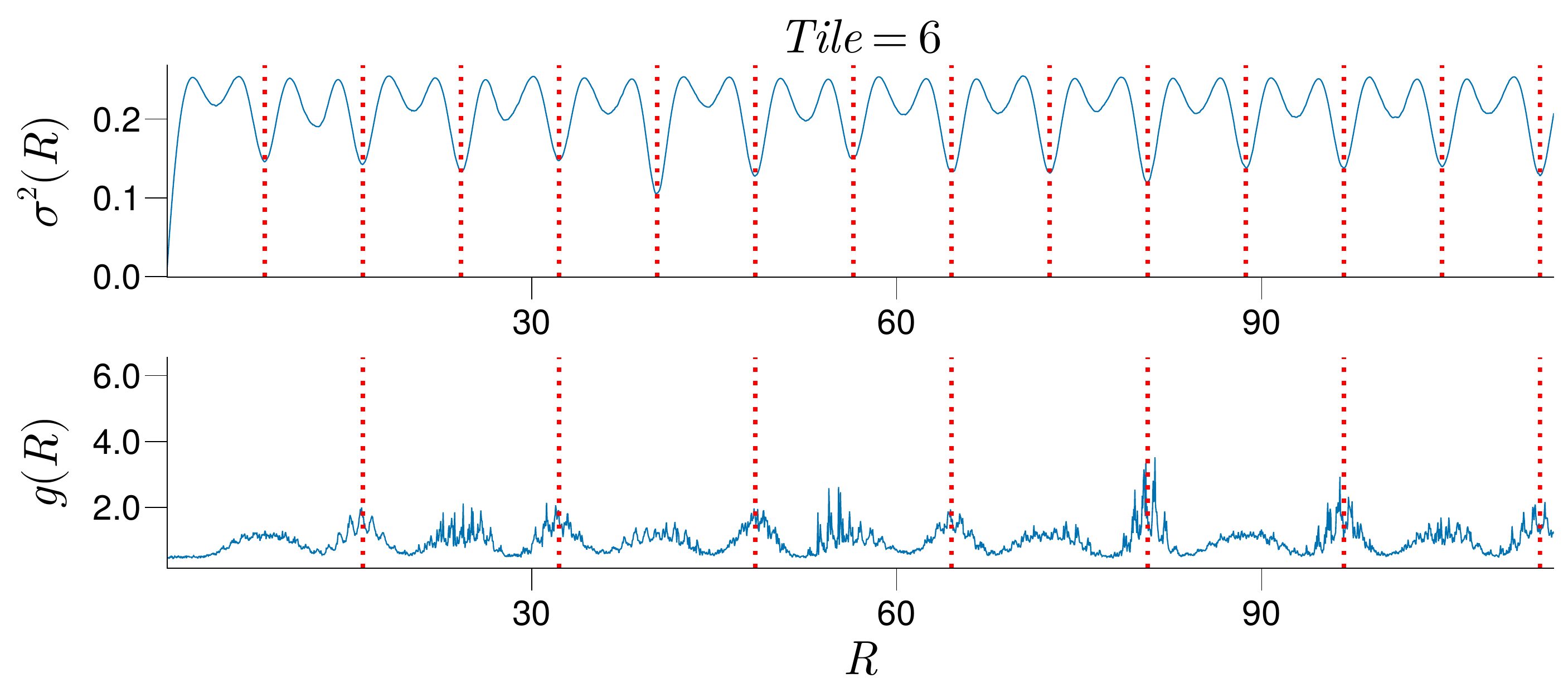}
    \end{subfigure}
    \begin{subfigure}{0.4\textwidth}
        \centering
        \includegraphics[width=\linewidth]{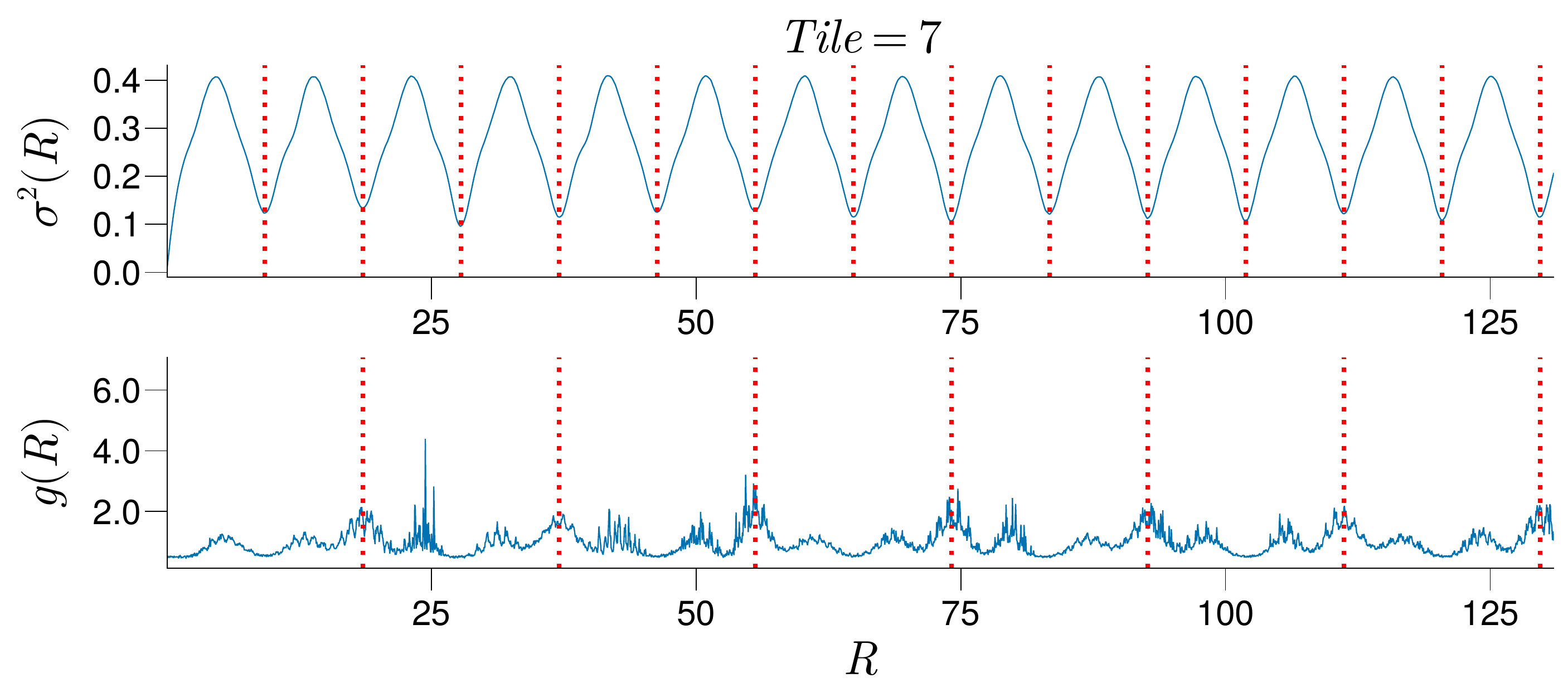}
    \end{subfigure}%
    \hfill
    \begin{subfigure}{0.4\textwidth}
        \centering
        \includegraphics[width=\linewidth]{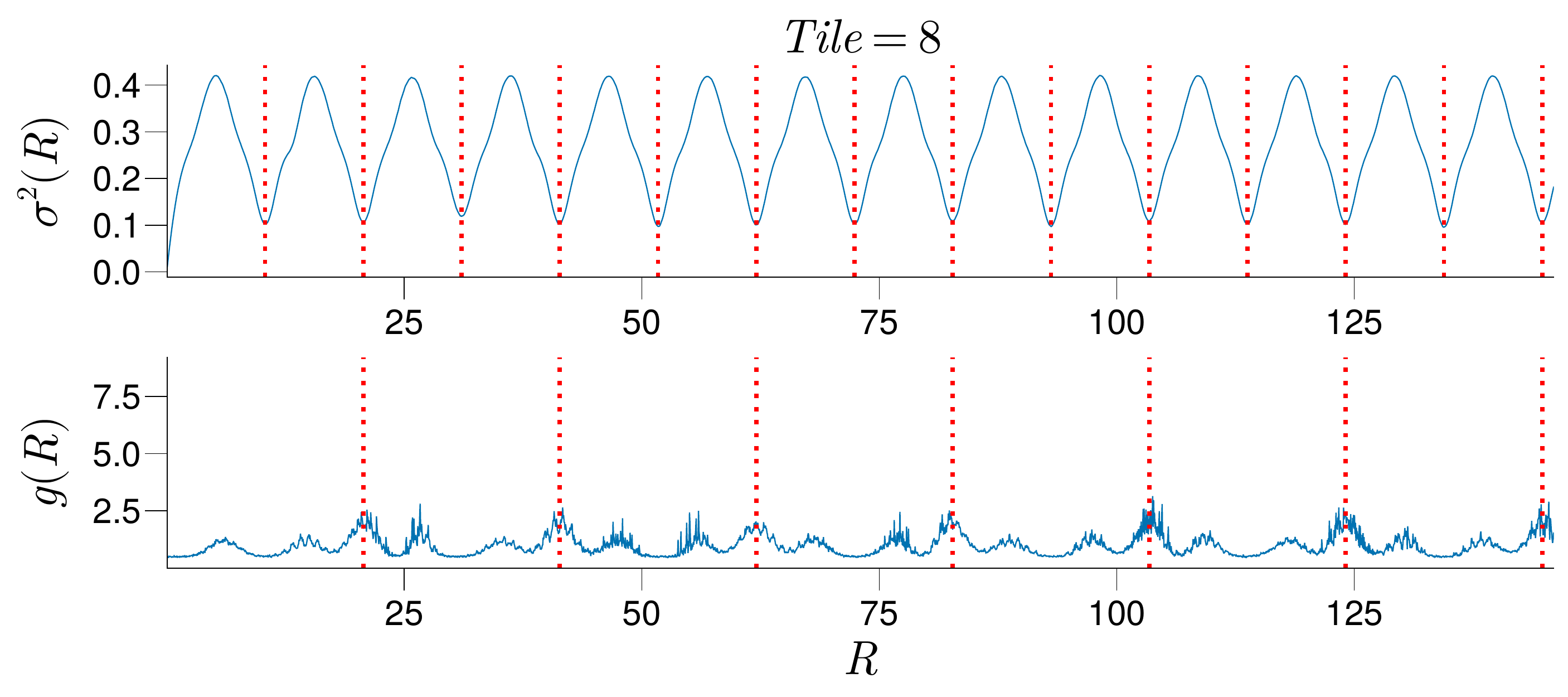}
    \end{subfigure}
    \begin{subfigure}{0.4\textwidth}
        \centering
        \includegraphics[width=\linewidth]{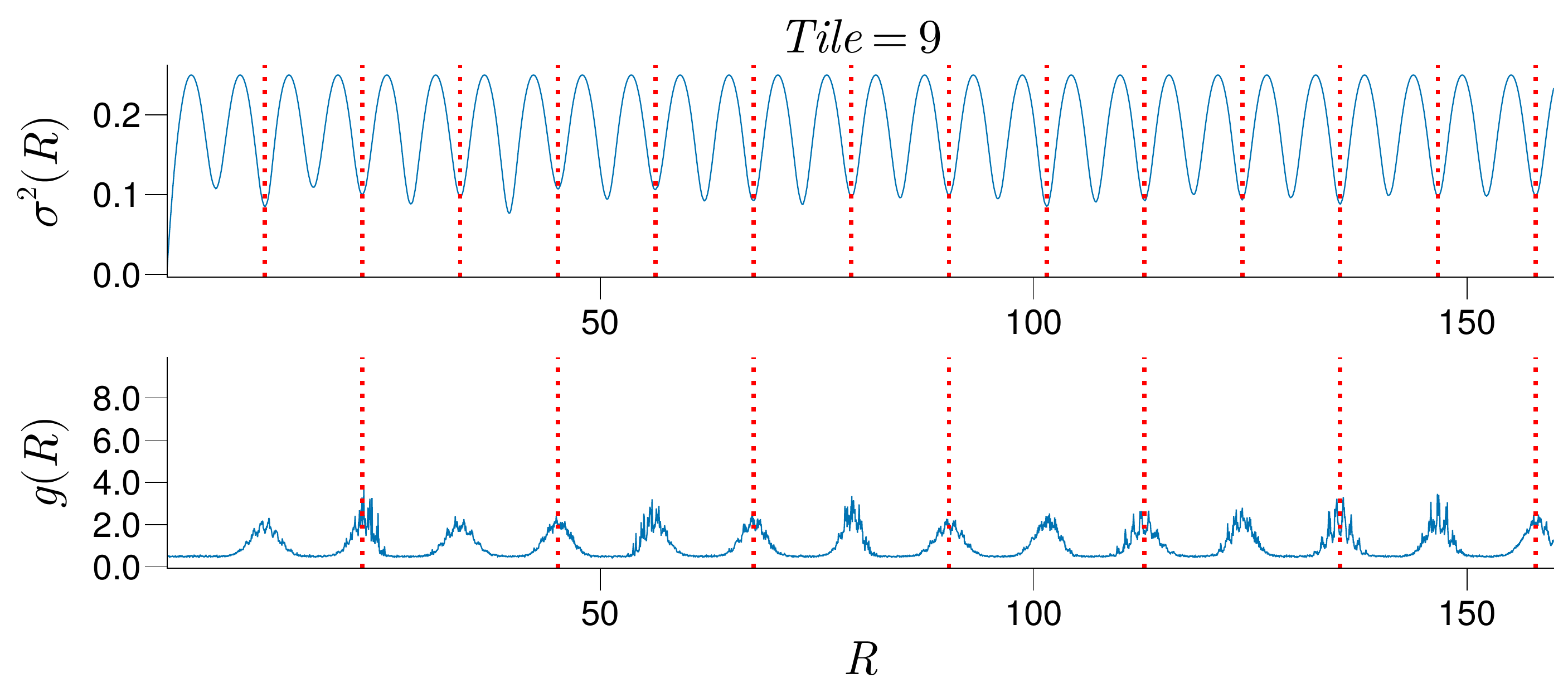}
    \end{subfigure}%
    \hfill
    \begin{subfigure}{0.4\textwidth}
        \centering
        \includegraphics[width=\linewidth]{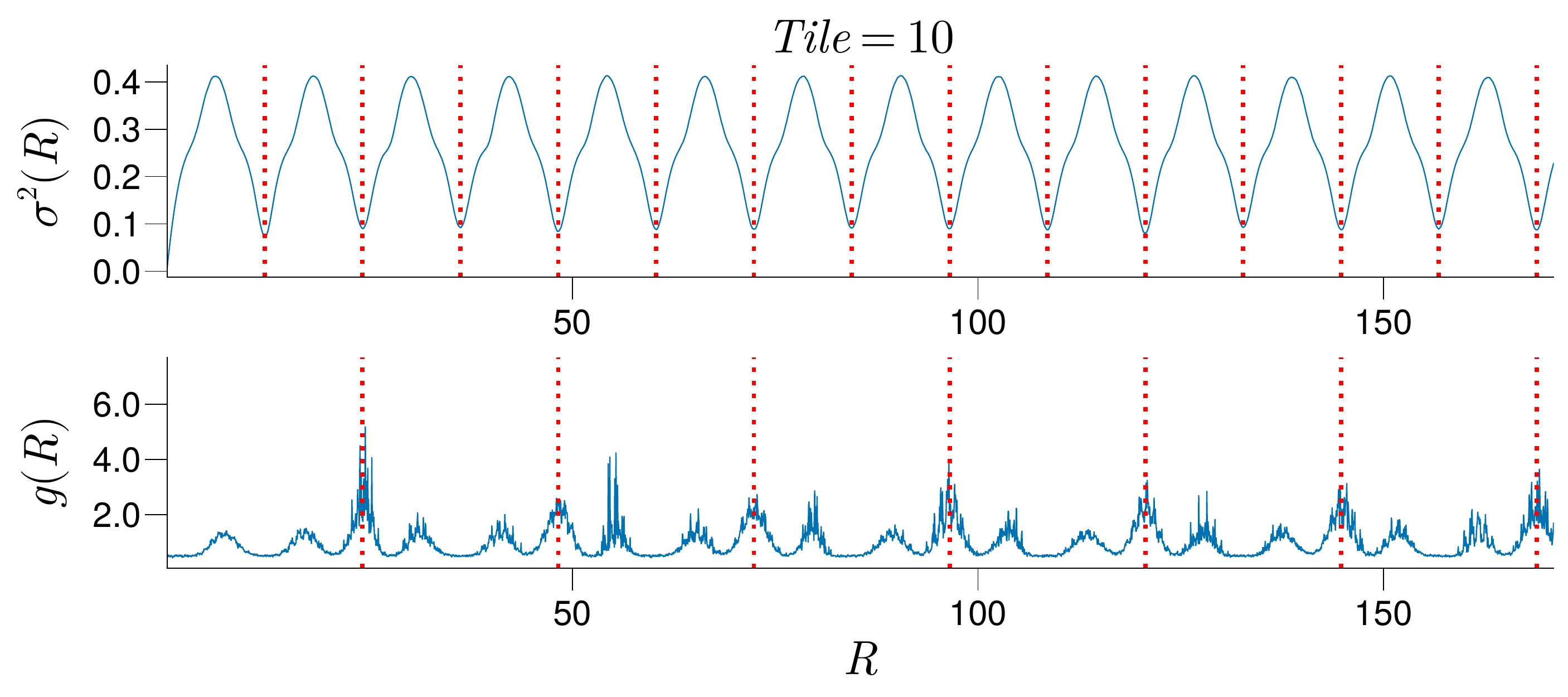}
    \end{subfigure}
    \begin{subfigure}{0.4\textwidth}
        \centering
        \includegraphics[width=\linewidth]{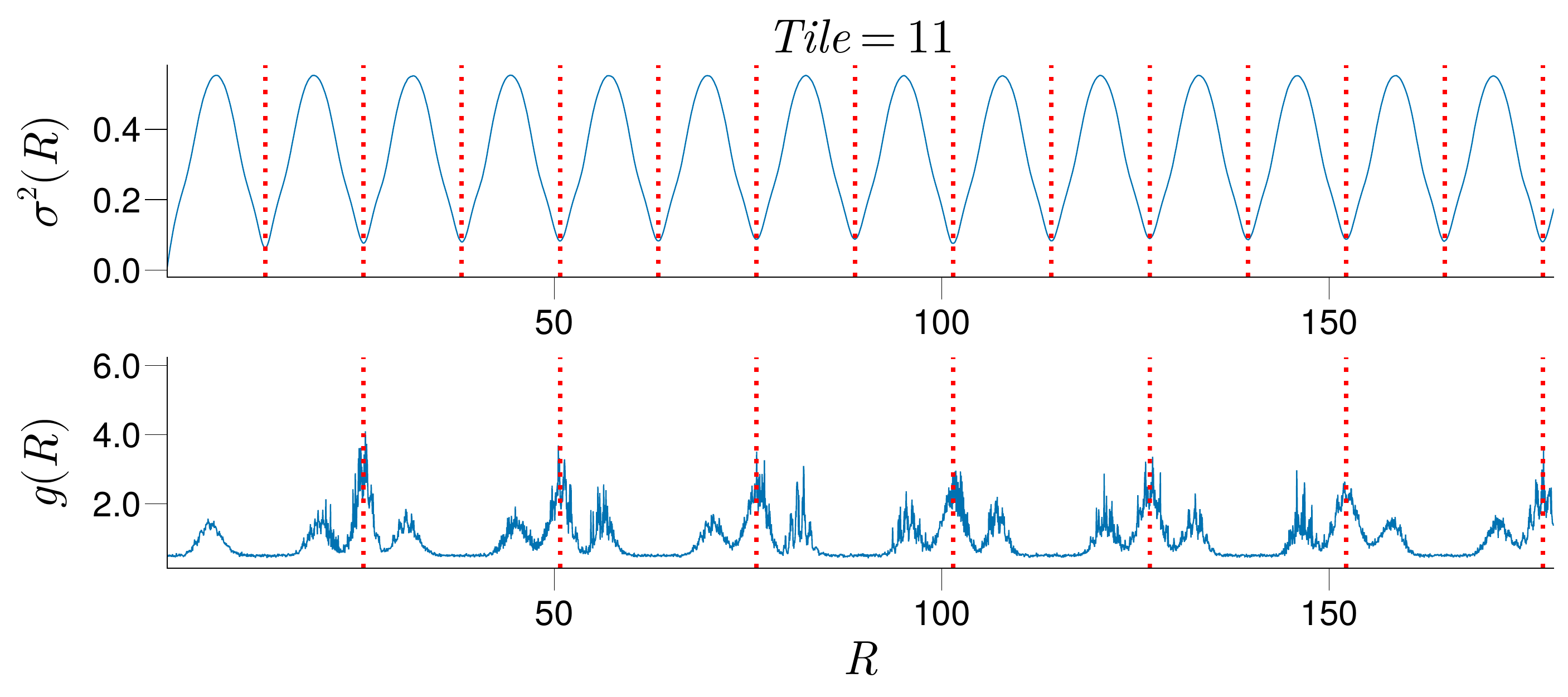}
    \end{subfigure}%
    \hfill
    \begin{subfigure}{0.4\textwidth}
        \centering
        \includegraphics[width=\linewidth]{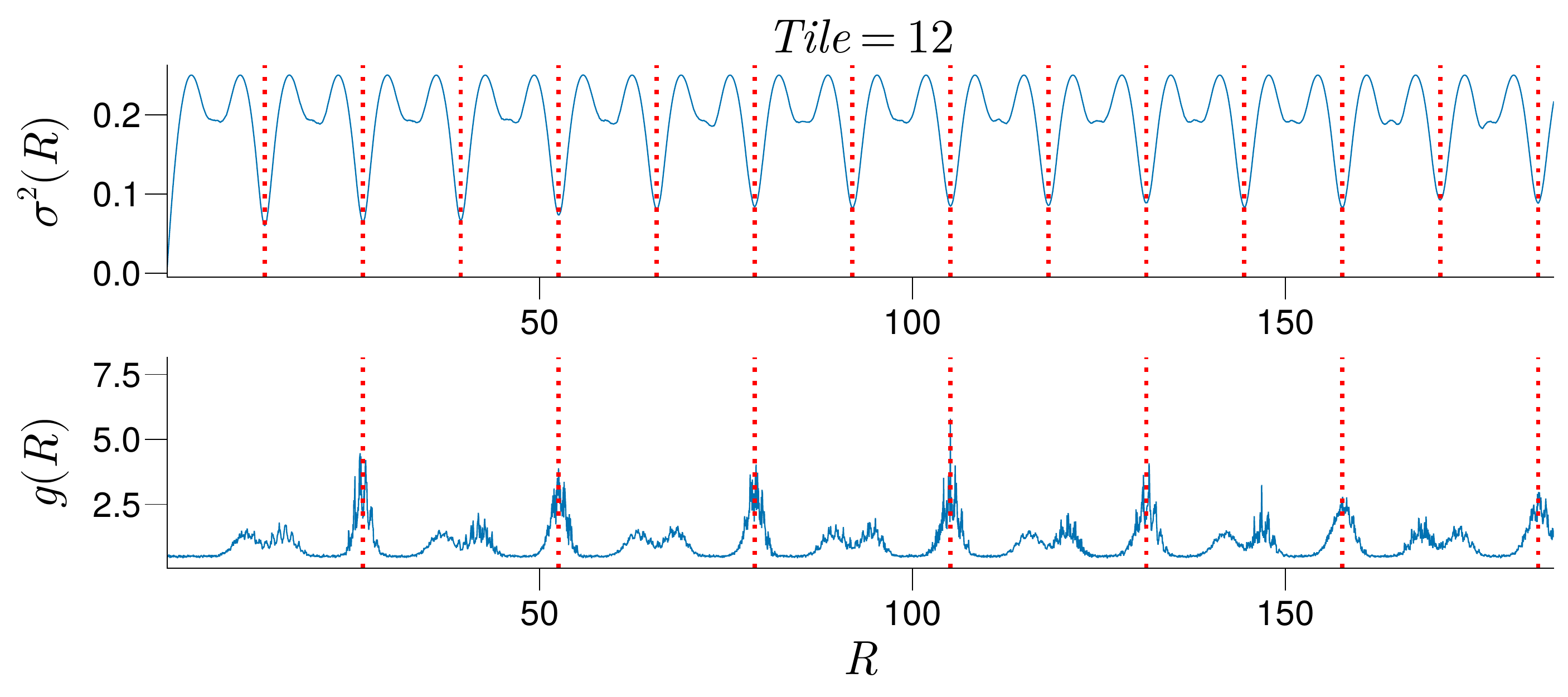}
    \end{subfigure}
    \begin{subfigure}{0.4\textwidth}
        \centering
        \includegraphics[width=\linewidth]{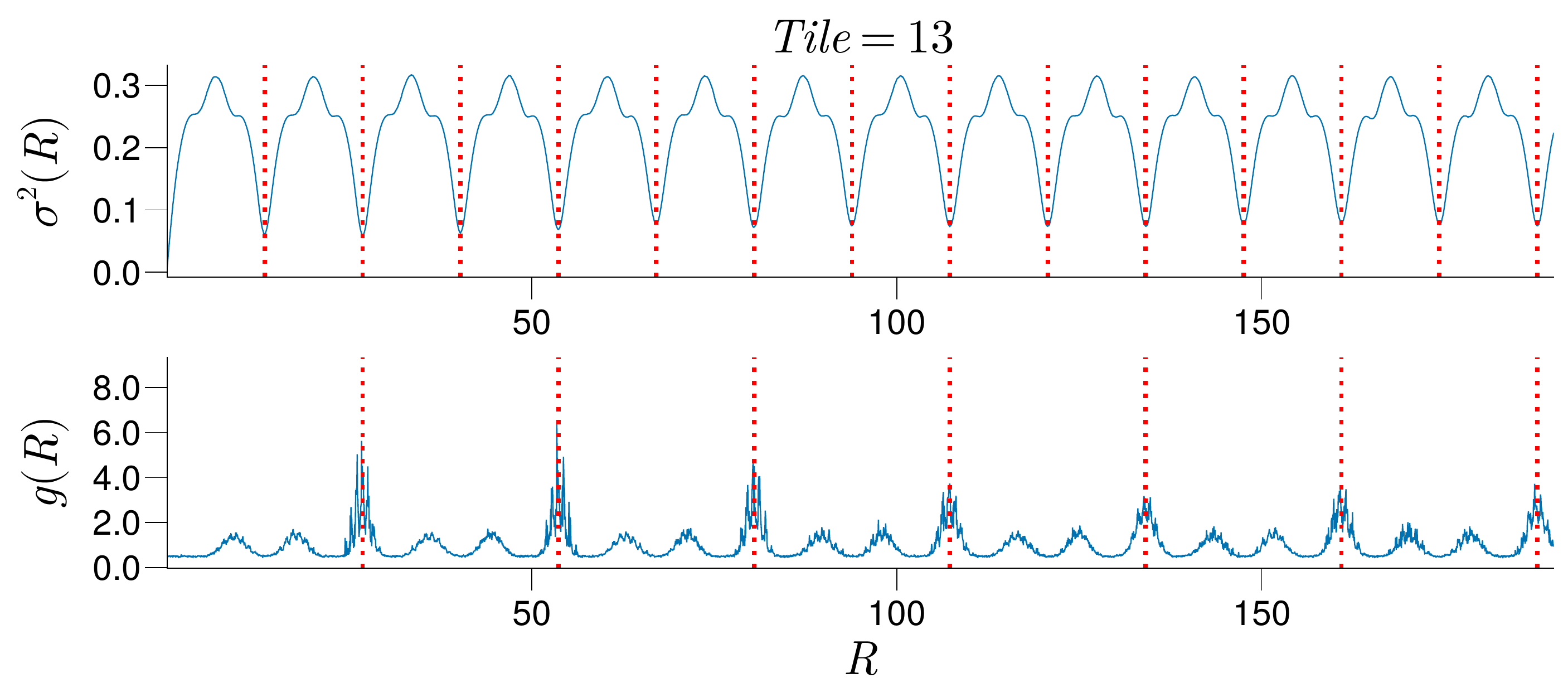}
    \end{subfigure}%
    \hfill
    \begin{subfigure}{0.4\textwidth}
        \centering
        \includegraphics[width=\linewidth]{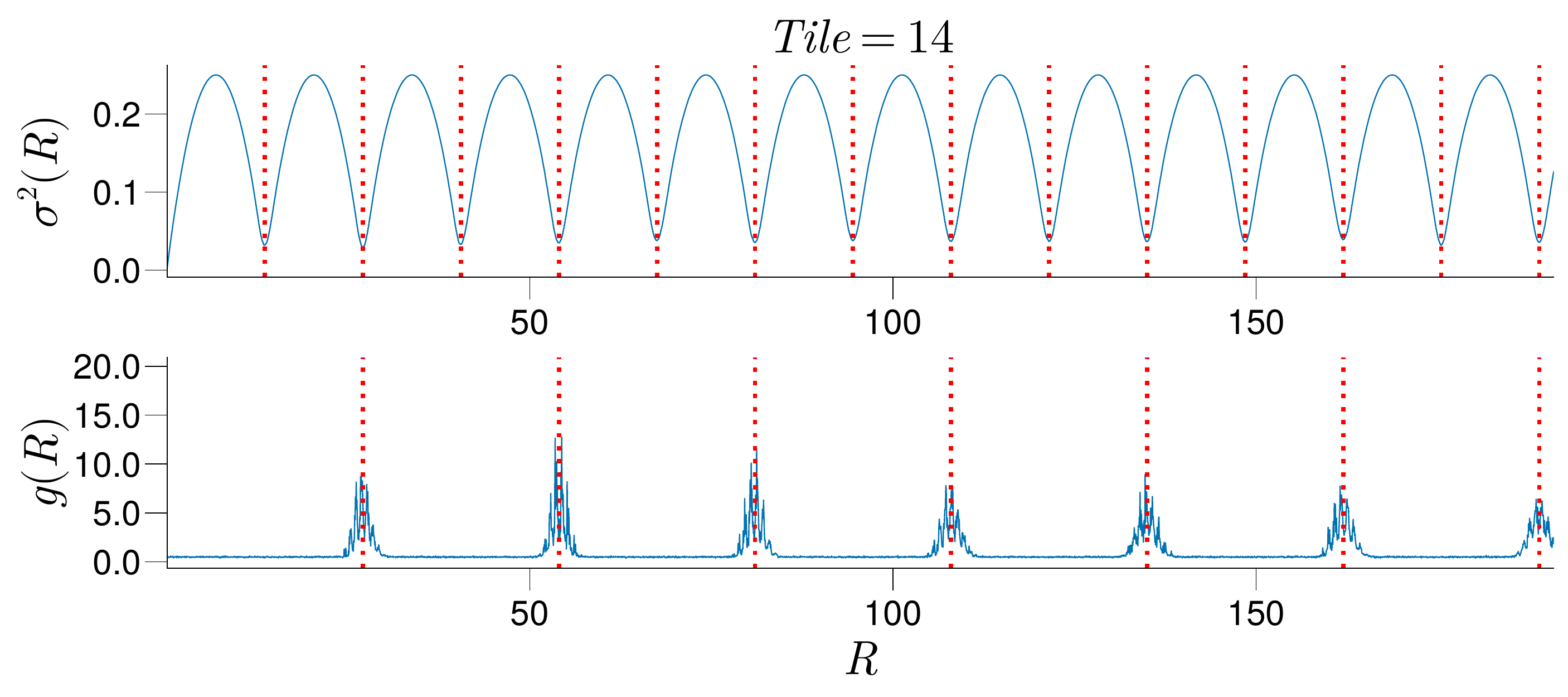}
    \end{subfigure}
    \caption{
    \textbf{Density fluctuations and pair correlations in one dimensions, resolved by prototile.}
    Number variance $\sigma^{2}(R)$ and pair correlation function $g(R)$ for a one-dimensional quasiperiodic system with $N = 27$, decomposed by prototiles. Vertical dashed lines mark integer multiples of $\lambda_N = \kappa_N$ in the $\sigma^{2}(R)$ panels and of $2\lambda_{N}$ in the $g(R)$ panels.
    }  
    \label{fig:1D_Sigma2_gR_PAG}
\end{figure}

The prototile decomposition presented in the main text for the two-dimensional $N = 23$ quasicrystal is extended here to one-dimensional systems. Figure~\ref{fig:1D_Sigma2_gR_PAG} shows $\sigma^2(R)$ and $g(R)$ for each prototile of the $N = 27$ quasicrystal, which comprises $\lfloor N/2 \rfloor + 1 = 14$ prototiles indexed in ascending order of size, with prototile $T=1$ denoting the tile with the smallest and prototile $T=14$ corresponding to the tile with the largest area.

The dominant length scale $\lambda_{N,T}$ associated with prototile $T$ is given by $\lambda_{N,T} = \frac{N}{2} \left\lvert \cos \left( \frac{2 \pi I_{T}}{N} \right) \right\rvert$, with 
$$
I_T = \begin{cases} \frac{1}{2}\!\left( \left\lfloor \tfrac{N}{2} \right\rfloor + T \right) & \text{if } T \text{ and } \left\lfloor \tfrac{N}{2} \right\rfloor \text{ have the same parity,} \\[6pt] \frac{1}{2}\!\left( \left\lceil \tfrac{N}{2} \right\rceil - T \right) & \text{otherwise.} \end{cases}
$$

In contrast to the two-dimensional case, the largest tile in one dimension exhibits the simplest, most regular density fluctuations, while smaller tiles display behavior closer to that of the full tiling.

\subsection{Determination of $\Lambda_{\infty}$}

The oscillatory behavior of $\sigma^2(R)$ raises the question of how its cycle-averaged amplitude $\Lambda_\infty$ depends on rotational symmetry $N$. We compute $\Lambda_\infty$ as the mean of $\sigma^2(R)$ over the interval $[R_i, R_f] = [\lambda_N,\, M\lambda_N]$, where $M = \lfloor R_\text{max}/\lambda_N \rfloor$ is the number of complete oscillation cycles within the available data.

Applying this procedure to all odd rotational symmetries from $N = 5$ to $N = 51$ yields the results shown in Fig.~\ref{fig:LambdaInf_vs_N_2D}b. The data are well described by the least-squares fit $\Lambda_\infty(N) = -0.608 + 0.0848\,N^{1.425}$, with error bars indicating the standard deviation of each data point.


\bibliography{apssamp} 

@article{Park,
  title = {High Frictional Anisotropy of Periodic and Aperiodic Directions on a Quasicrystal Surface},
  author = {Park, J. Y. and Ogletree, D. F. and Salmeron, M. and Ribeiro, R. A. and Canfield, P. C. and Jenks, C. J. and Thiel, P. A.},
  journal = {Science},
  volume = {309},
  number = {5739},
  pages = {1354--1356},
  year = {2005},
  publisher = {American Association for the Advancement of Science},
  doi = {10.1126/science.1113239},
  url = {https://science.org}
}

@article{Jang,
  title     = {Superior room-temperature ductility of typically brittle quasicrystals at small sizes},
  author    = {Jang, J. I. and Li, M. and Bhattacharya, S. and {others}},
  journal   = {Nature Communications},
  volume    = {7},
  pages     = {12261},
  year      = {2016},
  publisher = {Nature Publishing Group},
  doi       = {10.1038/ncomms12261},
  url       = {https://www.nature.com/articles/ncomms12261}
}

@article{Trebin1998,
  title = {Crack Propagation in Quasicrystals},
  author = {Mikulla, R. and Stadler, J. and Krul, F. and Trebin, H.-R. and Gumbsch, P.},
  journal = {Phys. Rev. Lett.},
  volume = {81},
  issue = {15},
  pages = {3163--3166},
  numpages = {0},
  year = {1998},
  month = {Oct},
  publisher = {American Physical Society},
  doi = {10.1103/PhysRevLett.81.3163},
  url = {https://link.aps.org/doi/10.1103/PhysRevLett.81.3163}
}

@article{Lu2007,
  title = {Decagonal and Quasi-Crystalline Tilings in Medieval Islamic Architecture},
  author = {Lu, Peter J. and Steinhardt, Paul J.},
  journal = {Science},
  volume = {315},
  number = {5815},
  pages = {1106--1110},
  year = {2007},
  doi = {10.1126/science.1135491},
  url = {https://science.org}
}

@book{Baake2013, place={Cambridge}, series={Encyclopedia of Mathematics and its Applications}, title={Aperiodic Order}, publisher={Cambridge University Press}, author={Baake, Michael and Grimm, Uwe}, year={2013}, collection={Encyclopedia of Mathematics and its Applications}}

@article{phason1,
  title = {Elasticity and Dislocations in Pentagonal and Icosahedral Quasicrystals},
  author = {Levine, Dov and Lubensky, T. C. and Ostlund, Stellan and Ramaswamy, Sriram and Steinhardt, Paul Joseph and Toner, John},
  journal = {Phys. Rev. Lett.},
  volume = {54},
  issue = {14},
  pages = {1520--1523},
  numpages = {0},
  year = {1985},
  month = {Apr},
  publisher = {American Physical Society},
  doi = {10.1103/PhysRevLett.54.1520},
  url = {https://link.aps.org/doi/10.1103/PhysRevLett.54.1520}
}

@article{marginal_corwin1,
  title = {Vibrational Properties of Hard and Soft Spheres Are Unified at Jamming},
  author = {Arceri, Francesco and Corwin, Eric I.},
  journal = {Phys. Rev. Lett.},
  volume = {124},
  issue = {23},
  pages = {238002},
  numpages = {6},
  year = {2020},
  month = {Jun},
  publisher = {American Physical Society},
  doi = {10.1103/PhysRevLett.124.238002},
  url = {https://link.aps.org/doi/10.1103/PhysRevLett.124.238002}
}

@article{marginal_liu1,
  title = {Vibrations and Diverging Length Scales Near the Unjamming Transition},
  author = {Silbert, Leonardo E. and Liu, Andrea J. and Nagel, Sidney R.},
  journal = {Phys. Rev. Lett.},
  volume = {95},
  issue = {9},
  pages = {098301},
  numpages = {4},
  year = {2005},
  month = {Aug},
  publisher = {American Physical Society},
  doi = {10.1103/PhysRevLett.95.098301},
  url = {https://link.aps.org/doi/10.1103/PhysRevLett.95.098301}
}

@article{marginal_wyart,
  title = {Marginal Stability Constrains Force and Pair Distributions at Random Close Packing},
  author = {Wyart, Matthieu},
  journal = {Phys. Rev. Lett.},
  volume = {109},
  issue = {12},
  pages = {125502},
  numpages = {5},
  year = {2012},
  month = {Sep},
  publisher = {American Physical Society},
  doi = {10.1103/PhysRevLett.109.125502},
  url = {https://link.aps.org/doi/10.1103/PhysRevLett.109.125502}
}

@article{
glass_spectrum_franz,
author = {Silvio Franz  and Giorgio Parisi  and Pierfrancesco Urbani  and Francesco Zamponi },
title = {Universal spectrum of normal modes in low-temperature glasses},
journal = {Proceedings of the National Academy of Sciences},
volume = {112},
number = {47},
pages = {14539-14544},
year = {2015},
doi = {10.1073/pnas.1511134112},
URL = {https://www.pnas.org/doi/abs/10.1073/pnas.1511134112}
}

@article{glasses_Lerner,
    author = {Lerner, Edan and Bouchbinder, Eran},
    title = {Low-energy quasilocalized excitations in structural glasses},
    journal = {The Journal of Chemical Physics},
    volume = {155},
    number = {20},
    pages = {200901},
    year = {2021},
    month = {11},
    issn = {0021-9606},
    doi = {10.1063/5.0069477},
    url = {https://doi.org/10.1063/5.0069477}
}

@article{glass_ediger2000,
  title={Spatially heterogeneous dynamics in supercooled liquids},
  author={Ediger, Mark D},
  journal={Annual Review of Physical Chemistry},
  volume={51},
  number={1},
  pages={99--128},
  year={2000},
  publisher={Annual Reviews 4139 El Camino Way, PO Box 10139, Palo Alto, CA 94303-0139, USA}
}

@inbook{glass_biroli,
  author    = {Biroli, Giulio and Bouchaud, Jean-Philippe},
  publisher = {John Wiley \& Sons, Ltd},
  address   = {Hoboken, NJ},
  isbn      = {9781118202470},
  title     = {The Random First-Order Transition Theory of Glasses: A Critical Assessment},
  booktitle = {Structural Glasses and Supercooled Liquids},
  chapter   = {2},
  pages     = {31--113},
  doi       = {10.1002/9781118202470.ch2},
  year      = {2012},
}

@article{sosa2022efficient,
  title={Efficient algorithm for simulating particles in true quasiperiodic environments},
  author={Sosa, Alan Rodrigo Mendoza and Kraemer, Atahualpa S},
  journal={Journal of Physics A: Mathematical and Theoretical},
  volume={55},
  number={24},
  pages={245001},
  year={2022},
  publisher={IOP Publishing}
}

@article{sosa2023structural,
  title={Structural studies of local environments in high-symmetry quasicrystals},
  author={Sosa, Alan Rodrigo Mendoza and Kraemer, Atahualpa S and O{\u{g}}uz, Erdal C and Schmiedeberg, Michael},
  journal={Scientific Reports},
  volume={13},
  number={1},
  pages={16696},
  year={2023},
  publisher={Nature Publishing Group UK London}
}

@article{oguz2017hyperuniformity,
  title={Hyperuniformity of quasicrystals},
  author={O{\u{g}}uz, Erdal C and Socolar, Joshua ES and Steinhardt, Paul J and Torquato, Salvatore},
  journal={Physical review B},
  volume={95},
  number={5},
  pages={054119},
  year={2017},
  publisher={APS}
}

@article{koga2024hyperuniformity,
  title={Hyperuniformity in two-dimensional periodic and quasiperiodic point patterns},
  author={Koga, Akihisa and Sakai, Shiro},
  journal={Physical Review E},
  volume={109},
  number={4},
  pages={044103},
  year={2024},
  publisher={APS}
}

@article{DeBruijn1981algebraic,
  title={Algebraic theory of Penrose's non-periodic tilings of the plane. I, II: dedicated to G. P{\'o}lya},
  author={De Bruijn, Nicolaas Govert},
  journal={Indagationes mathematicae},
  volume={43},
  number={1},
  pages={39--66},
  year={1981},
  publisher={Elsevier BV}
}

@article{torquato2018hyperuniform,
  title={Hyperuniform states of matter},
  author={Torquato, Salvatore},
  journal={Physics Reports},
  volume={745},
  pages={1--95},
  year={2018},
  publisher={Elsevier}
}

@article{socolar1986quasicrystals,
  title={Quasicrystals. II. Unit-cell configurations},
  author={Socolar, Joshua ES and Steinhardt, Paul J},
  journal={Physical Review B},
  volume={34},
  number={2},
  pages={617},
  year={1986},
  publisher={APS}
}

@article{shechtman1984metallic,
  title={Metallic phase with long-range orientational order and no translational symmetry},
  author={Shechtman, Dan and Blech, Ilan and Gratias, Denis and Cahn, John W},
  journal={Physical review letters},
  volume={53},
  number={20},
  pages={1951},
  year={1984},
  publisher={APS}
}

@article{levine1984quasicrystals,
  title={Quasicrystals: a new class of ordered structures},
  author={Levine, Dov and Steinhardt, Paul Joseph},
  journal={Physical review letters},
  volume={53},
  number={26},
  pages={2477},
  year={1984},
  publisher={APS}
}

@article{goldman1993quasicrystals,
  title={Quasicrystals and crystalline approximants},
  author={Goldman, AI and Kelton, RF},
  journal={Reviews of modern physics},
  volume={65},
  number={1},
  pages={213},
  year={1993},
  publisher={APS}
}

@article{matsubara2024aperiodic,
  title={Aperiodic approximants bridging quasicrystals and modulated structures},
  author={Matsubara, Toranosuke and Koga, Akihisa and Takano, Atsushi and Matsushita, Yushu and Dotera, Tomonari},
  journal={Nature Communications},
  volume={15},
  number={1},
  pages={5742},
  year={2024},
  publisher={Nature Publishing Group UK London}
}

@article{PhysRevLett.79.3363,
  title = {Quasiperiodic Optical Lattices},
  author = {Guidoni, L. and Trich\'e, C. and Verkerk, P. and Grynberg, G.},
  journal = {Phys. Rev. Lett.},
  volume = {79},
  issue = {18},
  pages = {3363--3366},
  numpages = {0},
  year = {1997},
  month = {Nov},
  publisher = {American Physical Society},
  doi = {10.1103/PhysRevLett.79.3363},
  url = {https://link.aps.org/doi/10.1103/PhysRevLett.79.3363}
}

@article{Jagannathan_2013,


doi = {10.1209/0295-5075/104/66003},
url = {https://doi.org/10.1209/0295-5075/104/66003},
year = {2014},
month = {jan},
publisher = {EDP Sciences, IOP Publishing and Società Italiana di Fisica},
volume = {104},
number = {6},
pages = {66003},
author = {Jagannathan, Anuradha and Duneau, Michel},
title = {An eightfold optical quasicrystal with cold atoms},
journal = {Europhysics Letters}
}

@article{mikhael2010proliferation,
  title={Proliferation of anomalous symmetries in colloidal monolayers subjected to quasiperiodic light fields},
  author={Mikhael, Jules and Schmiedeberg, Michael and Rausch, Sebastian and Roth, Johannes and Stark, Holger and Bechinger, Clemens},
  journal={Proceedings of the National Academy of Sciences},
  volume={107},
  number={16},
  pages={7214--7218},
  year={2010},
  publisher={National Academy of Sciences}
}

@article{burns1990optical,
  title={Optical matter: crystallization and binding in intense optical fields},
  author={Burns, Michael M and Fournier, Jean-Marc and Golovchenko, Jene A},
  journal={Science},
  volume={249},
  number={4970},
  pages={749--754},
  year={1990},
  publisher={American Association for the Advancement of Science}
}

@article{mikhael2008archimedean,
  title={Archimedean-like tiling on decagonal quasicrystalline surfaces},
  author={Mikhael, Jules and Roth, Johannes and Helden, Laurent and Bechinger, Clemens},
  journal={Nature},
  volume={454},
  number={7203},
  pages={501--504},
  year={2008},
  publisher={Nature Publishing Group UK London}
}

@article{fischer2011colloidal,
  title={Colloidal quasicrystals with 12-fold and 18-fold diffraction symmetry},
  author={Fischer, Steffen and Exner, Alexander and Zielske, Kathrin and Perlich, Jan and Deloudi, Sofia and Steurer, Walter and Lindner, Peter and F{\"o}rster, Stephan},
  journal={Proceedings of the National Academy of Sciences},
  volume={108},
  number={5},
  pages={1810--1814},
  year={2011},
  publisher={National Academy of Sciences}
}

@article{talapin2009quasicrystalline,
  title={Quasicrystalline order in self-assembled binary nanoparticle superlattices},
  author={Talapin, Dmitri V and Shevchenko, Elena V and Bodnarchuk, Maryna I and Ye, Xingchen and Chen, Jun and Murray, Christopher B},
  journal={Nature},
  volume={461},
  pages={964--967},
  year={2009},
  publisher={Nature Publishing Group UK London}
}

@article{Fan2026,
  title={Ideal non-crystals as a distinct form of ordered states without symmetry breaking},
  author={Fan, X. and Xu, D. and Zhang, J. and Hu, H. and Tan, P. and Xu, N. and Tanaka, H. and Tong, H.},
  journal={Nat. Mater.},
  volume={25},
  number={},
  pages={1020–1027},
  year={2026},
  publisher={}
}

@article{Wang2025,
  title={Hyperuniform disordered solids with crystal-like stability},
  author={Wang, Y. and Qian, Z. and Tong, H. and Tanaka, H.},
  journal={Nat Commun},
  volume={16},
  number={},
  pages={1398},
  year={2025},
  publisher={}
}

@article{Corwin2026,
  title={Ideal Glass and Ideal Disk Packing in Two Dimensions},
  author={Viola M. Bolton-Lum and R. Cameron Dennis and Peter K. Morse and Eric I. Corwin},
  journal={Phys. Rev. Lett.},
  volume={136},
  number={},
  pages={058201},
  year={2026},
  publisher={}
}

@article{Uri2023,
  title={Superconductivity and strong interactions in a tunable moiré quasicrystal},
  author={Aviram Uri and Sergio C. de la Barrera and Mallika T. Randeria and Daniel Rodan-Legrain and Trithep Devakul and Philip J. D. Crowley and Nisarga Paul and Kenji Watanabe and Takashi Taniguchi and Ron Lifshitz and Liang Fu and Raymond C. Ashoori and Pablo Jarillo-Herrero},
  journal={Nature},
  volume={620},
  number={},
  pages={762–767},
  year={2023},
  publisher={}
}

@article{Casiulis2025,
  title={Gyromorphs: A New Class of Functional Disordered Materials},
  author={Mathias Casiulis and Aaron Shih and Stefano Martiniani},
  journal={Phys. Rev. Lett.},
  volume={135},
  number={},
  pages={196101},
  year={2025},
  publisher={}
}

@article{Anderson72,
  title={Anomalous low-temperature thermal properties of glasses and spin glasses},
  author={P. W. Anderson and B. I. Halperin and C. M. Varma},
  journal={Philosophical Magazine},
  volume={25},
  number={},
  pages={1-9},
  year={1972},
  publisher={}
}

@article{Phillips72,
  title={Tunneling states in amorphous solids},
  author={W. A. Phillips},
  journal={Journal of Low Temperature Physics},
  volume={7},
  number={},
  pages={351–360},
  year={1972},
  publisher={}
}

@article{Mirkin2024,
  title={Colloidal quasicrystals engineered with DNA},
  author={Wenjie Zhou and Yein Lim and Haixin Lin and Sangmin Lee and Yuanwei Li and Ziyin Huang and Jingshan S. Du and Byeongdu Lee and Shunzhi Wang and Ana Sánchez-Iglesias and Marek Grzelczak and Luis M. Liz-Marzán and Sharon C. Glotzer and Chad A. Mirkin},
  journal={Nat. Mater.},
  volume={23},
  number={},
  pages={424-428},
  year={2024},
  publisher={}
}

@article{Marrows2018,
  title={Frustration and thermalization in an artificial magnetic quasicrystal},
  author={Dong Shi and Zoe Budrikis and Aaron Stein and Sophie A. Morley and Peter D. Olmsted and Gavin Burnell and Christopher H. Marrows},
  journal={Nature Phys},
  volume={14},
  number={},
  pages={309–314},
  year={2018},
  publisher={}
}

@article{Pinto2025,
  author    = {Pinto, Diogo E. P. and {\v{S}}ulc, Petr and Sciortino, Francesco and Russo, John},
  title     = {Automating Blueprints for the Assembly of Colloidal Quasicrystal Clusters},
  journal   = {ACS Nano},
  year      = {2025},
  volume    = {19},
  number    = {1},
  pages     = {512--519},
  doi       = {10.1021/acsnano.4c10434},
  month     = jan,
}

@article{Noya2025,
  author    = {Noya, Eva G. and Doye, Jonathan P. K.},
  title     = {A One-Component Patchy-Particle Icosahedral Quasicrystal},
  journal   = {ACS Nano},
  year      = {2025},
  volume    = {19},
  number    = {14},
  pages     = {13714--13722},
  doi       = {10.1021/acsnano.4c14885},
  month     = apr,
  pmid      = {40168641},
  pmcid     = {PMC12004934},
}

@article{Kamiya2018,
  author  = {Kamiya, K. and Takeuchi, T. and Kabeya, N. and Wada, N. and Ishimasa, T. and Ochiai, A. and Deguchi, K.
  and Imura, K. and Sato, N. K.},
  title   = {Discovery of superconductivity in quasicrystal},
  journal = {Nature Communications},
  volume  = {9},
  pages   = {154},
  year    = {2018},
  doi     = {10.1038/s41467-017-02667-x}
}

@article{Smallenburg2024,
  author    = {Fayen, Etienne and Filion, Laura and Foffi, Giuseppe and Smallenburg, Frank},
  title     = {Quasicrystal of Binary Hard Spheres on a Plane Stabilized by Configurational Entropy},
  journal   = {Physical Review Letters},
  year      = {2024},
  volume    = {132},
  number    = {4},
  pages     = {048202},
  doi       = {10.1103/PhysRevLett.132.048202},
}

@article{Dotera2014,
  author    = {Dotera, T. and Oshiro, Tammy Y. and Ziherl, P.},
  title     = {Mosaic two-lengthscale quasicrystals},
  journal   = {Nature},
  year      = {2014},
  volume    = {506},
  number    = {7487},
  pages     = {208--211},
  doi       = {10.1038/nature12938},
}

@article{Engel2015,
  author    = {Engel, Michael and Damasceno, Pablo F. and Phillips, Carolyn L. and Glotzer, Sharon C.},
  title     = {Computational self-assembly of a one-component icosahedral quasicrystal},
  journal   = {Nature Materials},
  year      = {2015},
  volume    = {14},
  number    = {1},
  pages     = {109--116},
  doi       = {10.1038/nmat4152},
}

@article{Freedman2006,
  author    = {Freedman, Barak and Bartal, Guy and Segev, Mordechai and Lifshitz, Ron and Christodoulides, Demetrios N. and Fleischer, Jason W.},
  title     = {Wave and defect dynamics in nonlinear photonic quasicrystals},
  journal   = {Nature},
  year      = {2006},
  volume    = {440},
  number    = {7088},
  pages     = {1166--1169},
  doi       = {10.1038/nature04722},
}

@article{Ahn2018,
  author    = {Ahn, Sung Joon and Moon, Pilkyung and Kim, Tae-Hoon and Kim, Hyun-Woo and Shin, Ha-Chul and Kim, Eun Hye and Cha, Hyun Woo and Kahng, Se-Jong and Kim, Philip and Koshino, Mikito and Son, Young-Woo and Yang, Cheol-Woong and Ahn, Joung Real},
  title     = {Dirac electrons in a dodecagonal graphene quasicrystal},
  journal   = {Science},
  year      = {2018},
  volume    = {361},
  number    = {6404},
  pages     = {782--786},
  doi       = {10.1126/science.aar8412},
}

@Article{LinCorrigendum2017,
  author    = {Lin, C and Steinhardt, P J and Torquato, S},
  journal   = {Journal of Physics: Condensed Matter},
  title     = {Corrigendum: Hyperuniformity variation with quasicrystal local isomorphism class (2017 J. Phys. Condens. Matter 29 204003)},
  year      = {2017},
  issn      = {1361-648X},
  month     = Nov,
  number    = {47},
  pages     = {479501},
  volume    = {29},
  doi       = {10.1088/1361-648x/aa8430},
  publisher = {IOP Publishing},
}

@Article{VanderPlas2018,
  author    = {VanderPlas, Jacob T.},
  journal   = {The Astrophysical Journal Supplement Series},
  title     = {Understanding the {L}omb–{S}cargle Periodogram},
  year      = {2018},
  issn      = {1538-4365},
  month     = May,
  number    = {1},
  pages     = {16},
  volume    = {236},
  doi       = {10.3847/1538-4365/aab766},
  publisher = {American Astronomical Society},
}

@article{klatt2022,
  author    = {Michael A. Klatt and Paul J. Steinhardt and Salvatore Torquato},
  title     = {Wave propagation and band tails of two-dimensional disordered systems in the thermodynamic limit},
  journal   = {Proceedings of the National Academy of Sciences},
  volume    = {119},
  number    = {52},
  pages     = {e2213633119},
  year      = {2022},
  doi       = {10.1073/pnas.2213633119},
}

@article{Torquato2015,
  author    = {Torquato, S. and Zhang, G. and Stillinger, F. H.},
  title     = {Ensemble Theory for Stealthy Hyperuniform Disordered Ground States},
  journal   = {Phys. Rev. X},
  volume    = {5},
  issue     = {2},
  pages     = {021020},
  year      = {2015},
  month     = {May},
  doi       = {10.1103/PhysRevX.5.021020},
  publisher = {American Physical Society}
}

@article{Steinhardt1996,
  author    = {Steinhardt, Paul J. and Jeong, Hyeong-Chai},
  title     = {A simpler approach to Penrose tiling with implications for quasicrystal formation},
  journal   = {Nature},
  volume    = {382},
  pages     = {431--433},
  year      = {1996},
  month     = {August},
  doi       = {10.1038/382431a0}
}

\end{document}